\documentclass[extra] {gji}

\pdfoutput=1

\usepackage{xcolor}
\usepackage[colorlinks=True,%
            urlcolor=blue,%
            linkcolor=blue,%
            citecolor= blue,%
            pdfauthor={O. Barrois},%
            pdftitle={},%
            pdftex]{hyperref}
            
\usepackage{timet}
\usepackage{graphicx}
\usepackage{amsmath,amsfonts,amssymb}
\usepackage{longtable}
\usepackage{caption}
\usepackage{bm}

 \usepackage{lineno}

\renewcommand{\cite}{\citet}

\newcommand{\new}[1]{#1}
\newcommand{\newr}[1]{#1}

\title[QG, Hybrid and 3D core convection models]
	{Comparison of Quasi-Geostrophic, Hybrid and 3D models of planetary core convection}
\author[Barrois et al.]
  {O. Barrois$^1$, T. Gastine$^2$, C.~C. Finlay$^1$ \\
  $^1$ Division of Geomagnetism, DTU Space, Technical University of Denmark, Lyngby DK-2800, Denmark.\\
  $^2$ Universit\'e de Paris, Institut de Physique du Globe de Paris, CNRS, F-75005 Paris, France.}
\date{Received XXX; in original form XXX}
\pagerange{\pageref{firstpage}--\pageref{lastpage}}
\volume{XXX}
\pubyear{2022}

\begin{document}

\label{firstpage}

\maketitle

\begin{summary}

We present investigations of rapidly-rotating convection in a thick spherical shell geometry relevant to planetary cores, comparing results from Quasi-Geostrophic (QG), 3D and hybrid QG-3D models.
The $1\new{70}$ reported calculations span Ekman numbers, $Ek$, between $10^{-4}$ and $10^{-10}$, Rayleigh numbers, $Ra$, between $2$ and $150$ times supercritical, and Prandtl numbers, $Pr$, between $10$ and $10^{-2}$.
The default boundary conditions are no-slip at both the ICB and the CMB for the velocity field, with fixed temperatures at the ICB and the CMB.
Cases driven by both homogeneous and inhomogeneous CMB heat flux patterns are also explored, the latter including lateral variations, as measured by $Q^*$, the peak-to-peak amplitude of the pattern divided by its mean, taking values up to $5$.
The Quasi-Geostrophic (QG) model is based on the open-source {\tt pizza} code.  We extend this in a hybrid approach to include the temperature field on a 3D grid.
In general, we find convection is dominated by zonal jets at mid-depths in the shell, with thermal Rossby waves prominent close to the outer boundary when the driving is weaker.
For the thick spherical shell geometry studied here the hybrid method is best suited for studying convection at modest forcing, $Ra \leq 10 \, Ra_c$ \new{when $Pr=1$}, and departs from the 3D model results at higher $Ra$, displaying systematically lower heat transport characterized by lower Nusselt and Reynolds numbers.
\new{We find that the lack of equatorially anti-symmetric and \newr{$z$-correlations between temperature and velocity} in the buoyancy force contributes to the weaker flows in the hybrid formulation.}
On the other hand, the QG models yield broadly similar results to the 3D models\new{, for the specific aspect ratio and range of Rayleigh numbers explored here\newr{. 
W}e cannot point to major disagreements between these two datasets \newr{at $Pr \geq 0.1$, although} the QG model is effectively more strongly driven than the hybrid case due to its cylindrically-averaged thermal boundary conditions.}
\new{When $Pr$ is decreased, the range of agreement between the Hybrid and 3D models expands, {\it e.g.} up to $Ra \leq 15 \, Ra_c$ at $Pr=0.1$, indicating the hybrid method may be better suited to study convection in the regime \newr{$Pr\ll 1$}.}
\newr{We effectively observe two regimes: (i) at $Pr \geq 0.1$ the QG and 3D models agree in the studied range of $Ra/Ra_c$ while the hybrid model fails when $Ra> 10\,Ra_c$; (ii) at $Pr = 0.01$ the QG and 3D disagree above $Ra/Ra_c = 10$ while the hybrid and 3D models agree fairly well up to $Ra \sim 20\,Ra_c$.}
Models that include laterally-varying heat flux at the outer boundary reproduce regional convection patterns that compare well with those found in similarly forced 3D models.
Previously proposed scaling laws for rapidly-rotating convection are tested; our simulations are overall well described by a triple balance between Coriolis, inertia and Archimedean (CIA) forces with the length-scale of the convection following the diffusion-free Rhines-scaling.
The Prandtl number, $Pr$, 
affects the number and the size of the jets with larger structures obtained at lower $Pr$; higher velocities and lower heat transport are also seen on decreasing $Pr$.
The scaling behaviour of the convective velocity shows a strong dependence on $Pr$.
This study is an intermediate step towards a hybrid model of core convection also including 3D magnetic effects.

\end{summary}


\section{Introduction}
\label{sec:state_of_art}

Many celestial bodies such as rocky and gas planets of the Solar system are rapidly rotating.
The effects of rotation on fluid systems have been widely studied \--- it impedes the onset of the convection \citep{chandrasekhar1961hydrodynamic}, constrains heat transport \citep{rossby1969study} and shapes the convection into the form of thin columns nearly invariant along the rotation axis \citep{busse1970thermal}.
Such convective flows are subject to a zeroth order Geostrophic balance between the Coriolis force and the pressure gradient that arises when $Ek \ll 1$ and $Ro \ll 1$ \citep{julien2012heat}\new{, and when the typical time-scale of the convection is much longer that the rotation period}, where $Ek = \nu / \Omega d^2$ measures the viscous effects compared to the Coriolis force and $Ro = Re Ek$ measures the nonlinear inertial effects compared to the Coriolis force.
Here $Re = U d / \nu$ is the Reynolds number with $\nu$ the kinematic viscosity of the fluid, $\Omega$ the angular velocity of the planet, $d$ the typical size of the fluid container and $U$ the typical fluid velocity.

In this study we focus on the convective dynamics relevant for the Earth's outer core.
It is expected to be in a strongly-driven state, 
with $Re \gg 1$ and turbulent convection \citep[see ][for a review]{roberts2013genesis}.
Such a regime is extremely challenging to explore both experimentally and numerically, because of the important non-linearities and the necessity of resolving fast rotational dynamics while wishing to track the evolution of long-lived jets and vortices \citep[{\it e.g.},][]{stellmach2014approaching,aurnou2015rotating,gastine2016scaling}.
The relevant dimensionless parameters for the Earth's core are thought to be $Ek \sim 10^{-15}$, $Re \sim 10^9$ and $Ra/Ra_c \gg 10^3$ where $Ra$ is the Rayleigh number and $Ra_c$ is the critical Rayleigh number for the onset of convection. The most ambitious 3D simulations and experiment\new{al} studies are only able to reach $Ek \sim 10^{-7}$; $Ra/Ra_c \sim 10^{3}$ and $Re \sim 5\times 10^3$ \citep{aubert2015geomagnetic,schaeffer2017turbulent,sheyko2018scale}.

One alternative avenue for studying this challenging regime is to use reduced quasi-geostrophic (QG) convection models.
In their classical form QG models consider perturbations about a leading order balance between Coriolis and Pressure gradient forces, whose axial vorticity is invariant along the rotation axis.
The dynamics is then essentially confined to the equatorial plane. \citet{busse1970thermal} initially developed QG models in an annulus geometry assuming a small boundary slope.
The QG framework was later modified and extended to better account for spherical geometry and to include phenomenon such as Ekman pumping by \citet{cardin1994chaotic,aubert2003quasigeostrophic,schaeffer2005quasigeostrophic,gillet2006quasi,calkins2012influence}.
With such QG models it has been possible to investigate rotating convection for Ekman numbers as low as $Ek = 10^{-11}$ close to the onset of convection \citep{guervilly2019turbulent}.

Such QG models are essentially a 2D approximation of the real 3D situation.
The 2D treatment of temperature often used in QG models is not rigorously justified \citep[see, {\it e.g.},][]{gillet2006quasi} and fails to capture thermal wind contributions.
Furthermore such classical QG models focus on an axially-invariant approximation of the axial vorticity and on related horizontal flows in the equatorial plane. In spherical geometry they perform worst close to the outer boundary where the boundary slope becomes large, or when the forcing becomes large enough that the vertical velocity becomes significant, and 3D motions set in \citep{calkins2013three}.
More advanced extensions of the QG framework have recently been proposed where the full velocity field is better accounted for by projecting onto a QG basis \citep{labbe2015magnetostrophic,maffei2017characterization,gerick2020pressure} or by $z$-averaging before taking the curl \citep{jackson2020plesio}.
In this work, we follow \citet{gastine2019pizza} and use the QG formulation proposed by \citet{schaeffer2006quasi} that was expanded in a hybrid approach by \citet{guervilly2016subcritical,guervilly2017multiple} to also include a 3D temperature field.
Our numerical implementation of this hybrid QG-3D method \new{(or simply Hybrid)} is an extension of the {\tt pizza} code by \citet{gastine2019pizza} to include a 3D  temperature field in a spherical shell geometry.
Here we explore advantages and limitations of QG and hybrid QG-3D models compared with full 3D core convection models.

Thermal boundary conditions may play an important role in convection in planetary cores. Strictly speaking these boundary conditions are not fixed but time-dependent, and coupling to compositional effects should be considered \citep[][]{glatzmaier1996evolutionary}.
In practice however, when considering Earth's core, fixed heat-flux conditions at the core-mantle boundary and fixed temperature conditions at the inner core boundary are often argued to be relevant \citep[{\it e.g.},][]{gubbins2003thermal}.
Early studies with heat-flux boundary conditions suggested these might promote slightly longer wavelengths and larger convective flows \citep{gibbons2007convection,sakuraba2009generation} although such discrepancies have more recently been attributed to different levels of forcing \citep{yadav2016effect,schwaiger2020relating}.
This is consistent with an asymptotic equivalence between heat flux and temperature boundary conditions \citep{calkins2015equivalence} \new{with standard universal scaling laws retrieved far from the onset in both cases \citep{clarte2021effects}}.
More dramatic effects are possible when the heat flux boundary conditions vary laterally, which in some locations will enhance heat-transport and can result in a preservation of large-scale downwelling systems \citep{mound2017heat,long2020scaling,sahoo2020convection}.
The above statements are primarily based on studies carried out at moderate Ekman numbers ($Ek\geq2\times 10^{-6}$) that were often weakly or moderately driven.
Here, although the majority of our simulations use fixed temperature boundary conditions, we report results of a number of calculations with imposed heat flux outer boundary condition, including inhomogeneous cases where this varies laterally.
We examine whether inhomogeneous boundary conditions continue to impact the convective pattern in more strongly-driven cases at Ekman numbers slightly smaller than those considered in previous studies ($Ek\geq 10^{-6}$ with our hybrid approach and $Ek\geq 10^{-7}$ with our QG approach) 
and investigate whether QG and hybrid QG-3D models can capture relevant aspects of convection in such cases.

Scaling laws describe how global quantities characterizing the convection, such as the convective length scale, flow speed and heat transport, vary with the control parameters based on the underlying dynamics \citep[{\it e.g.}, ][]{gillet2006quasi,king2013flow,gastine2016scaling}.
Two theoretical scaling laws have attracted much attention for describing the properties of rapidly-rotating convective flows: one based on a triple balance between the Coriolis, Inertia and Archimedean forces -- called the CIA-scaling \citep{ingersoll1982motion,cardin1994chaotic} -- and the other based on a triple balance between Viscous, Archimedean and Coriolis forces -- called the VAC-scaling \citep{king2013flow}.
Early studies at modest rotation rates had difficulty in distinguishing between the two scalings \citep{aubert2001systematic,gillet2006quasi,king2013flow}, but more recent investigations have shown a preference for the CIA balance in the bulk of the fluid, away from viscous boundary layers \citep{gastine2016scaling,long2020scaling,schwaiger2020relating}.
An impressive convergence towards the viscous-free scaling in the limit of low viscosity and close to the onset of convection has also recently been described by \cite{guervilly2019turbulent} in the context of fluids with Prandtl numbers $Pr < 1$.
Here our main goal is to complement these studies, using QG, 3D and hybrid QG-3D simulations in a thick spherical shell geometry, focusing on relatively strongly-driven cases (high $Ra/Ra_c$) and exploring the role of the Prandtl number, which 
may influence the typical size of the convective pattern \citep{calkins2012influence,king2013turbulent, guervilly2016subcritical}.

The article is organised as follows:
Section \ref{sec:Method} presents the equations and the methodology of our QG, hybrid QG-3D, and 3D models.
Section \ref{sec:Results} presents results obtained using our models 
focusing on comparisons between QG, Hybrid and 3D calculations, and including cases with inhomogoneous heat flux boundary conditions.
We also describe the impact of Prandtl number on the form of convection at low Rossby number and examine how well our results satisfy convective scaling laws.
We conclude with a discussion and a summary of our findings in Section \ref{sec:Conclusion}.


\section{Methodology}
\label{sec:Method}

\subsection{Quasi-Geostrophic model formulation}
\label{sec:QG_equations}

In this section we first describe the basic QG model employed before moving on to the new 3D modifications we have implemented.
We use the same QG model formulation and notation as \citet{gastine2019pizza}, who followed closely the approach set out by \citet{schaeffer2005quasigeostrophic} and \citet{gillet2006quasi}.
We work in cylindrical coordinates $\left( s, \phi, z \right)$ in a spherical shell between the inner radius $s_i$ and the outer radius $s_o$, rotating about the $z$-axis with a constant angular velocity $\Omega$.
The horizontal components of the velocity field ${\bm u}_\perp$, perpendicular to the rotation axis, are assumed to be invariant along the rotation axis, {\it i.e.} ${\bm u}_\perp=(u_s, u_\phi, 0)$\new{, where $u_s$ and $u_\phi$ are respectively the radial and azimuthal velocities}. 
It is further assumed that the dynamics is encapsulated by the evolution of the axial vorticity averaged in the $z$ direction, such that the dynamics is restricted to that in the equatorial plane of the spherical shell \citep{maffei2017characterization}.
Below we refer to this as the classical QG model in order to distinguish it from recently developed variants \citep{labbe2015magnetostrophic,gerick2020pressure, jackson2020plesio}.  

Non-dimensionalization is carried out using the shell thickness $d = s_o - s_i$ as the reference length-scale, the viscous diffusion time $d^2 / \nu$ as the reference time-scale, and the temperature contrast between the boundaries $\Delta T = T_i - T_o = T(s_i) - T(s_o)$ as the reference for temperature.
Throughout this study, we adopt $\eta = s_i / s_o = 0.35$ suitable for a thick shell such as the Earth's outer core.
The gravity ${\bm g}$ is assumed to be linear with respect to the cylindrical radius such that $g(s) \propto s$ and it is non-dimensionalized based on its value at the outer boundary $g_o = g(s_o)$.

Following \citet{schaeffer2005quasigeostrophic} and \citet{gastine2019pizza} it is assumed that the axial velocity $u_z$ varies linearly with $z$ in the direction of the rotation axis, including contributions from \new{the radial velocity} $u_s$ and the Ekman pumping \new{${\cal P}$}, {\it i.e.}
\begin{align}
\label{eq:uz_linearity}
u_z(s,\phi,z) = z \left[ \beta u_s + \mathcal{P}(Ek, {\bm u}_\perp, \omega_z) \right]\,,
\end{align}
with ${\cal P}(Ek, {\bm u}_\perp, \omega_z)$ the Ekman pumping term deduced from Greenspan's formula in a rigid sphere (see Eq.~\ref{eq:Ek_pump}), 
\begin{align}
\label{eq:beta}
\beta = \dfrac{1}{h}\dfrac{\mathrm{d} h}{\mathrm{d} s} = - \dfrac{s}{h^2}\,,
\end{align}
and $h \equiv \sqrt{s_o^2 - s^2}$, the half-height of a cylinder aligned with the rotation axis at a radius $s$.
The spherical-QG continuity equation then reads
\begin{align}
\label{eq:QG_approximation}
\dfrac{\partial (s u_s)}{\partial s} + \dfrac{\partial u_\phi}{\partial \phi} + \beta s u_s = 0\,.
\end{align}

We represent the non-axisymmetric QG-velocity by a streamfunction $\psi$ such that
\begin{align}
\label{eq:QG_streamfunct}
u_s = \dfrac{1}{s}\dfrac{\partial \psi}{\partial \phi}\,, \; \; u_\phi = \overline{u_\phi} - \dfrac{\partial \psi}{\partial s} - \beta \psi\,,
\end{align}
which ensures that (\ref{eq:QG_approximation}) is satisfied.  $\overline{u_\phi}$ is the remaining axisymmetric zonal flow component. The overbar $\overline{x}$ denotes the azimuthal average of any quantity $x$, such that
\begin{align}
\label{eq:phi_average}
\overline{x} \equiv \dfrac{1}{2 \pi} \displaystyle\int_{0}^{2\pi}  x\, \mathrm{d}\phi\,.
\end{align}
The axial vorticity $\omega_z = {\bm e}_z \cdot {\bm \nabla} \times {\bm u}$ is then 
\begin{align}
\label{eq:QG_vortz-psi}
\omega_z = \dfrac{1}{s}\dfrac{\partial (s \overline{u_\phi})}{\partial s} - \nabla^2 \psi - \dfrac{1}{s}\dfrac{\partial (\beta s \psi)}{\partial s}\,.
\end{align}
The dynamics of the axial vorticity can then be described by the axial component of the curl of the momentum equation in cylindrical coordinates, averaged in the $z$-axis direction, which due to the assumed 2D form of the contributing fields, may be written
\begin{align}
\label{eq:momentum_QG}
\dfrac{\partial \omega_z}{\partial t} + {\bm \nabla}_\perp \cdot \left( {\bm u}_\perp \omega_z \right) - \dfrac{2}{Ek} \beta u_s = \nabla_\perp^2 \omega_z - \dfrac{Ra}{Pr} \dfrac{1}{s_o} \dfrac{\partial \vartheta_{2D}}{\partial \phi} \\
+ \mathcal{P}(Ek, {\bm u}_\perp, \omega_z)\,, \nonumber
\end{align}
where the subscript $_\perp$ corresponds to the horizontal part of the operators -- {\it e.g.}, ${\bm \nabla}_\perp
= \left( {\bm e}_s \cdot \partial_s  + r^{-1} {\bm e}_\phi \cdot \partial_\phi \right)\,,  \; \; {\bm u}_\perp = \left( u_s, u_\phi, 0 \right)$ -- and $\mathcal{P}(Ek, {\bm u}_\perp, \omega_z)$
corresponds to the Ekman-pumping contribution \citep{schaeffer2005quasigeostrophic} for the non-axisymmetric motions
\begin{align}
\label{eq:Ek_pump}
\mathcal{P}(Ek, {\bm u}_\perp, \omega_z) = - \Pi \left[ \omega_z - \dfrac{\beta}{2} u_\phi + \beta \dfrac{\partial}{\partial \phi} u_s - \dfrac{5 s_o}{2h} u_s \right]\,,
\end{align}
with
\begin{align}
\label{eq:Ek_pump_fac}
\Pi = \left(\dfrac{s_o}{Ek}\right)^{1/2} \dfrac{1}{h^{3/2}}\,.
\end{align}
The non-dimensional control parameters, the Ekman number, the Rayleigh number and the Prandtl number are respectively defined by
\begin{align}
\label{eq:adim_par}
Ek = \dfrac{\nu}{\Omega d^2}\,, \; Ra = \dfrac{\alpha_T g_o \Delta T d^3}{\kappa \nu}\,, \; Pr = \dfrac{\nu}{\kappa}\,,
\end{align}
where $\alpha_T$ is the thermal expansion coefficient, $\nu$ is the kinematic viscosity, and $\kappa$ is the thermal diffusivity.

The $z$-averaged axial vorticity equation (\ref{eq:momentum_QG}) has to be supplemented by an equation to account for the axisymmetric motions.
This is obtained by taking the $\phi$-average of the azimuthal component of the Navier-Stokes equation and reads
\begin{align}
\label{eq:zonal_QG}
\dfrac{\partial \overline{u_\phi}}{\partial t} + \displaystyle\overline{u_s \dfrac{\partial u_\phi}{\partial s}} + \displaystyle\overline{\dfrac{u_s u_\phi}{s}} = \nabla_\perp^2 \overline{u_\phi} - \dfrac{1}{s^2} \overline{u_\phi} - \Pi\,\overline{u_\phi}\,,
\end{align}
where the last term of the right-hand-side is the Ekman-pumping contribution for the axisymmetric motions.
The boundary conditions for the velocity field are described in detail in \S\ref{sec:BCs}.

The other coupled prognostic equation used to complete the system is the QG-temperature perturbation equation, 
\begin{align}
\label{eq:heat_QG}
\dfrac{\partial \vartheta_{2D}}{\partial t} + {\bm \nabla}_\perp  \cdot ({\bm u}_\perp \vartheta_{2D}) + \beta u_s \vartheta_{2D} + u_s \dfrac{\mathrm{d} T_{2D}^{\text{cond}}}{\mathrm{d} s} = \dfrac{1}{Pr} \nabla_\perp^2 \vartheta_{2D}\,,
\end{align}
where the temperature is written as a perturbation about a mean 2D conducting state, {\it i.e.} $T_{2D} = T_{2D}^{\text{cond}} + \vartheta_{2D}$, where $T_{2D}^{\text{cond}}$ is the conducting background state, a solution of $\nabla^2 T_{2D}^{\text{cond}} = 0$ subject to the chosen boundary conditions.
For fixed-temperature boundary conditions at $s_i$ and $s_o$ this yields
\begin{align}
\label{eq:T_2D_cond}
T_{2D}^{\text{cond}} = \dfrac{1}{\ln \eta} \ln \left[(1 - \eta) s \right], \; \dfrac{\mathrm{d} T_{2D}^{\text{cond}}}{\mathrm{d} s} = \dfrac{1}{s \ln \eta}\,.
\end{align}
Further details on the boundary conditions for the temperature field, including the possibility of heat flux boundary conditions are given in \S\ref{sec:BCs}.

\subsection{\new{Extended} hybrid QG-3D model}
\label{sec:Hyb_equations}

In this study we follow \cite{guervilly2010thesis} and \cite{guervilly2016subcritical} and go beyond the classical QG model presented in the previous section to develop a hybrid QG-3D model in which the QG perturbation temperature equation (\ref{eq:heat_QG}) is replaced by the full 3D temperature equation
\begin{align}
\label{eq:heat_3D}
\dfrac{\partial \vartheta_{3D}}{\partial t} + {\bf u}_{3D} \cdot {\bm \nabla} \vartheta_{3D} + u_r \dfrac{\mathrm{d} T^{\text{cond}}_{3D}}{\mathrm{d}r} =
\dfrac{1}{Pr} \nabla^2 \vartheta_{3D}\,,
\end{align}
where $r$ is the spherical radius and ${\bm u}_{3D} = \left( u_r, u_\theta, u_{\phi 3D} \right)$ is the 3D-velocity in spherical coordinates.
Similarly to the QG case, we write the temperature as a perturbation temperature about the conducting background state, {\it i.e.} $T_{3D} = T_{3D}^{\text{cond}} + \vartheta_{3D}$.
We compute the conducting temperature profile, or dimensionless radial temperature profile, $T_{3D}^{\text{cond}}$, as the solution of $\nabla^2 T_{3D}^{\text{cond}} = 0$, which for a fixed temperature contrast between $r_i$ and $r_o$ without internal heating, yields
\begin{align}
\label{eq:T_3D_cond_Dormy_3D}
T_{3D}^{\text{cond}}(r) = \dfrac{r_o r_i}{r} - r_i,
\; \dfrac{\mathrm{d} T_{3D}^{\text{cond}}}{\mathrm{d} r} = -\dfrac{r_i r_o}{r^2}\,
.
\end{align}
This is the full 3D version of Eq.~(\ref{eq:T_2D_cond}) in spherical geometry.
We reconstruct the 3D-velocity field, ${\bm u}_{3D}$ \new{using the conversion between cylindrical and spherical coordinate systems as} needed in this equation, from the QG velocity field\new{, such that}
\begin{equation}
\label{eq:vel_3D-interp}
\left\lbrace\begin{aligned}
u_r(r,\theta,\phi_{3D})& = \sin\theta\, u_s(s,\phi) + \cos\theta\, u_z(s,\phi,z)\,, \\
u_\theta(r,\theta,\phi_{3D}) &= \cos\theta\, u_s(s,\phi) - \sin\theta\, u_z(s,\phi,z)\,, \\
u_{\phi 3D}(r,\theta,\phi_{3D})& = u_\phi(s,\phi)\,,
\end{aligned}\right.
\end{equation}
where $u_z$ is proportional to $z$ and incorporates the effects of the Ekman pumping (see Eq.~\ref{eq:uz_linearity}) consistent with our initial quasi-geostrophic assumption (Eq.~\ref{eq:QG_approximation})\new{, and where the cylindrical quantities are cast onto the 3D-grid using a bi-linear extrapolation (see Appendix~\ref{sec:3D-z-func-bench} for more details).
Inside the Tangent cylinder, the velocities are set to zero and thus only temperature diffusion occurs in that region.}

Considering a 3D temperature field, Eq.~(\ref{eq:momentum_QG}) becomes
\begin{align}
\label{eq:momentum_QG-hyb}
\dfrac{\partial \omega_z}{\partial t} +{\bm \nabla}_\perp \cdot \left( {\bm u}_\perp \omega_z \right) - \dfrac{2}{Ek} \beta u_s = \nabla_\perp^2 \omega_z - \dfrac{Ra}{Pr} \left\langle\new{ \dfrac{1}{r_o}} \dfrac{\partial T_{3D}}{\partial \phi_{3D}} \right\rangle \\
+ \mathcal{P}(Ek, {\bm u}_\perp, \omega_z)\,, \nonumber
\end{align}
where the angular brackets $\langle x \rangle$ refer to the axial or $z$-average of any quantity $x$ defined by
\begin{align}
\label{eq:z_average}
\langle x \rangle \equiv \dfrac{1}{2h} \displaystyle\int_{-h}^{h}  x \, \mathrm{d}z\,.
\end{align}

The equation for the zonal motions (\ref{eq:zonal_QG}) is not modified as it does not involve the temperature.

Since we treat the temperature in 3D, we can now also take into account the thermal wind contribution to the velocity field which results in an extra term being added to $u_{\phi 3D}$, and which satisfies
\begin{align}
\label{eq:thw-u_phi3D_unint}
\dfrac{\partial u_{\phi 3D}}{\partial z} = \dfrac{Ra Ek}{2 Pr} \dfrac{g(r)}{r} \dfrac{\partial T_{3D}}{\partial \theta}\,.
\end{align}
Integrating between the position $z$ and the height of the column above the equator, $h$, we obtain
\begin{align}
\label{eq:thw-u_phi3D}
u_{\phi 3D}(r, \theta, \phi_{3D}) = u_{\phi 3D}(h, \theta, \phi_{3D}) - \dfrac{Ra Ek}{2 Pr} \int_{z}^{h} \dfrac{g(r)}{r}\dfrac{\partial T_{3D}}{\partial \theta} \mathrm{d} z' \,,
\end{align}
where $g(r)=r/r_o$ is the 3D gravity field. Here, for efficiency, the thermal wind contribution is assumed to be symmetric about the equatorial plane, although this condition can be relaxed if needed depending on the chosen boundary conditions.

\new{Because of the full 3D treatment of the heat equation, the consideration of the thermal wind effects and the fact that the thermal boundary conditions are the same as in the 3D case, it is natural to expect the hybrid QG-3D model to behave better than the classical QG model when compared with a full-3D model, a hypothesis that will be further assessed in the Results Section.}

\subsection{3D model formulation}
\label{sec:3D_equations}

In order to compare the results of our QG and hybrid QG-3D models, we also consider a purely 3D model.
Similarly to the two previous setups, we consider convection of a Boussinesq fluid enclosed in a spherical shell of inner radius $r_i$ and outer radius $r_o$ rotating about the $z$ axis.
The same scales and dimensionless parameters are used and thus the 3D Navier-Stokes equations read
\begin{align}
\label{eq:mass-conservation_3D}
{\bm \nabla} {\bm \cdot} {\bm u}_{3D} = 0\,,
\end{align}
\begin{align}
\label{eq:momentum_3D}
\dfrac{\partial {\bm u}_{3D}}{\partial t} + {\bm u}_{3D} \cdot {\bm \nabla} {\bm u}_{3D} + \dfrac{2}{Ek} {\bm e}_z \times {\bm u}_{3D} = - \nabla P \\
+ \dfrac{Ra}{Pr} \dfrac{r}{r_o} \vartheta_{3D} {\bm e}_r
+ {\bm \nabla^2} {\bm u}_{3D}\,, \nonumber
\end{align}
\begin{align}
\label{eq:temp_3D}
\dfrac{\partial \vartheta_{3D}}{\partial t} + {\bm u}_{3D} \cdot {\bm \nabla} \vartheta_{3D} + u_r \dfrac{\mathrm{d} T^{\text{cond}}_{3D}}{\mathrm{d}r} =
\dfrac{1}{Pr} \nabla^2 \vartheta_{3D}\,,
\end{align}

where $P$ is the pressure, and ${\bm e}_{r,z}$ are respectively the unit vectors in the radial and the axial directions.

The velocity field, ${\bm u}_{3D}$, is decomposed into poloidal ${\cal W}$ and toroidal  ${\cal Z}$ potentials following
\begin{align}
\label{eq:u_torpol}
{\bm u}_{3D} = \nabla \times (\nabla \times {\cal W} {\bm e}_r) + \nabla \times {\cal Z} {\bm e}_r\,.
\end{align}

\subsection{Boundary Conditions}
\label{sec:BCs}


Since our focus is on modelling the dynamics of the Earth's outer core, we treat the fluid shell as a container with rigid, impenetrable, and co-rotating walls.
This implies that in the rotating frame of reference all velocity components should vanish at $s_o$ and $s_i$ in the QG or Hybrid models and at $r_i$ and $r_o$ in the 3D calculations.


Imposing fixed temperature at the boundaries yields, respectively for the QG- and the 3D-temperature field
\begin{align}
\label{eq:bc_temp}
\vartheta_{2D} = 0, \; \text{at } s = \left\lbrace s_i\,, s_o \right\rbrace\,, \\
\vartheta_{3D} = 0, \; \text{at } r = \left\lbrace r_i\,, r_o \right\rbrace\,. \nonumber
\end{align}
In this study, the majority of our simulations are conducted under these boundary conditions, but we also consider an other set of thermal boundary conditions with a fixed temperature at the inner radius and an imposed flux at the outer boundary. The latter involves
\begin{align}
\label{eq:bc_heat-flux}
\dfrac{\partial \vartheta}{\partial r} = 0, \; \text{at } r = r_o, \; \vartheta = 0, \; \text{at } r = r_i\,.
\end{align}
where $\vartheta$ can either be $\vartheta_{2D}$ or $\vartheta_{3D}$.
The heat flux (or the temperature) may be spatially variable, and any combination of fixed temperature and fixed heat flux at either the inner or the outer boundary can be applied in our model.

Below we present a number of examples performed using a fixed heat flux at the outer boundary and a fixed temperature at the inner boundary (see \S\ref{sec:res-IHFBCs}).
With heat-flux boundary conditions at the outer boundary the radial conductive profiles become
\begin{align}
\label{eq:T_2D-3D_cond_Q}
T_{2D}^{\text{cond}-Q} = T_i + Q_o s_o \ln \left( \dfrac{s}{s_i} \right)\,, \\
T_{3D}^{\text{cond}-Q} = T_i + Q_o r_o^2 \ln \left( \dfrac{1}{r_i} - \dfrac{1}{r} \right)\,,
\end{align}
where $T_i$ and $Q_o$ are respectively the temperature at the inner boundary and the heat flux at the outer boundary.
The Rayleigh number should then be understood as a flux-based Rayleigh number, {\it i.e.}
\begin{align}
\label{eq:RaQ}
Ra_Q = \dfrac{\alpha_T g_o Q_o d^4}{\nu \kappa}\,. 
\end{align}
Lateral variations in the amplitude in the heat flux are then defined by
\begin{align}
\label{eq:q_star}
Q^* = \dfrac{|Q_{\text{max}} - Q_{\text{min}}|}{|Q_o|}\,.
\end{align}

\subsection{Numerics}
\label{sec:Numerics}

The calculations presented here were carried out using an extension of the open-source pseudo-spectral spherical QG code {\tt pizza} \citep{gastine2019pizza} -- freely available at \url{https://github.org/magic-sph/pizza} under the GNU GPL v3 license.
The {\tt pizza} \new{code} is written in Fortran, uses a Fourier decomposition in $\phi$ and either Chebyshev collocation \citep{glatzmaier1984numerical}, or a sparse Chebyshev integration method in $s$ \citep[{\it e.g.},][]{stellmach2008efficient,muite2010numerical,marti2016computationally}.
It also supports a number of implicit-explicit (IMEX) time-stepping schemes including multi-step methods \citep[{\it e.g.},][]{ascher1995implicit} and semi-implicit Runge-Kutta schemes \citep[{\it e.g.},][]{ascher1997implicit}.
The reader is invited to consult \citep{gastine2019pizza} for further details about the \new{original} implementation of {\tt pizza} and its parallelization.

The purely 3D simulations were computed with the open-source magnetohydrodynamics code {\tt MagIC} \citep{wicht2002inner,gastine2016scaling} -- available at \url{http://www.github.com/magic-sph/magic} under the GNU GPL v3 license.
Similarly to \texttt{pizza}, \texttt{MagIC} supports various multistep and Runge-Kutta IMEX time schemes.

In this study, 3D fields in \texttt{pizza} and \texttt{MagIC} are expanded in Spherical Harmonics up to the degree and order $\ell_\text{max}$ in the angular $(\theta,\phi)$ directions and in Chebyshev polynomials with $N_r$ collocation grid points in the radial direction.
The 2D quantities in \texttt{pizza} are expanded in Fourier series  up to the degree $N_m$ in the azimuthal direction and in Chebyshev polynomials up to degree $N_s$ in the radial direction. The open-source \texttt{SHTns}\footnote{\url{https://bitbucket.org/nschaeff/shtns}} library is employed in both codes to handle the Spherical Harmonic Transforms \citep{schaeffer2013efficient}.
Parallelisation of the hybrid QG-3D code relies on the Message Passing Interface ({\tt MPI}) library.

Our numerical implementation follows closely the approach of \citet{schaeffer2005quasigeostrophic,guervilly2010thesis,guervilly2016subcritical}, with an important difference that they employed finite differences in radius while we resort to a Chebyshev collocation method. In the hybrid setup, the $z$-extrapolation of the variables from the 2D-grid to the 3D-grid is computed using Eq.~(\ref{eq:vel_3D-interp}). 
Reduction of the quantities from the 3D grid back onto the 2D grid, and the computation of the thermal wind, relies on two $z$-integration functions  described in the Appendix~\ref{sec:3D-z-func-bench}.  For clarity, all 3D-quantities are labelled with a subscript $_{3D}$ (such as $u_{\phi 3D}$), QG quantities have no subscripts.

\subsection{Posterior diagnostics}
\label{sec:Posterior_diagnostic}

We next introduce the various diagnostics and notations that are used for the analysis of the simulations.
The Nusselt number $Nu$, which characterises the heat transport of the system, is defined here in the fixed temperature configuration as the ratio between the total heat flux and the heat carried by conduction, \textit{i.e.}
\begin{align}
\label{eq:nusselt_shell}
Nu = 1+\dfrac{\displaystyle\dfrac{\mathrm{d} \widehat{\vartheta}}{\mathrm{d} r} \vert_{r=r_o}}{\displaystyle\dfrac{\mathrm{d} T^{\text{cond}}}{\mathrm{d} r}\vert_{r=r_o}}\,,
\end{align}
where, the temperature perturbation $\vartheta$ can be either $\vartheta_{2D}$ or $\vartheta_{3D}$ and $T^{\text{cond}}$ is either $T_{2D}^{\text{cond}}$ (\ref{eq:T_2D_cond}) or $T_{3D}^{\text{cond}}$ (\ref{eq:T_3D_cond_Dormy_3D}) and in QG calculations $r$ and $r_o$ are replaced by $s, s_o$.
Note that in the heat flux boundary case, this definition of $Nu$ leads to $Nu=1$ because $\partial_r\vartheta (r_o) = 0$, so following \cite{goluskin2016internally} we instead use
\begin{align}
\label{eq:nusselt_delta}
Nu_\Delta = \dfrac{T^\text{cond}(r_i) - T^\text{cond}(r_o)}{\widehat{\vartheta}(r_i) - \widehat{\vartheta}(r_o)}\,.
\end{align}
In the above expressions, $\widehat{x}$ corresponds to the time average of any quantity $x$, such that
\begin{align}
\label{eq:time_average}
\widehat{x} \equiv \dfrac{1}{\tau}\displaystyle\int_{t_0}^{t_0+\tau} x \mathrm{d}t\,,
\end{align}
with $\tau$ the averaging time.

The dimensionless kinetic energy, $E_\text{kin}$ per unit volume, is defined by 
\begin{align}
\label{eq:E_kin}
E_\text{kin} =\dfrac{1}{2} \dfrac{1}{\cal V} \int_\mathcal{V} {\bf u}_\perp^2 \mathrm{d} \mathcal{V}'\,,
\end{align}
where $\mathcal{V}$ corresponds to the full spherical shell volume in the 3D configurations and the volume outside the tangent cylinder in the QG setups. In the QG case we thus have $\mathrm{d} \mathcal{V}' = h(s) s\,\mathrm{d} s\,\mathrm{d} \phi$.
From this expression, we define a diagnostic for the fluid velocity which characterises the average flow speed, based on the root-mean-square (r.m.s.) of the velocity, and which is denoted by the Reynolds number
\begin{align}
\label{eq:Reynolds}
Re = \displaystyle\widehat{\sqrt{2 E_\text{kin}}}\,.
\end{align}
We also define the time-averaged convective Reynolds number, where the axisymmetric zonal flow contribution has been removed, since it can represent a significant fraction of the total kinetic energy without directly contributing to the heat transfer \citep{gastine2016scaling},
\begin{align}
\label{eq:Reynolds_conv}
Re_c = \displaystyle\widehat{\sqrt{2(E_\text{kin} - E_\text{zon})}}\,,
\end{align}
where $E_\text{zon}$ is the dimensionless axisymmetric kinetic energy per unit volume, similar to (\ref{eq:E_kin}), that is defined by $E_{zon} = 1 /2 {\cal V} \int_\mathcal{V} \overline{ u_\phi}^2 \mathrm{d} \mathcal{V}'$ and is associated with the time-averaged zonal Reynolds number,
\begin{align}
\label{eq:Reynolds_zonal}
Re_\text{zon} = \displaystyle\widehat{\sqrt{2 E_\text{zon}}}\,.
\end{align}

Finally, for the typical flow length-scale we use the typical cylindrical radial velocity length-scale $\mathcal{L}_{u_s}^{-1}$, determined from the time-averaged $u_s$ energy spectrum
\begin{align}
\label{eq:lus_flow}
\mathcal{L}_{u_s}^{-1} = \displaystyle\widehat{\left( \dfrac{d \displaystyle\sum_{m=0}^{m_{\text{max}}} m {u_s}_m}{\pi \displaystyle\sum_{m=0}^{m_{\text{max}}} {u_s}_m} \right)}\,.
\end{align}

\section{Results}
\label{sec:Results}

We present here results of rapidly-rotating convection, focusing on a regime well above the onset of convection, {\it i.e.} $Ra \geqslant 5 \, Ra_c$.
We explore the Ekman numbers from $Ek=10^{-4}$ down to $Ek=10^{-10}$, and consider Prandtl numbers from $Pr=10$ down to $Pr=10^{-2}$, and moderate-to-high supercriticalities ranging from $Ra = 1.7 \, Ra_c$ up to $ = 157.3 \, Ra_c$, reaching $Ra$ as large as $4.83 \times 10^{14}$.
As well as hybrid QG-3D simulations we present a large number of purely QG cases; these are more computationally efficient to run and allow a more comprehensive exploration of the parameter space.
We also present a collection of fully 3D runs computed with {\tt MagIC} \citep{wicht2002inner} and which have been either specifically computed for this study or taken from \citep{schwaiger2020relating}.
The temporal convergence of the runs has been ensured by running each simulation long enough to obtain a statistical equilibrium of the diagnostics.
The numerical truncation ranges from $(N_s, N_m)/(N_r, \ell_{\text{max}}) = (97,96)/(97,96)$ for the highest/lowest $Ek/Ra$ of ($10^{-4}/2 \times 10^{6}$) up to $(N_s, N_m)/(N_r, \ell_{\text{max}}) = (9217,9216)/(-,-)$ for the lowest/highest $Ek/Ra$ numbers ($10^{-10}/4.83 \times 10^{14}$).
In total $1\new{44}$ runs have been performed and a list of their key diagnostics is given in Table~\ref{tab:run_list} in Appendix~\ref{sec:Append-A-Results}.

\begin{figure*}
\centering{
	\includegraphics[width=0.98\linewidth]{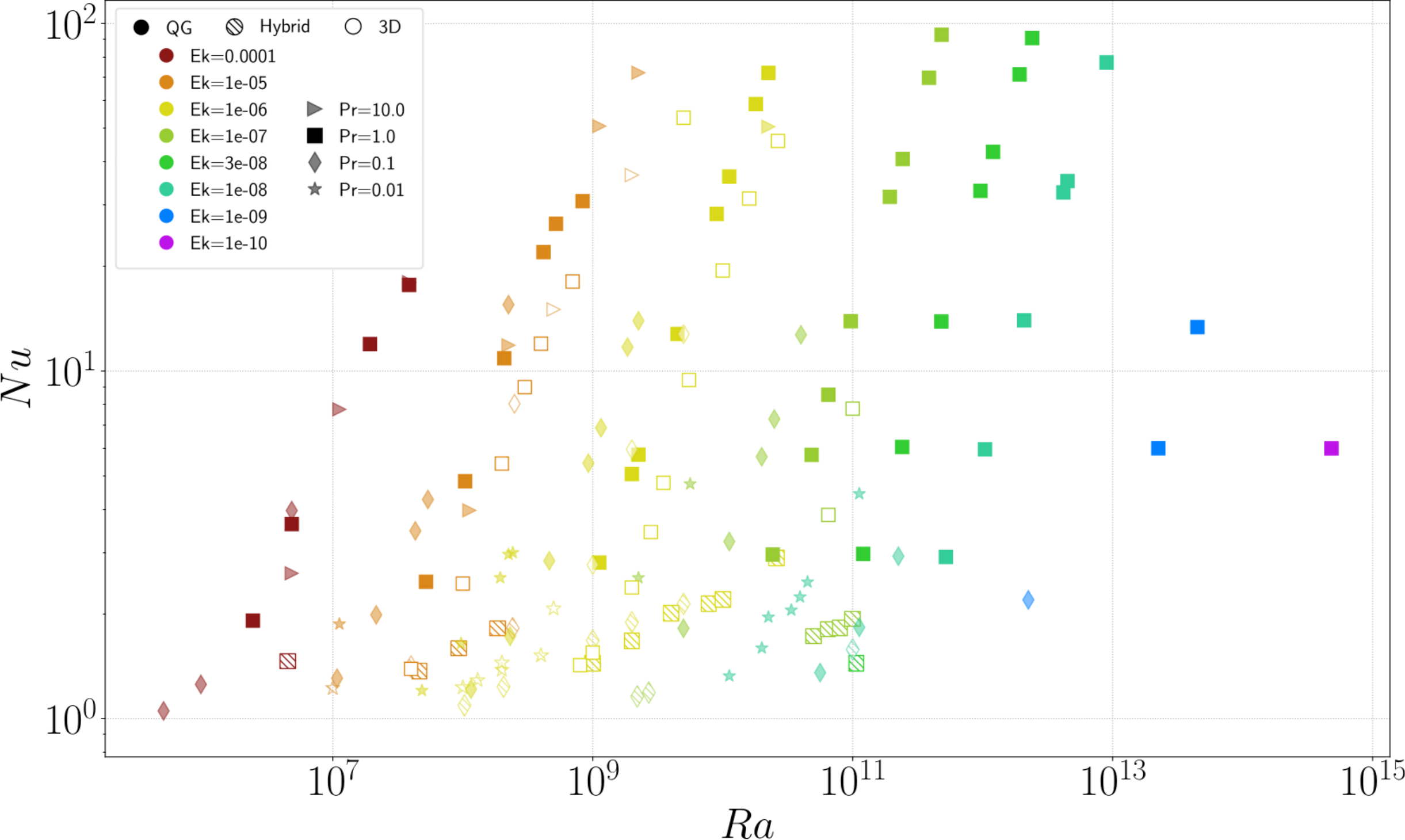}}
	\caption{
	Nusselt number, $Nu$ as a function of the Rayleigh number, $Ra$.
	Summary of all the runs \new{with fixed temperature contrast} considered in this study with the various Ekman and Prandtl numbers displayed with respectively different symbols and colors.
	The runs performed using the hybrid approach are represented with hatched symbols and the full 3D runs with empty symbols.
	}
	\label{fig:Summary-Laws_Nu-vs-Ra}
\end{figure*}

Figure~\ref{fig:Summary-Laws_Nu-vs-Ra} summarises all the runs we have carried out \new{with fixed temperature contrast}  for this study in terms of their heat transfer $Nu$ as a function of the applied forcing $Ra$.
The colours indicate the Ekman numbers while the different Prandtl numbers explored are indicated with different symbol shapes and transparency.
Hybrid QG-3D and purely 3D runs are marked with hatched and empty symbols respectively.

\subsection{Heat transfer}
\label{sec:res-pizza-hyb}

Previous work by \cite{guervilly2019turbulent} have explored the parameters at low Ekman numbers -- reaching down to $Ek = 10^{-11}$ -- and have validated the hybrid approach for a weakly supercritical $Ra$ and low $Pr$ setup \citep[{\it e.g.},][]{guervilly2017multiple,guervilly2019turbulent} in a full sphere geometry.

Here, we extend the Rayleigh number range to reach higher supercriticalities and restrict ourselves to higher Ekman numbers ($Ek \leq 10^{-10}$) with a focus around $Ek = 10^{-6}$ in order to facilitate comparisons with full-3D simulations.
All our runs \new{have} $Ro \ll 1$ as is appropriate for QG convection studies. \new{A small number of cases conducted at the highest Ekman numbers have a local Rossby numbers based on the length-scale of the flow which are up to $0.1$ (this is discussed further in \S\ref{sec:scaling_laws}).}


\begin{figure*}
\centering{
	\includegraphics[width=0.98\linewidth]{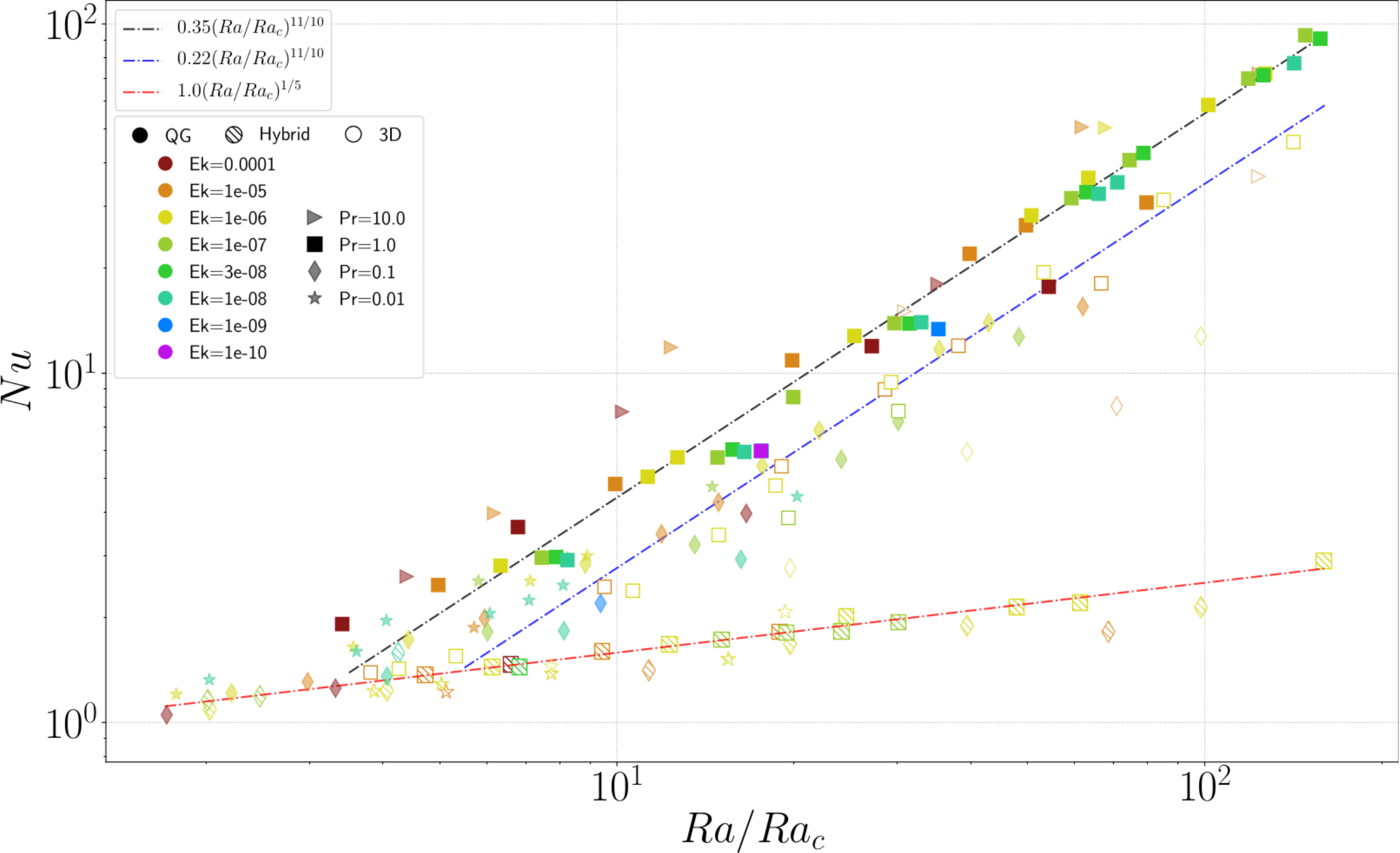}}
	\caption{
	Nusselt number, $Nu$ as a function of the supercriticality, $Ra / Ra_c$.
	Various Ekman and Prandtl numbers displayed with respectively different symbols and colors.
	The runs performed using the hybrid approach are represented with hatched symbols and the full 3D runs with empty symbols.
	}
	\label{fig:Summary_Nu-vs-Ra|Rac}
\end{figure*}

Figure~\ref{fig:Summary_Nu-vs-Ra|Rac} displays all the Nusselt numbers of our data set versus the supercriticality, {\it i.e.} $Ra / Ra_c$.
The critical Rayleigh numbers have been computed for each configuration using either the {\tt Linear Solver Builder package} \citep[{\tt LSB}, ][]{valdettaro2007convergence} for QG models, or the open-source linear solver \texttt{SINGE} \citep[\url{https://bitbucket.org/vidalje/singe}, see ][]{vidal2015singe} for 3D configurations, although we have used the asymptotic expansion by \cite{dormy2004onset} for $Pr=1$ when $Ek < 10^{-7}$ in this latter case.
\new{Concerning the hybrid QG-3D model, we have determined the onset for $3$ configurations -- at $Pr=1$ and $Ek=\lbrace 10^{-4}\,, 10^{-5}\,, 10^{-6} \rbrace$ -- by time-integrating the nonlinear equations (\ref{eq:zonal_QG}-\ref{eq:heat_3D}-\ref{eq:momentum_QG-hyb}) using the {\tt pizza} code with an initial sectorial temperature perturbation and by bracketing the Rayleigh number until the critical value is attained.
When $Ek \leq 10^{-7}$ or $Pr \neq 1$ we have assumed the same critical Rayleigh as for 3D configurations.}
In all cases, a simple extrapolation employing the asymptotic scaling for rotating convection $Ra_c \sim Ek^{-4/3}$ has been used whenever the aforementioned techniques could not be applied for practical reasons.
\new{Concerning $Pr=1$, the $Ra_c$ values obtained with {\tt LSB} and {\tt SINGE} methods agree within $\sim 6\%$ at all $Ek$ and follow the expected converging trend \citep{dormy2004onset}. The $Ra_c$ value of the hybrid model is $\sim 2\%$ lower than that of the 3D at $Ek=10^{-4}$ and $\sim 13\%$ lower at $Ek=10^{-6}$.
The $m_c$ values obtained with all methods agree with each other within a range of $m \pm 2$ in all configurations.}
See Appendix~\ref{sec:Append-C-rac} for more details about our estimates of $Ra_c$.

We can observe that for the lowest supercriticalities ($Ra \leq 10\, Ra_c$) all the points in the weakly non-linear regime follow a power law of the form $Nu-1 \sim Ra/Ra_c-1$.
For stronger forcing with $Ra > 10\, Ra_c$, the numerical models seem to approach an asymptotic behaviour of the form $Nu \propto (Ra/Ra_c)^\alpha$ (black and blue dotted-line for the QG and 3D runs, respectively) with no additional dependence on the Ekman number.
A simple polynomial fit suggests a power law with a slope of about $\alpha \simeq 1.1$, an exponent in line with previous findings of rotating convection in spherical shells with $r_i/r_o=0.35$ and fixed temperature boundary conditions \citep[{\it e.g.},][]{yadav2016effect}. This is somewhat lower than the theoretical asymptotic scaling for rapidly-rotating convection $Nu \sim Ra^{3/2} Ek^{2} Pr^{-1}$ put forward by \cite{julien2012heat} and retrieved in the 3D spherical shell computations by \cite{gastine2016scaling} when $r_i/r_o=0.6$.
The QG runs are slightly offset compared to 3D cases towards larger Nusselt numbers for the same supercriticality ($Ra/Ra_c$).
Strikingly however, this asymptotic scaling is followed only by the QG and 3D simulations, while the hybrid runs follow a much shallower slope $Nu \sim (Ra/Ra_c)^{1/5}$.
Several outliers also appear in the QG and 3D configurations:
the series of QG points at $Pr=10$ seem to follow a different slope with $Nu$ values considerably higher than the values obtained at $Pr \neq 10$ for the same $Ra/Ra_c$ ratio.
All the QG and 3D runs at $Pr=0.1$ (filled and empty diamonds) lie below the mean trend, suggesting that the heat transport is less efficient for the same supercriticality when $Pr < 1$.
Thus, the purely 3D and QG cases are generally in agreement and we find two different behaviours (splitting around $Ra/Ra_c \sim 10$) with a weakly non-linear regime and a regime with steep scaling at higher forcing levels, while the hybrid QG-3D setup starts to significantly depart from the 3D configuration for $Ra>10\, Ra_c$ \new{(at $Pr=1$)} and follows a much shallower scaling behaviour. 
\new{\newr{In constrast, when $Pr = 10^{-2}$ the QG and 3D models disagree above $Ra/Ra_c = 10$, while t}he range of accordance between the hybrid QG-3D and the 3D configurations seems to extend to higher $Ra$, up to $Ra > 15-20\, Ra_c$ at $Pr=0.1-0.01$, suggesting the range of agreement between the two \newr{latter} models may be larger at low $Pr$ and low $Ra/Ra_c$ \citep[as suggested in previous studies, {\it e.g.} in][]{guervilly2016subcritical,guervilly2017multiple,guervilly2019turbulent}.}

\subsection{Comparison of convective planforms}

\subsubsection{Comparison at modest driving}
\label{sec:res-pizza-hyb_low-Ra}

To further investigate the features observed in Fig.~\ref{fig:Summary_Nu-vs-Ra|Rac}, we begin by comparing results from our hybrid QG-3D model with 3D and QG simulations at modest driving, which we define here to be the parameter regime where $Ra \leq 20\,Ra_c$, for a case at $Ek = 10^{-6}$, $Pr = 1$, and $Ra = 2 \times 10^{9}$ ($= 10.6 \, Ra_c$ for the 3D, \new{$= 12.3 \, Ra_c$} for the Hybrid, and $= 11.3 \, Ra_c$ for the QG setups).

\begin{figure*}
\centering{
    \includegraphics[width=0.76\textwidth]{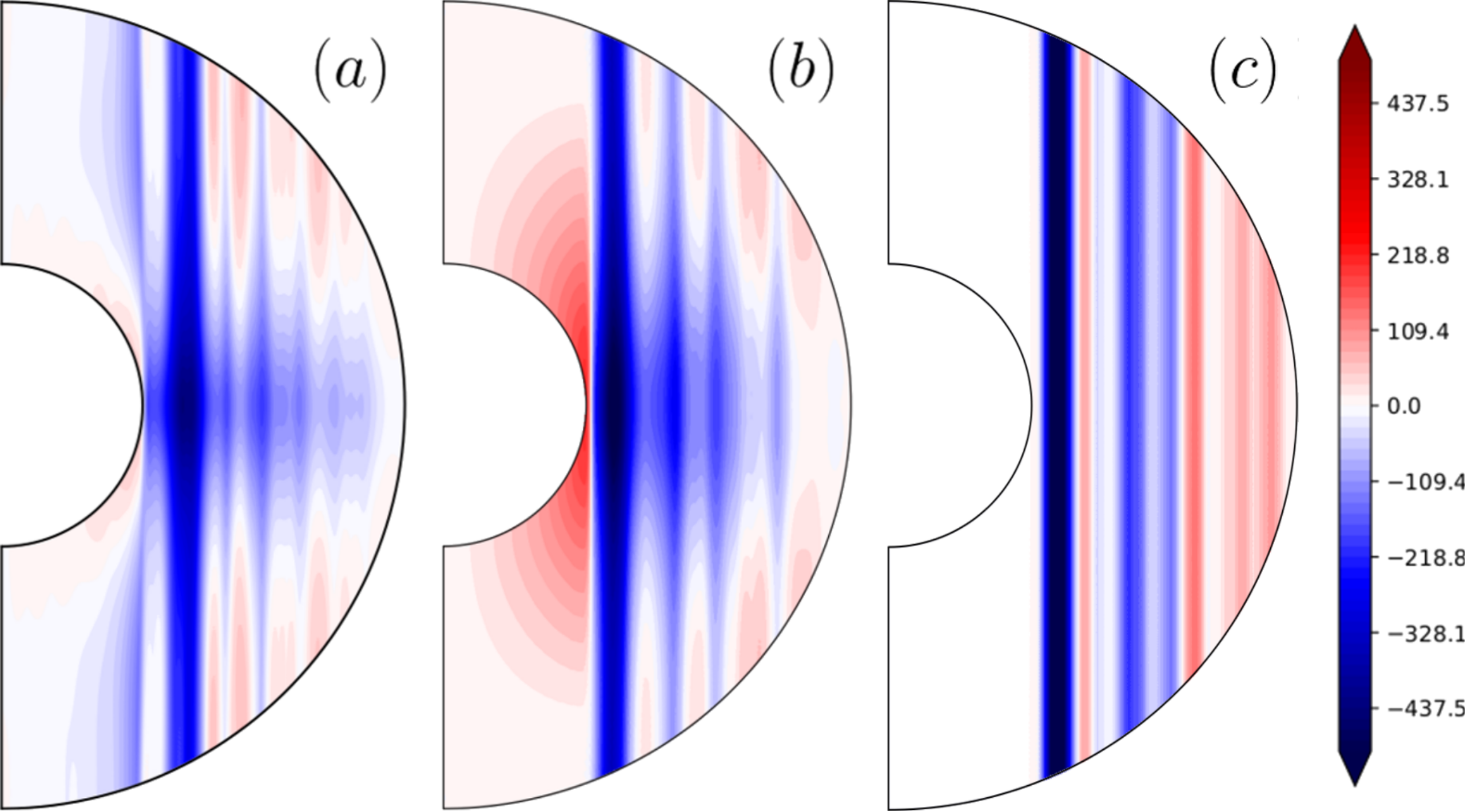}}
    \caption{
    Meridional section of the $\phi$-averaged azimuthal velocity $\overline{u_{\phi 3D}}$ contribution including the effect of the 3D thermal wind for the 3D (a) and the Hybrid (b) cases.
    A 3D extension of the purely  QG $\phi$-averaged azimuthal velocity (using Eq.~\newr{(}\ref{eq:vel_3D-interp}\newr{)} is also presented (c).
    The three computations have been carried out at the same parameters $Ek = 10^{-6}$, $Pr = 1$, and $Ra = 2 \times 10^{9}$ \new{($= 10.6 \, Ra_c$ for the 3D, $= 12.3 \, Ra_c$ for the Hybrid, and $= 11.3 \, Ra_c$ for the QG setups)}.
    The resolution is $(N_r, \ell_{\text{max}}) = (129, 341)$ in the 3D case, $(N_s , N_m)/(N_r, \ell_{\text{max}}) = (257, 256)/(257,256)$ in the hybrid case and $(N_s, N_m) = (385, 384)$ in the QG case.
    The same colorbar is used in all cases and is saturated to highlight finer structures.
    }
    \label{fig:Comp_Hyb-3D_uphi}
\end{figure*}

In Fig.~\ref{fig:Comp_Hyb-3D_uphi} we present a comparison of meridional sections of the $\phi$-averaged azimuthal velocity $\overline{u_{\phi 3D}}$ obtained in a 3D simulation 
(a), from our hybrid QG-3D model (b) and from the QG model (c).
The resolution of the QG case is $(N_s, N_m) = (385, 384)$, the hybrid case uses a spatial resolution of $(N_s , N_m)/(N_r, \ell_{\text{max}}) = (257, 256)/(257, 256)$, and the resolution in the 3D case is $(N_r, \ell_{\text{max}}) = (129, 341)$.

\begin{figure*}
\centering{
    \includegraphics[width=0.64\textwidth]{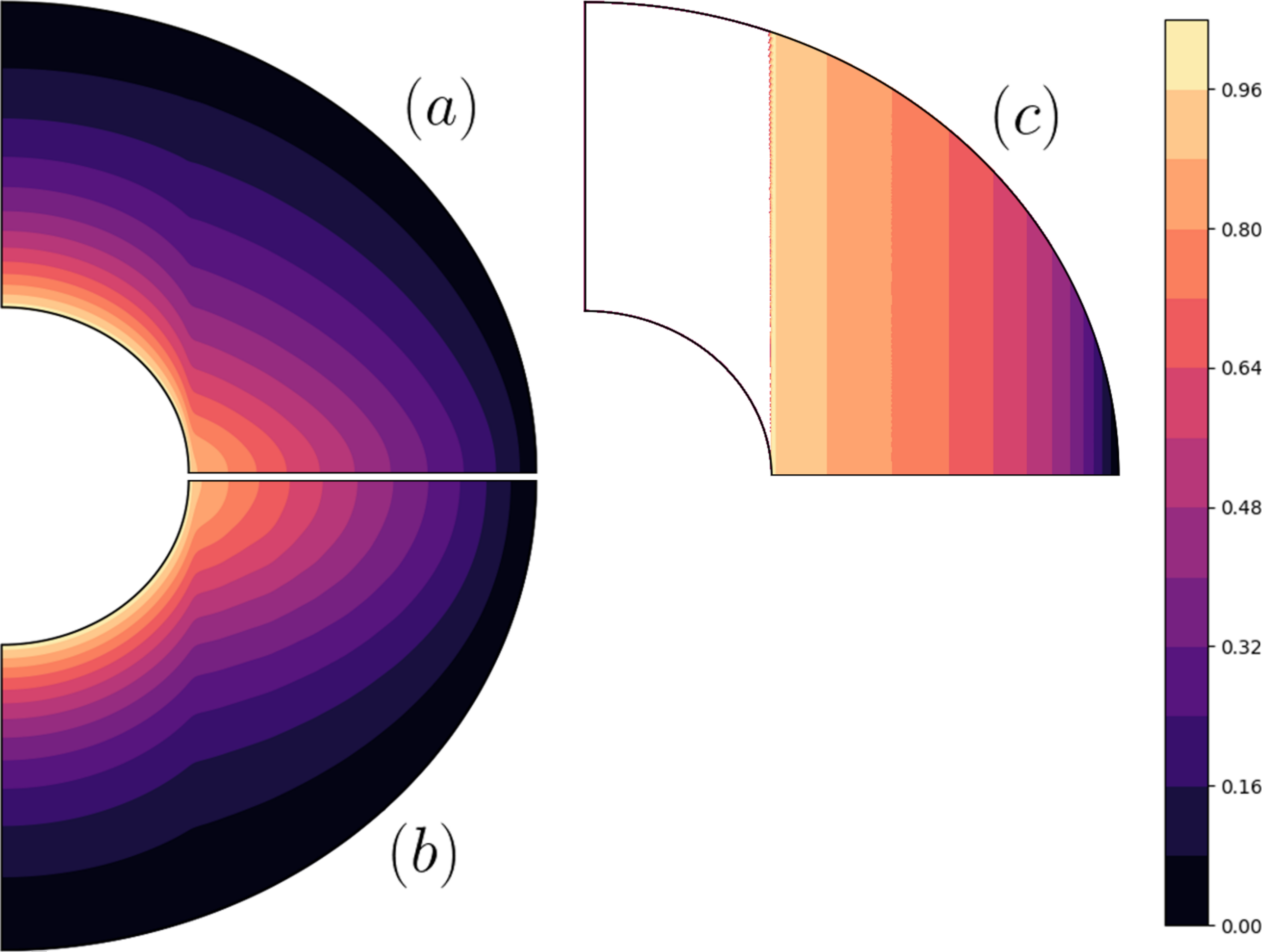}}
\centering{
    \includegraphics[width=0.77\textwidth]{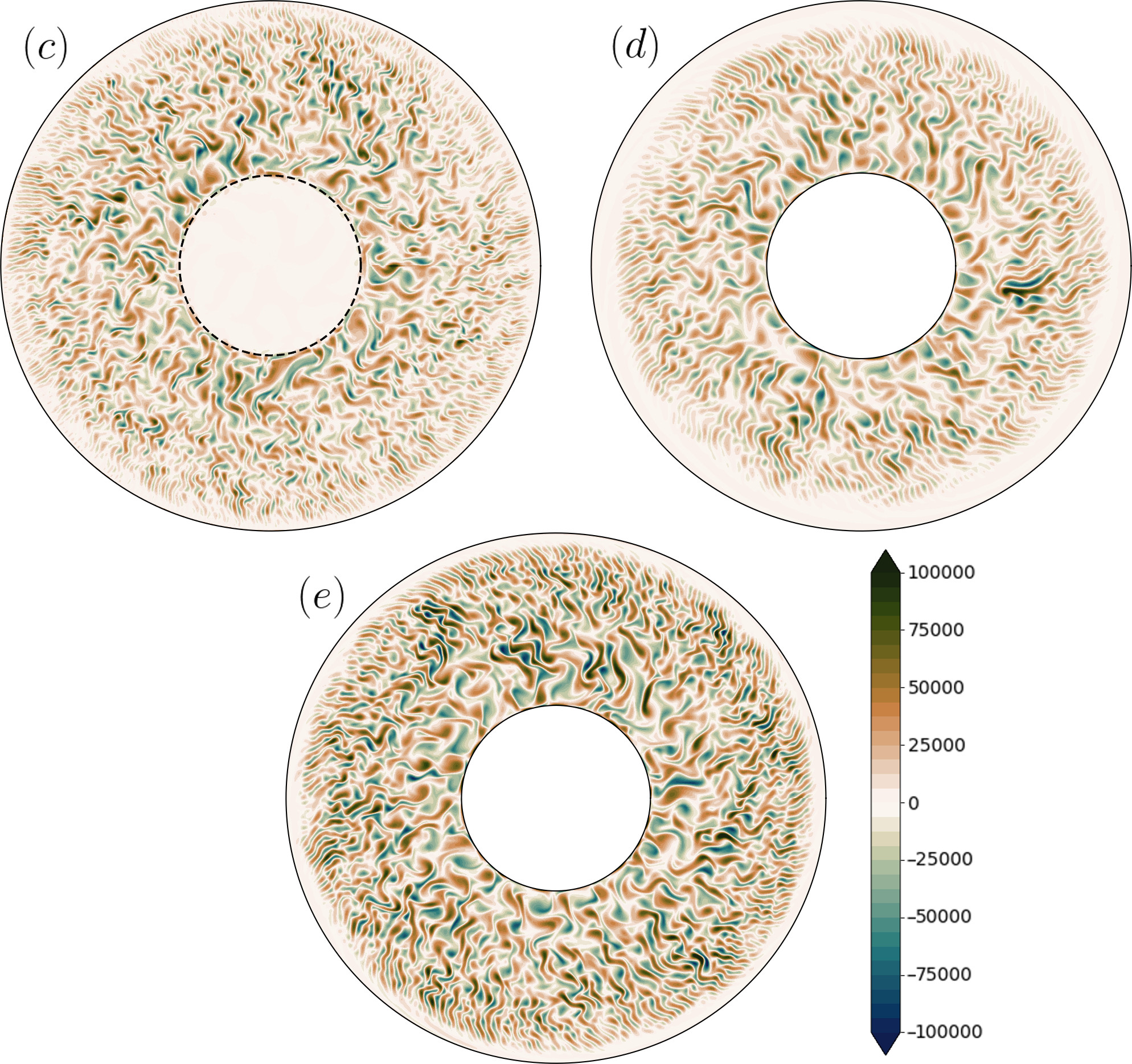}}
	\caption{
    Top panel: Comparison of the meridional section of the $\phi$-averaged of the temperature field $\overline{T_{3D}}$ for the 3D (a), the Hybrid (b) \new{and the QG case (c).
    The QG temperature field has been extended in $z$ using the conversion between cylindrical and spherical coordinate systems.}
    Bottom panel: $z$-averaged vorticity for the 3D simulation (d), and equatorial section of the axial vorticity $\omega_z$ for the hybrid QG-3D simulations (e), and the QG-simulation (f).
    The three computations have been carried out at the same parameters $Ek = 10^{-6}$, $Pr = 1$, and $Ra = 2 \times 10^{9}$.
    The spatial resolution in the 3D case is $(N_r, \ell_{\text{max}}) = (129, 341)$; in the hybrid case is $(N_s , N_m)/(N_r, \ell_{\text{max}}) = (257, 256)/(257,256)$; and in the QG case is $(N_s, N_m) = (385, 384)$ (bottom).
    For the three temperature and the three vorticity plots respectively, the same colorscales are used.
    Note that the colorscale for the vorticity is saturated to highlight the fine structure of the flows.
    }
    \label{fig:Comp_Hyb-3D_Ra=2e9_temp-vortz}
\end{figure*}

Fig.~\ref{fig:Comp_Hyb-3D_Ra=2e9_temp-vortz} additionally displays the $\phi$-averaged temperature field $\overline{T_{3D}}$ (top panel) in the purely 3D (a), Hybrid (b) and QG (c) cases as well as equatorial sections of the $z$-averaged vorticity $\omega_z$ (middle panel) in 3D (d), hybrid (e) cases, and in the purely QG case (bottom - f).
Note that in the 3D configuration, this involves a $z$-averaging procedure -- we average over the North-Hemisphere only inside the TC -- while this is straightforward in QG and hybrid cases.
As previously stated the control parameters are strictly the same.

Considering the meridional profiles of the zonal velocities (Fig.~\ref{fig:Comp_Hyb-3D_uphi}) we observe that the innermost retrograde jet near the tangent cylinder is slightly offset outwards in the 3D case while it is very close to the inner boundary in the hybrid case and QG case, creating an artificially strong shear at the tangent cylinder.
In both the 3D and Hybrid cases, these two jets display a similar columnar structure that span the entire height of the shell with the strongest velocity amplitude (compared to the other jets) and that do not vary much with $z$.
In the bulk, beside this fairly geostrophic jet, we find several thinner and weaker jets which are ageostrophic and demonstrate that the thermal wind has an important effect here; these features are reproduced in the hybrid case \new{but not in the QG case}. 
\new{In the QG case, we retrieve the strongest jet near the tangent cylinder, followed by perfectly geostrophic jets of alternating sign, with prograde jets dominating near the CMB.}
It is worth noting that $Re_\text{zon} \ll Re_c$ in all three cases (the exact values are given at the end of this section) which \new{is consistent with} the relatively weak zonal jets \new{found} in the bulk.
Near the \new{equator}, the amplitude of the azimuthal velocity is \new{slightly }larger in the 3D case compared with the hybrid case.
Since the velocity inside the tangent cylinder is set to zero before applying the thermal wind approximation in our hybrid approach, significant differences with the 3D models are visible in that region.
Overall however, the hybrid case qualitatively reproduces much of the zonal flow dynamics that happens in the bulk of the 3D case, although there are discrepancies towards the inner and outer boundaries.

Turning to the azimuthally-averaged temperature fields (Fig.~\ref{fig:Comp_Hyb-3D_Ra=2e9_temp-vortz}(a-b-c)), we find that the profiles in the 3D and hybrid QG-3D cases are very similar with isothermal lines that are bent across the tangent cylinder and that extend in the equatorial plane.
These isothermal lines are slightly more squeezed towards the equatorial plane in the 3D case compared to the hybrid case and there is a difference in the spacing of the isotherms in the $z$-direction, likely due to the simple relationship we have used to reconstruct $u_z$.
For these parameters, the 3D temperature profile is rather well retrieved in the hybrid case\new{, in contrast to the QG case which does not have the correct temperature profile and displays a largely homogeneous temperature in the bulk and a sharp drop toward the outer boundary}.

Figure~\ref{fig:Comp_Hyb-3D_Ra=2e9_temp-vortz}(c-d-e) shows a
comparison of the axial vorticity $\omega_z$ between the purely 3D, the Hybrid, and the purely QG cases.
Considering the shell from the inner core to mid-depths, we find it hard to distinguish the three planforms of convection which all display filaments of vorticity \new{of similar amplitude and length-scales}, wider near the inner boundary, and sheared in the azimuthal direction with a gradual reduction of the convective cells size with increasing $s$.
Closer to the outer boundary, the convective pattern changes in all cases with the filaments becoming more radially elongated.
This transition occurs at about the same radius $s \sim s_o -  1 /3$ in each case.
Obvious differences are seen approaching the outer boundary of the container.
In both the Hybrid and the purely QG cases the velocity field transitions into elongated azimuthal structures typical of thermal Rossby waves.
The vorticity in the 3D case, on the other hand, becomes almost perpendicular to the outer boundary with very thin and radially-elongated filaments.
The discrepancy may reflect a fundamental difference in the boundary geometry between the different configurations: in both the QG and the hybrid model the slope of the container $|\beta|$ (\ref{eq:beta}) \new{increases with the cylindrical radius.}
This treatment impedes radial motions 
\new{and favors the propagation of thermal Rossby waves over the advective processes as} the stretching term due to $\beta u_s$ becomes the dominant source of axial vorticity because of the steepening of the \new{slope at large radii; a phenomenon expected to weaken with an increasing forcing} \citep{guervilly2017multiple}.
Other QG implementations that also incorporate the horizontal components of vorticity have recently been developed and may perform better in this low latitude region \citep[see, {\it e.g.},][]{labbe2015magnetostrophic,maffei2017characterization,gerick2020pressure}.
	
Global diagnostics in the 3D and hybrid cases are rather similar with a convective Reynolds $Re_c$, a zonal Reynolds $Re_\text{zon}$ and a Rossby number $Ro$ of respectively $448.1$, $97.1$ and $4.58 \times 10^{-4}$ in the 3D-case, and $404.4$, $86.9$ and $4.14 \times 10^{-4}$ in the hybrid case.
The Nusselt number $Nu$ differs more strongly with values of $2.38$ in the 3D case and of $1.67$ in the hybrid case.
The same diagnostics obtained for the purely QG simulation are $550.4$, $126.6$ and $5.65 \times 10^{-4}$, respectively for $Re_c$, $Re_\text{zon}$ and $Ro$ while $Nu = 5.05$.
This example indicates how the hybrid approach is capable of accounting the 3D convective dynamics happening in the fluid bulk at modest driving, {\it i.e.} here with $Ra = 10.6\, Ra_c$.

\subsubsection{Limitations of the hybrid approach}
\label{sec:res-pizza-hyb_high-Ra}

We now compare results from our hybrid QG-3D model with 3D and QG simulations for a more strongly driven case at $Ek = 10^{-6}$, $Pr = 1$, and $Ra = 10^{10}$ ($= 53.2\, Ra_c$ for the 3D, \new{$= 61.3\, Ra_c$} for the Hybrid, and $= 63.4\, Ra_c$ for the QG setup).
Below we use the term 'strong driving' to refer to the parameter regime where $Ra > 20\, Ra_c$.

\begin{figure*}
\centering{
    \includegraphics[width=0.64\textwidth]{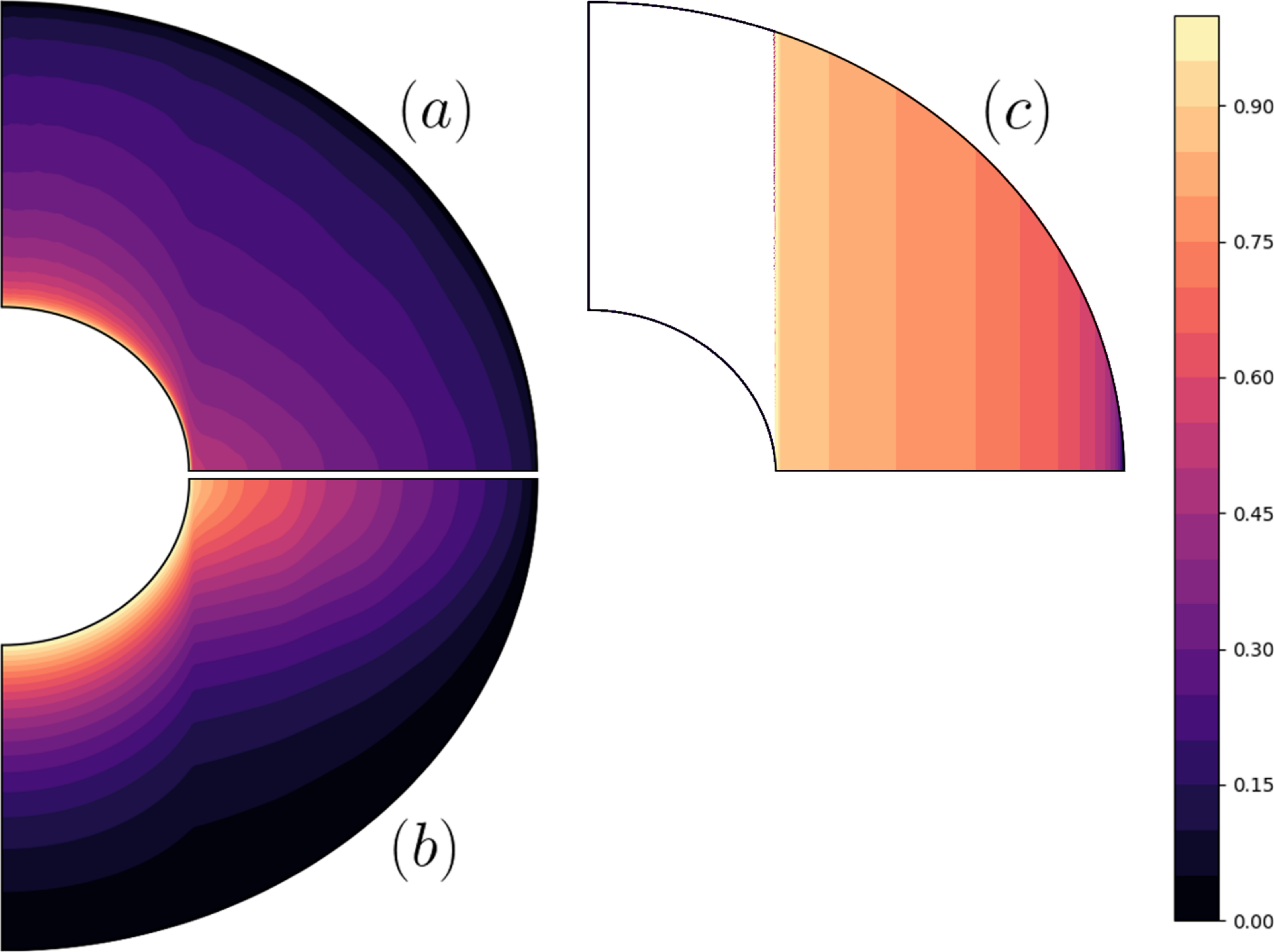}}
    
\centering{
    \includegraphics[width=0.77\textwidth]{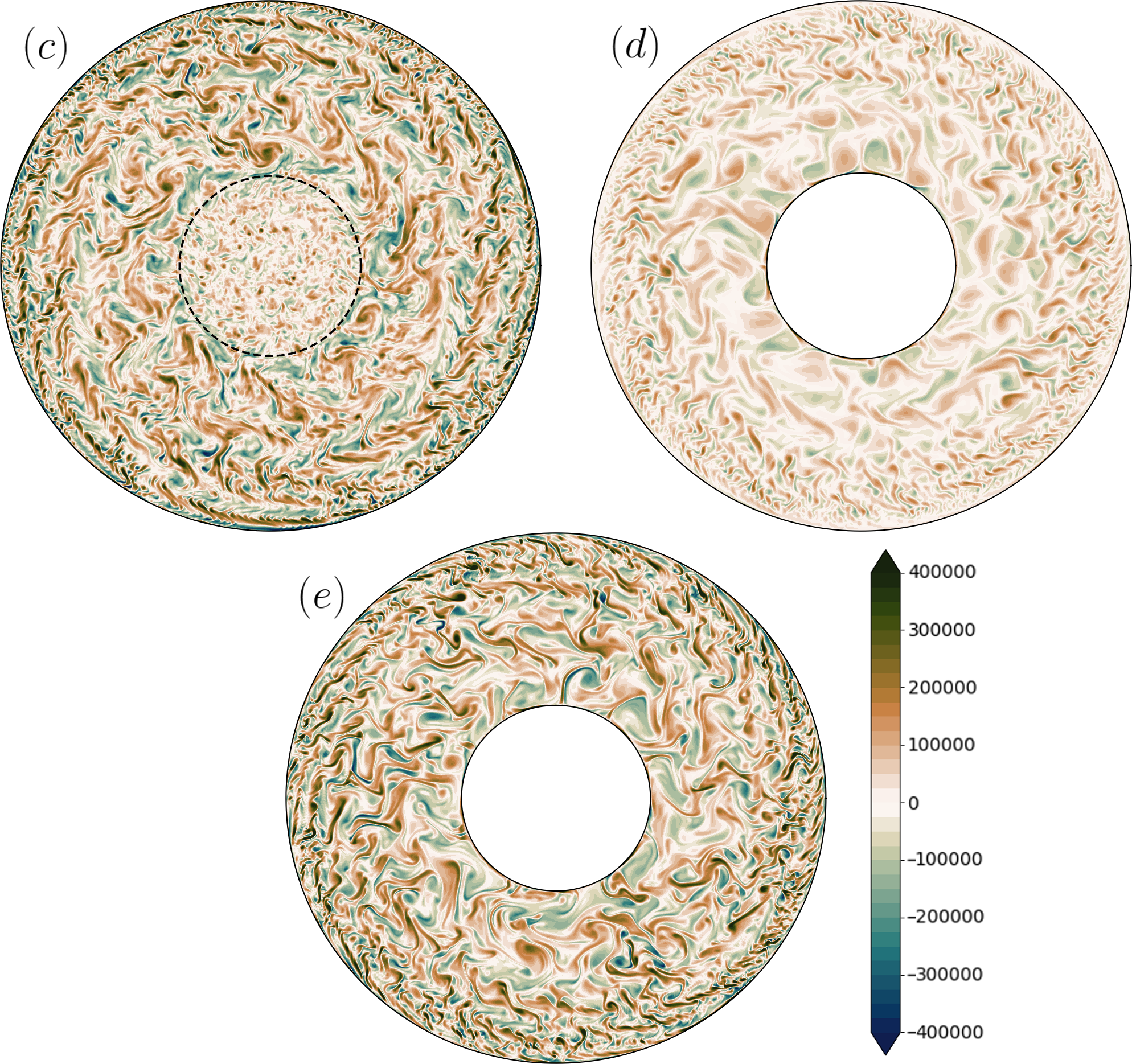}}
	\caption{
    Top panel: comparison of the meridional section of the $\phi$-averaged of the temperature field $\overline{T_{3D}}$ for the 3D (a), the Hybrid (b) \new{and the QG case (c).
    The QG temperature field has been extended in $z$ using the conversion between cylindrical and spherical coordinate systems.}
    Bottom panel: $z$-averaged vorticity for the 3D simulation (d), and equatorial section of the axial vorticity $\omega_z$ for the hybrid QG-3D simulations (e), and the QG-simulation (f).
    The three computations have been carried out at the same parameters $Ek = 10^{-6}$, $Pr = 1$, and $Ra = 1 \times 10^{10}$ \new{($= 53.2\, Ra_c$ for the 3D, $= 61.3\, Ra_c$ for the Hybrid, and $= 63.4\, Ra_c$ for the QG setup)}.
    The resolution in the 3D, hybrid and QG cases is respectively $(N_r, \ell_{\text{max}}) = (321, 682)$, $(N_s, N_m)/(N_r, \ell_{\text{max}}) = (513, 512)/(513,341)$ and $(N_s, N_m) = (577, 576)$.
    For the three temperature and the three vorticity plots respectively, the same colorscales are used.
    Note that the colorscale for the vorticity is saturated to highlight the fine structure of the flows.
    }
    \label{fig:Comp_Hyb-3D_Ra=1e10_temp-vortz}
\end{figure*}

In Fig.~\ref{fig:Comp_Hyb-3D_Ra=1e10_temp-vortz} we present the $\phi$-averaged temperature field $\overline{T_{3D}}$ (top panel) in the purely 3D (a), Hybrid (b) and QG (c) cases as well as the $z$-averaged vorticity $\omega_z$ in  the 3D case (d), and equatorial sections of the axial vorticity in the hybrid (e), and the QG cases (f).
The resolution in the 3D, hybrid and QG cases is respectively $(N_r, \ell_{\text{max}}) = (321, 682)$, $(N_s, N_m)/(N_r, \ell_{\text{max}}) = (513, 512)/(513,341)$ and $(N_s, N_m) = (577, 576)$.

Compared to the previously moderately-forced case, the meridional sections of the temperature field now significantly differ.
The hybrid temperature stays similar to the previous case at $Ra = 2 \times 10^{9}$, while in the purely 3D case we find the temperature is better mixed with isotherms \new{further away from each others} and less contrast \new{in the fluid bulk}.
\new{The QG temperature profile still does not present the correct temperature variation across the bulk but is more homogeneous and displays a sharper contrast toward the CMB when compared with the lower forced case.}
Rapid variations of the isotherms close to the boundaries in the 3D \new{and QG} cases indicate the formation of thermal boundary layers\new{, whereas the hybrid model has not developed such layers}.
Similarly, looking at sections of the axial vorticity, the hybrid and 3D cases now show significant differences.
The vorticity in the 3D case is much stronger than in the hybrid case, with filaments that are more sheared in the azimuthal direction and with significantly-perturbed thin Rossby waves near the outer boundary.
The hybrid case has in contrast not departed far from the previously moderately-driven case, the main difference being that the convective motions now span the entire shell with larger convective cells.
We also observe that convection has started inside the tangent cylinder in the 3D configuration and is already vigorous at these parameters.
Interestingly, the QG case seems to be closer to the 3D case than the Hybrid, with filaments of vorticity strongly sheared in the $\phi$-direction and a vigorous convective pattern degenerating into thin Rossby waves towards the outer boundary.
Overall, the purely 3D and QG cases have reached a regime of vigorous convection with a well-mixed temperature background while the hybrid case displays much weaker convection, comparable to the modest driving regime.
These differences are also observed in the global diagnostics with $Re_c$, $Re_\text{zon}$, $Ro$ and $Nu$ 
that are respectively equal to $3339.8$, $2405.2$, $4.12 \times 10^{-3}$, and $19.5$ in the 3D case; $1455.9$, $829.5$, $1.68 \times 10^{-3}$, and $2.20$ in the hybrid case; and $2584.4$, $1845.5$, $3.18 \times 10^{-3}$, and $36.3$ in the QG case.
Between the $Nu$ and $Re_c$ numbers of the Hybrid and 3D models, we have observed a relation of the form $Nu^{Hyb}/Nu^{3D} \propto (Re_c^{Hyb}/Re_c^{3D})^{2/5}$.

\begin{figure}
\centering
    \includegraphics[width=0.97\linewidth]{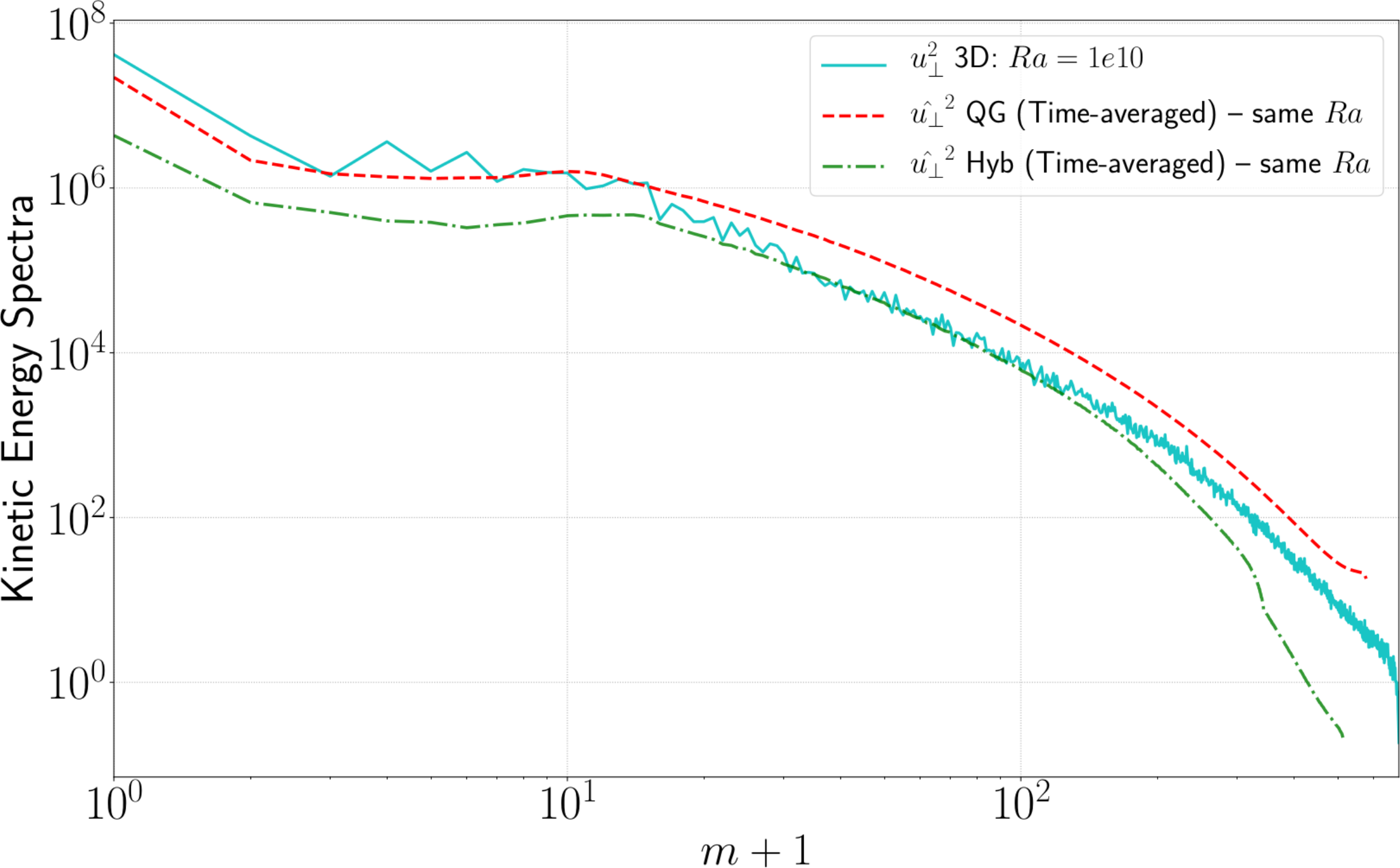}
	\caption{
    Kinetic energy spectra of the vertically-averaged total velocity, $\frac{1}{2}{\bm u}_\perp^2 = \frac{1}{2}\left(u_s^2 + u_\phi^2\right)$, as a function of the order $m+1$, for the 3D (cyan plain line), Hybrid (green dot-dashed line) and QG (orange dotted line) models for a configuration conducted with the same parameters in the three cases $Ek = 10^{-6}$, $Pr = 1$, and $Ra = 1 \times 10^{10}$.
    The resolution in the 3D, hybrid and QG cases is respectively $(N_r, \ell_{\text{max}}) = (321, 682)$, $(N_s, N_m)/(N_r, \ell_{\text{max}}) = (513, 512)/(513,341)$ and $(N_s, N_m) = (577, 576)$.
    }
    \label{fig:Comp_QG-Hyb-3D_Ra=1e10_spectra}
\end{figure}

\new{Examining the kinetic energy spectra for the horizontal velocity, shown in Fig.~\ref{fig:Comp_QG-Hyb-3D_Ra=1e10_spectra}, we find that the QG, 3D and hybrid models show similar decreasing slopes up to $m\sim 200$, although there is less energy in the hybrid model.  The spectrum for the hybrid configuration also shows more steeply decreasing slope at large $m$, confirming the lack of power at small length-scales already seen in the convective planforms.}

\begin{figure}
\centering{
	\includegraphics[width=.97\linewidth]{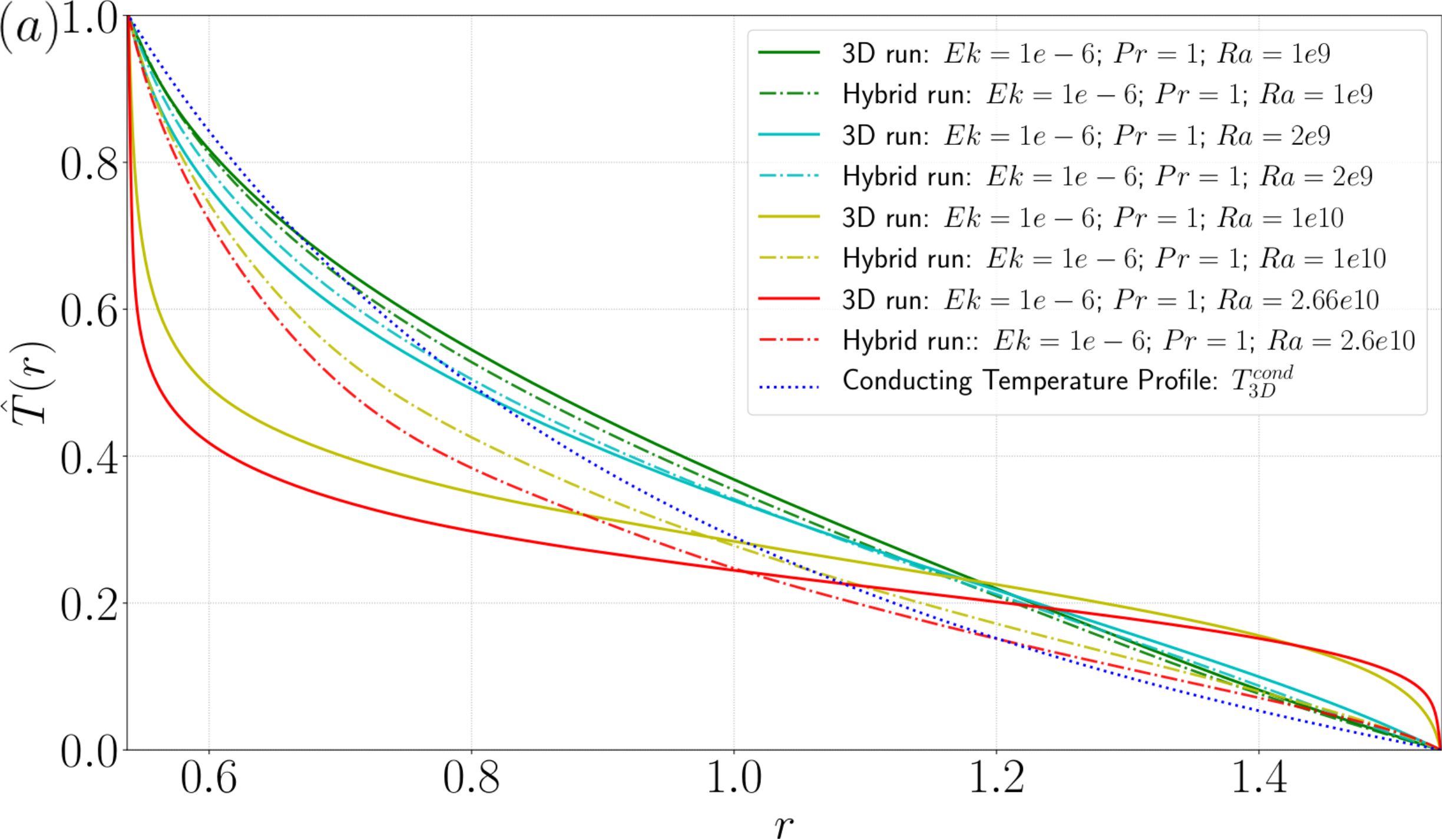}
	
	\includegraphics[width=.97\linewidth]{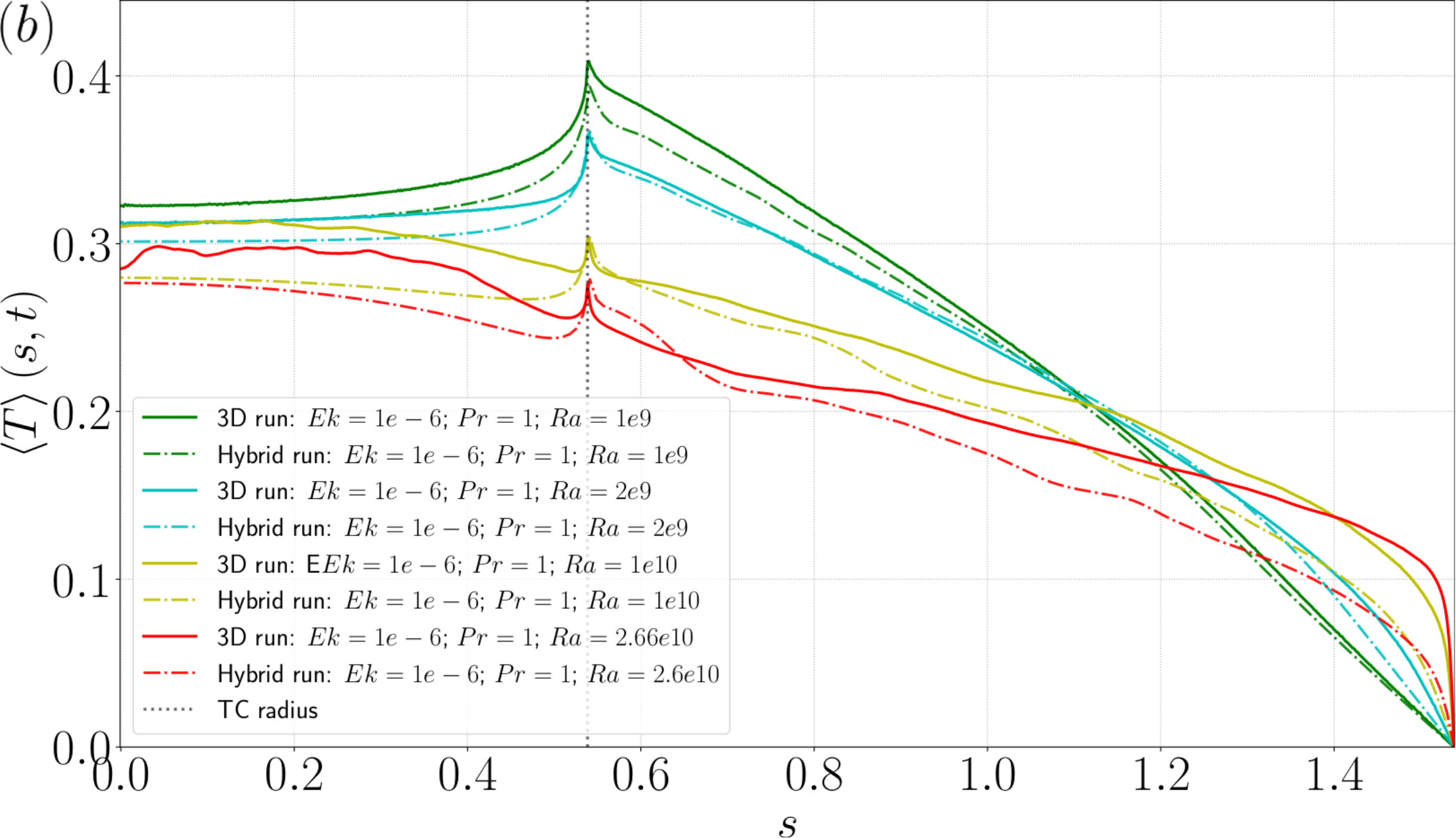}}
	\caption{
	 Top: spherical time-averaged radial temperature profiles $\hat{T}_{3D}(r)$ of 4 cases conducted at the same parameters at $Ek = 10^{-6}$ and $Pr = 1$ using our Hybrid method (dotted lines) or a 3D model (full lines) varying $Ra$ from $Ra = 10^9$ (green curves) up to $Ra = 2.66 \times 10^{10}$ (red curves).
	Bottom: Same as above but for the cylindrical radial $z$-averaged temperature profiles $\left< T_{3D} \right>(s)$. Note that the latter profiles are not time-averaged and were derived from snapshots.
	}
	\label{fig:temperature-vs-R-and-S}
\end{figure}

Fig.~\ref{fig:temperature-vs-R-and-S} shows the time-averaged radial temperature profiles $\widehat{T}_{3D}(r)$ (a) and example snapshots of $z$-averaged cylindrical temperature profiles $\left< T_{3D} \right>(s)$ (b) obtained with the purely 3D and the hybrid setups at $Ek = 10^{-6}$ and $Pr = 1$ for a series of increasing supercriticalities, ranging from $Ra = 10^{9} = 5.3\, Ra_c$ up to $Ra = 2.66 \times 10^{10} = 141.5\, Ra_c$.
We retrieve the fact that the profiles are fairly similar when $Ra \leq 2 \times 10^{9} \sim 10\, Ra_c$ (green and cyan curves), while the Hybrid and purely 3D temperature profiles diverge significantly as $Ra$ is further increased.
We observe the formation of thermal boundary layers at both spherical shell boundaries with a well-mixed interior in the 3D case while the hybrid temperature profiles do not vary much (red and yellow curves) and stay close to the radial conducting state (blue dashed-curve).
The same conclusions can be drawn when looking at the $z$-averaged cylindrical radial temperature profiles (Fig.~\ref{fig:temperature-vs-R-and-S}(b)) although here we can observe that the inner boundary temperature decreases with increasing $Ra$ in both setups with a slightly larger decrease of $\left< T_{3D} \right> (s_i)$ in the 3D case compared to the hybrid case.
Note the increased activity inside the tangent cylinder in the 3D case when $Ra \geq 10^{10}$ which is not accounted for in our hybrid QG-3D model.
The evolution of $\left< T_{3D} \right> (s_i)$ in both the 3D and hybrid cases is important: the temperature at the ICB drops significantly when $Ra$ is increased, suggesting that approximating the fixed 3D temperature at $r_i$ by a fixed temperature at $s_i$ in the QG-approach (about $T_{2D}(s_i) \sim 0.445$ at all $Ra$) is rather crude (again see Fig.~\ref{fig:Comp_Hyb-3D_Ra=1e10_temp-vortz}(c)).
Because of this reduction of temperature with increasing $Ra$ at the inner boundary in cylindrical coordinates, we expect that the driving in the purely QG configuration will eventually \new{disagree with} the 3D setup at \new{very} large Rayleigh numbers \new{\newr{before} the 3D convection \newr{reaches} the non-rotating regime}\newr{.}

\new{
\begin{figure*}
\centering{
	\includegraphics[width=\linewidth]{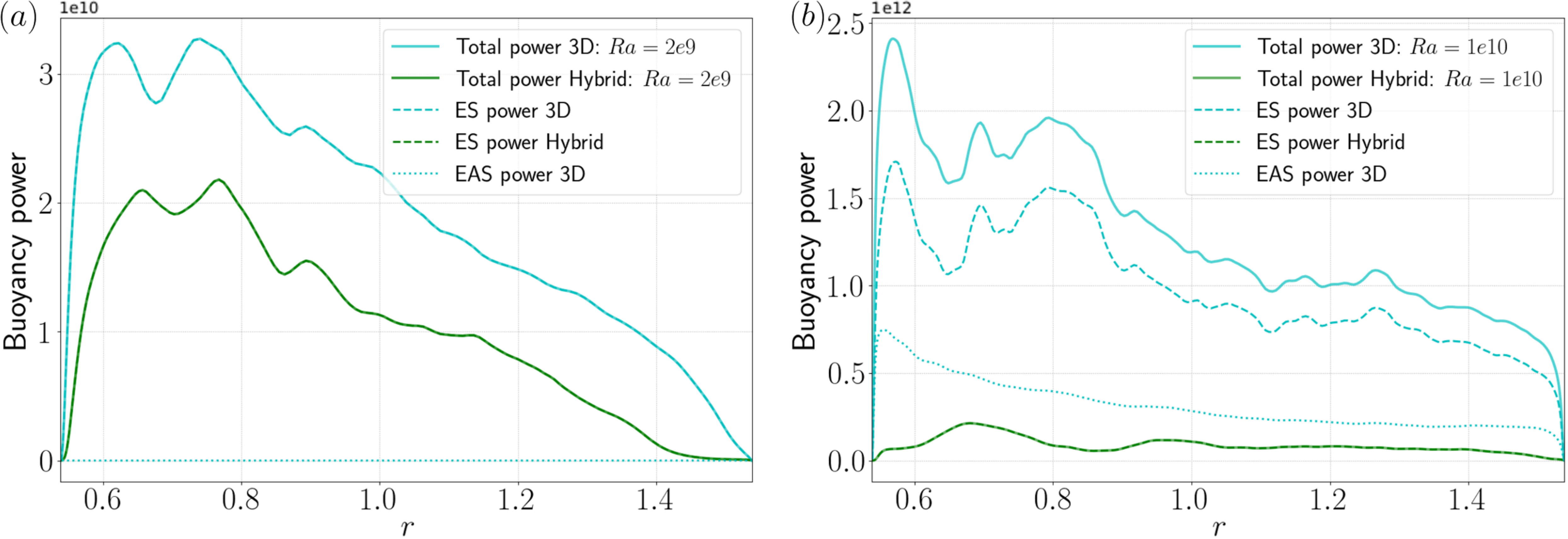}}
	\caption{
	 Buoyancy power $\frac{Ra}{Pr}u_r g T_{3D}$, decomposed into its equatorially-symmetric (ES, $u_r^{ES} g T_{3D}^{ES}$) and equatorially anti-symmetric (EAS, $u_r^{EAS} g T_{3D}^{EAS}$) for the 3D (cyan curves) and the Hybrid (green curves) models for two different simulations carried out at $Ek = 10^{-6}$, $Pr = 1$, and $Ra = 2 \times 10^{9}$ (a), and at $Ek = 10^{-6}$, $Pr = 1$, $Ra = 1 \times 10^{10}$ (b).
	}
	\label{fig:Comp_Hyb-3D_Ra2e9-1e10_buopower}
\end{figure*}

The fundamental difference between the 3D and the QG/Hybrid models lies in the assumed QG nature of the velocity field and in the use of $z$-averaging to represent the convective dynamics.  This implies that both the equatorially anti-symmetric parts and the $z$-component of the buoyancy force are missing in the QG and hybrid models compared to the full 3D setup.  An important difference between the Hybrid and the QG setups is that the boundary conditions at the inner core are applied over the entire tangent cylinder in the QG case, artificially providing more power to the QG setup compared to the Hybrid or the 3D configurations.  This enables the QG model to transition more quickly towards a turbulent state as $Ra$ is increased.  

Convective power is locally given by the quantity $\frac{Ra}{Pr} u_r g T_{3D}$, which we can decompose into its equatorially-symmetric (ES) and equatorially anti-symmetric (EAS) components.  Fig.~\ref{fig:Comp_Hyb-3D_Ra2e9-1e10_buopower} presents results concerning the integrated ES and EAS convective power profiles as a function of the spherical radius obtained with the 3D and the hybrid models for the two cases of \S\ref{sec:res-pizza-hyb_low-Ra} and \S\ref{sec:res-pizza-hyb_high-Ra}, {\it i.e.} at $Ek = 10^{-6}$, $Pr = 1$, $Ra = 2 \times 10^{9}$ and at $Ek = 10^{-6}$, $Pr = 1$, $Ra = 1 \times 10^{10}$, respectively.  At $Ra = 2 \times 10^{9}$, we retrieve a similar buoyancy power profile for the hybrid and 3D models, although the hybrid model has less energy especially towards the CMB. EAS modes become noticeable between $Ra=2\times 10^{9}$ (where they are almost zero) and $Ra=5.5\times 10^{9}$ (where they account for $9\%$ of the total power) and grow increasingly strong, reaching $23\%$ of the convective power by $Ra = 10^{10}$.
In addition, we find that $10\%$ of the power is driven by the $\frac{Ra}{Pr} u_z g_z T_{3D}$ 
at $Ra=2\times 10^{9}$ which grows to $30\%$ at $Ra = 10^{10}$.

At $Ra=2\times 10^{9}$, we find that the peak-to-peak ratio between the convective power in the 3D and hybrid models is around $1.5$ from which (based on the IAC scaling, see Sec.~\ref{sec:scaling_laws} and Fig.~\ref{fig:Scaling-Laws_Lld-Rec_CIA})) we expect a ratio between the velocities of $1.5^{2/5} \sim 1.17$,  close to the actual ratio of $Re_c^{3D}/Re_c^{Hyb}=448.1/404.4 \sim 1.11$. At $Ra = 1 \times 10^{10}$, we  have $Re_c^{3D}/Re_c^{Hyb}=3339.9/1455.9 \sim 2.29$ which requires a ratio of $2.29^{5/2} \sim 7.97$ in terms of the convective power.
The missing power due to the EAS modes and the $z$-component of the buoyancy force are alone not sufficient to explain all of this difference.
In the strongly driven regime, it is possible that the lack of convection inside the tangent cylinder (although this represents only $15-20\%$ of the total volume of the shell) and the enforced linearity of $u_z$ (see Eq.~\ref{eq:uz_linearity}) may also contribute to missing power at high $Ra/Ra_c$, but it is difficult to separate these contributions given the models in our database.

To summarize, our results suggest that the differences in the transition to turbulence in the 3D, QG and Hybrid come from differences in the underlying convective power.  The Hybrid and QG models lack the equatorially anti-symmetric and the $z$-component of the convective power.  This leads to a delayed transition to turbulent flow in the hybrid model.   Differences are less noticeable in the QG case, likely because the inner thermal boundary condition is applied over the entire tangent cylinder. At strong forcing the lack of convection inside the tangent cylinder and assumed linearity of $u_z$ in the hybrid model may also play a role.
}

\new{The above limitations result in }the hybrid setup remaining in the weakly non-linear regime with only a small increase of the heat transport and of the velocity with increasing $Ra$.
This is in line with the previously-observed discrepancies in the heat transport (see Fig.~\ref{fig:Summary_Nu-vs-Ra|Rac}) with Nusselt numbers that stay on a lower slope in the hybrid case than in the purely QG and the 3D cases.
In the hybrid configurations, the thermal-boundary layers do not fully develop, and the temperature profiles do not significantly depart from the background conducting state, 
which translates into weaker convection when the forcing is increased.
\new{In the QG configuration the missing buoyancy power is partly offset by additional power provided by the cylindrical boundary conditions.} In practice, this means the hybrid QG-3D method has a range of good agreement with full 3D computations which is limited to $Ra \leq 10\, Ra_c$\new{, at $Pr=1$}, and has a decreasing predictive capacity with increasing $Ra$.
\new{On the contrary, we cannot point to important disagreements between the QG and 3D datasets, suggesting that the QG model retains its predictive power, even at high $Ra/Ra_c$, at least in this particular configuration.}
The impact of that conclusion on the whole dataset will be further discussed in \S\ref{sec:scaling_laws}.

\subsection{Influence of laterally-varying heat flux Boundary conditions}
\label{sec:res-IHFBCs}

In this section we present examples of rapidly-rotating convection with an imposed laterally-varying heat flux condition at the outer boundary and a fixed temperature condition at the inner boundary.
Under these conditions, $\new{26}$ additional runs have been performed and key control parameters are given in Table~\ref{tab:run_ihbc} in Appendix~\ref{sec:Append-B-res-HFB}.

\begin{figure*}
\centering{
    \includegraphics[width=0.97\textwidth]{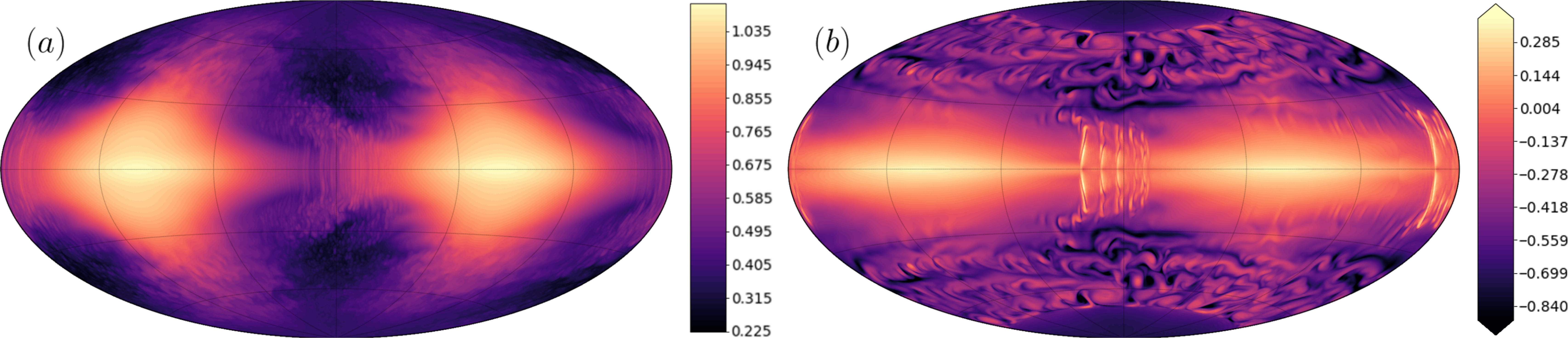}}
    \centering{
    \includegraphics[width=0.79\textwidth]{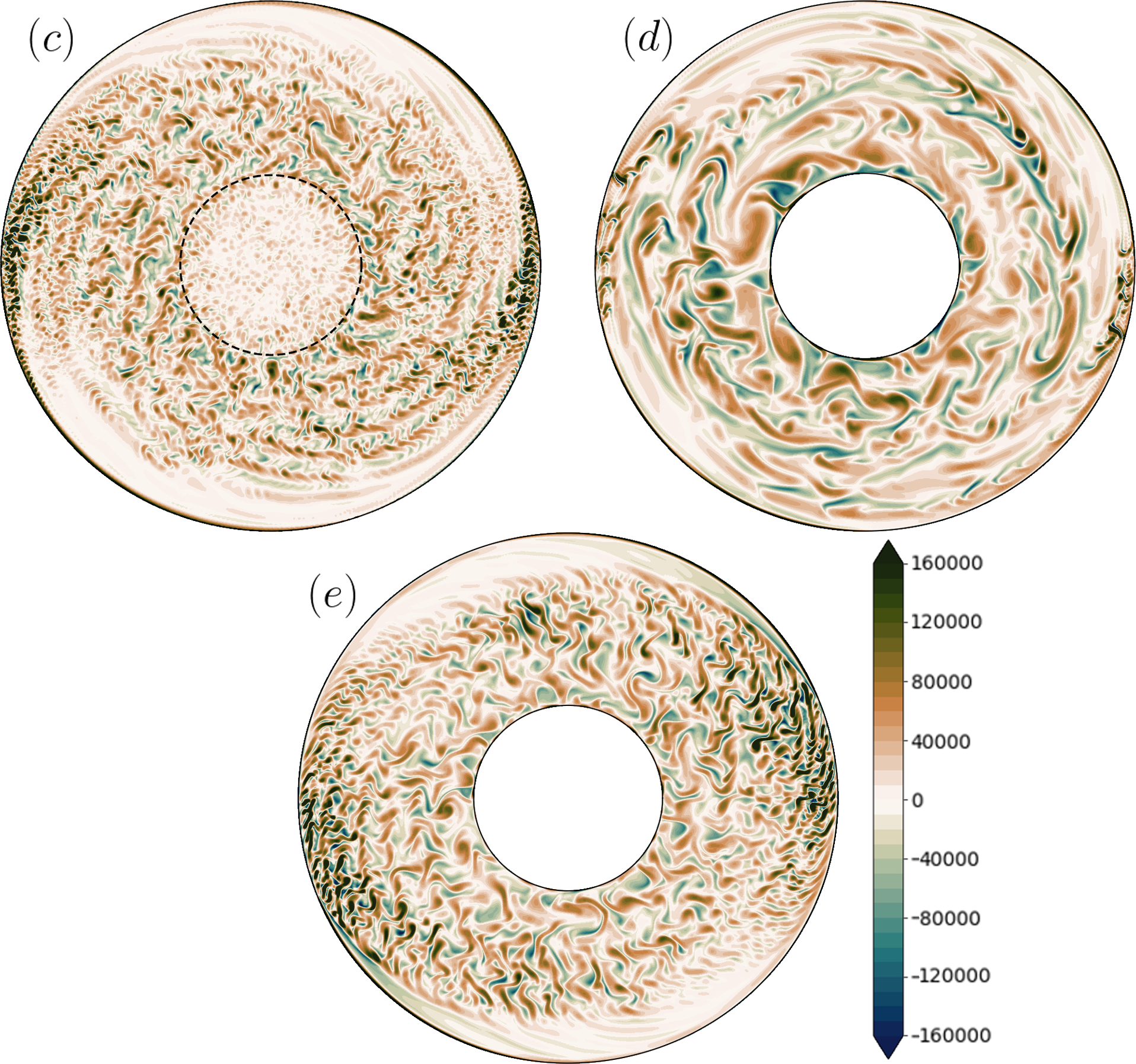}}
	\caption{
    Top panel: Comparison of the temperature field at the CMB $T_{3D}(r_o)$ for (a) the 3D case and (b) the Hybrid case.
    \new{Bottom panel: $z$-averaged vorticity for the 3D simulation (c), and equatorial section of the axial vorticity $\omega_z$ for the hybrid QG-3D simulations (d), and the QG-simulation (e).}
    The three computations have been carried out at the same parameters $Ek = 10^{-6}$, $Pr = 1$, and $Ra_Q = 8 \times 10^{9}$ imposing a $\ell=m=2, \; Q^* = 3$ lateral variation around the imposed heat-flux condition.
    The resolutions in the 3D and hybrid cases are respectively $(N_r, \ell_{\text{max}}) = (321, 682)$, and $(N_s, N_m)/(N_r, \ell_{\text{max}}) = (385, 416)/(385,416)$ \new{ and is $(N_s, N_m) = (513,512)$ in the QG case}.
    For the three vorticity plots, the same colorscale is used and is saturated to highlight the fine structure of the flows.
    }
    \label{fig:Comp_Hyb-3D_Raf=8e9_IHBC}
\end{figure*}

We first present in detail a comparison of results obtained from QG, 3D and hybrid QG-3D calculations carried out at the same parameters $Ek = 10^{-6}$, $Pr = 1$, and $Ra_Q = 8 \times 10^{9}$ with a $\ell=m=2, \; Q^* = 3$ lateral variation about the imposed heat-flux condition.
The resolutions in the 3D, Hybrid and QG cases are respectively $(N_r, \ell_{\text{max}}) = (321, 341)$, $(N_s, N_m)/(N_r, \ell_{\text{max}}) = (385, 416)/(385,416)$ and $(N_s, N_m) = (513, 512)$.

Fig.~\ref{fig:Comp_Hyb-3D_Raf=8e9_IHBC} displays example snapshots of the 3D temperature field at the CMB $T_{3D}(r_o)$ using heat-flux boundary conditions for the 3D model in (a) and the hybrid model in (b).
Both cases show the expected $\ell=m=2$ variation due to regions of higher and lower heat flux, and we also see the imprint of the underlying convection linked to the regions of enhanced heat-flux at the equator alternating with large quiescent regions of high temperature associated with lower heat-flux.
The amplitude of the temperature anomalies in the hybrid case is however larger with temperature variations in the hybrid and 3D cases spanning $-1.49 \leq T_{3D}(r_o) \leq 0.38$ and $0.23 \leq T_{3D}(r_o) \leq 1.11$, respectively.
The imprint of the underlying convection is more clearly seen in the larger eddies evident in the hybrid case, especially at mid-to-high latitudes.
This is consistent with the weaker convective forcing that occurs in the Hybrid compared with the 3D configuration, as discussed in section \S\ref{sec:res-pizza-hyb_high-Ra}. This is also reflected in global diagnostics, with lower $Nu_\Delta$ values of $2.03$ in the hybrid case compared with $7.15$ in the 3D case.
The sharp transition from convective to diffusive-only dynamics inside the tangent cylinder due to our hybrid implementation is again obvious in Fig.~\ref{fig:Comp_Hyb-3D_Raf=8e9_IHBC}(b).

Turning to the convective dynamics of the column-averaged axial vorticity (Fig.~\ref{fig:Comp_Hyb-3D_Raf=8e9_IHBC}(c-d-e)) we observe significant differences between the three cases.
The vorticity structures up to $r \sim 2/3 r_o$ are qualitatively similar in the 3D and the QG cases (c and e) while the hybrid case (d) displays larger scale vortices and a less turbulent structure, consistent with it being less strongly driven.
Close to the outer boundary, all cases show the expected $m=2$ lateral variation with alternating regions of weak and enhanced convection, although there are differences in the exact locations and morphologies of these regions in the presented snapshots.


At these parameters the hybrid approach fails to retrieve the small length-scale convective structures at high latitudes found in the 3D case and has a noticeable temperature offset at the outer boundary.
On the other hand it does reproduce similar, albeit larger scale, structures at mid-to-low latitudes.
This is evident for example in the plots of the temperature anomaly in the top panels of Fig.~\ref{fig:Comp_Hyb-3D_Raf=8e9_IHBC}, towards the center of the images and close to the equator, where signatures of vigorous convection and related wave structures are seen. 
The observed differences in the bottom panels of Fig.~\ref{fig:Comp_Hyb-3D_Raf=8e9_IHBC} can largely be attributed to differences of the buoyancy power in the three cases.
At these parameters our hybrid approach is unable to drive convection which is as turbulent as that seen in the 3D case, while the purely QG is slightly over-driven compared with the 3D case.
Note that standard global diagnostics can be misleading here as they involve averages over the entire shell.

\begin{figure*}
	\centering{
	\includegraphics[width=.97\linewidth]{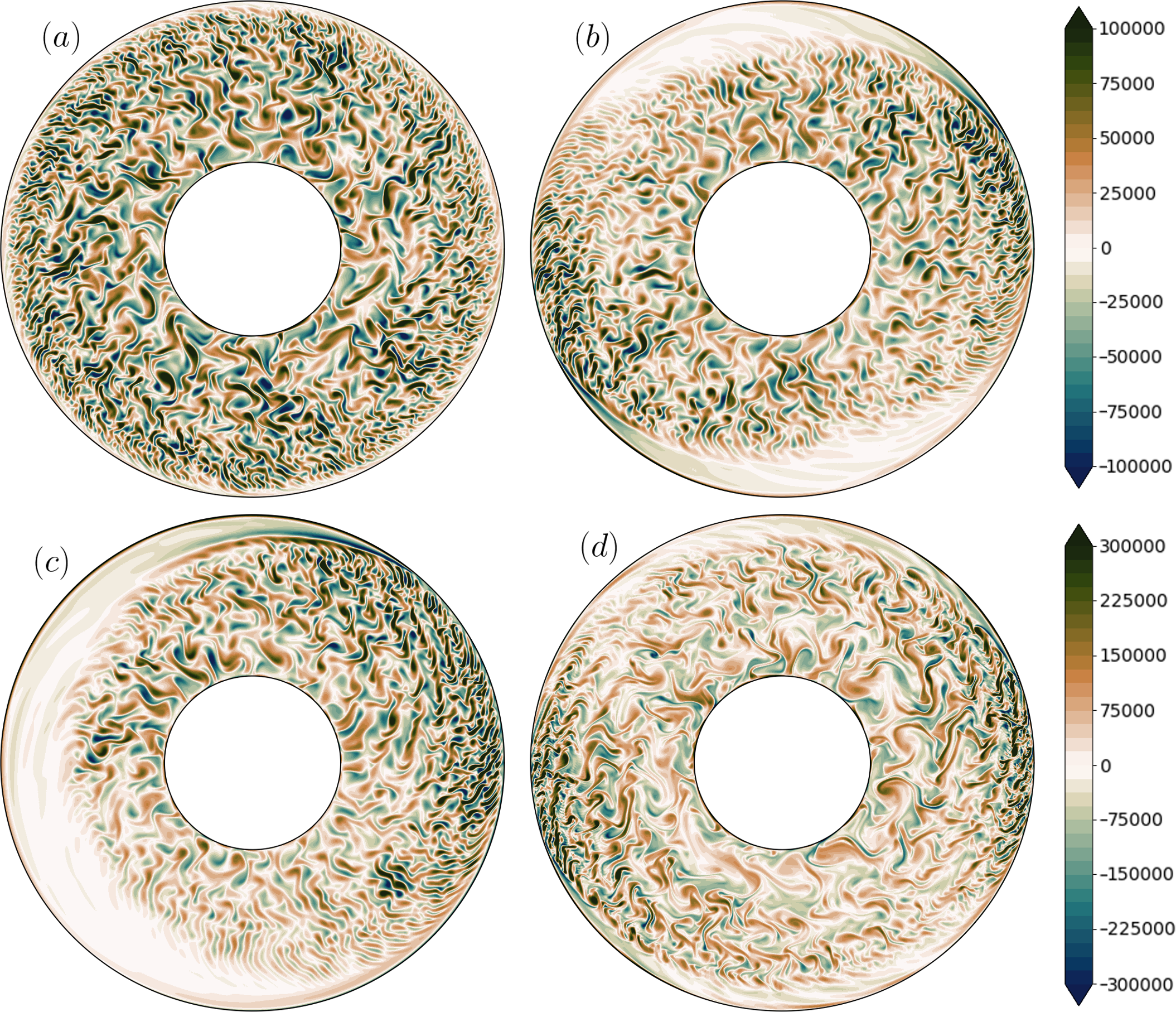}}
	\caption{
	Equatorial snapshots of axial vorticity, $\omega_z$, purely QG with heat-flux imposed at the top and a fixed temperature imposed at the bottom: $Ek = 10^{-6}, \; Ra_Q = 4 \times 10^{9}, \; Pr = 1$ with no lateral variations (a); same parameters ($Ek = 10^{-6}, \; Ra_Q = 4 \times 10^{9}, \; Pr = 1$) but with a $m=2, \; Q^* = 3$ lateral variation (b); same parameters ($Ek = 10^{-6}, \; Ra_Q = 4 \times 10^{9}, \; Pr = 1$) but with a $m=1, \; Q^* = 3$ lateral variation and  (c); and $Ek = 10^{-6}, \; Ra_Q = 3.6 \times 10^{10}, \; Pr = 1$ but with a $m=2, \; Q^* = 3$ lateral variation (d).
	The cases (a-b-c) have the same colorscale displayed on the top right and the case (d) has its own colorscale displayed on the bottom right.
	Note that the colorscales are saturated to highlight the fine structure of the flows.
	The four cases are purely QG and their spatial resolution is $(N_s , N_m) = (513, 512)$ in all cases.
	}
	\label{fig:QG-IHBC_Ek-6_Raf-4e9-4e10}
\end{figure*}

Fig.~\ref{fig:QG-IHBC_Ek-6_Raf-4e9-4e10} explores further the impact of laterally-varying heat flux boundary conditions, showing results from a more extreme convective regime using the QG model which is computationally least expensive.
It presents examples of equatorial snapshots of the axial vorticity $\omega_z$ in four cases with top heat flux/bottom-temperature imposed boundary conditions: (a) $Ek = 10^{-6}, \; Pr = 1, \; Ra_Q = 4 \times 10^{9}$ with no lateral variations; (b) same parameters ($Ek = 10^{-6}, \; Pr = 1, \; Ra_Q = 4 \times 10^{9}$) but with a $m=2, \; Q^* = 3$ lateral variation; (c) same parameters ($Ek = 10^{-6}, \; Pr = 1, \; Ra_Q = 4 \times 10^{9}$) but with a $m=1, \; Q^* = 3$ lateral variation; and $Ek = 10^{-6}, \; Pr = 1, \; Ra_Q = 3.6 \times 10^{10}$ but with a $m=2, \; Q^* = 3$ lateral variation (d).
Cases (a-b-c) have thus the same parameters but different lateral heat-flux conditions applied while case (d) has the same lateral heat-flux conditions as case (c) but is approximately ten times more supercritical.
The case without lateral variations in (a) features very similar convective patterns compared with fixed temperature boundary conditions setups (see {\it e.g.} Fig.~\ref{fig:Comp_Hyb-3D_Ra=2e9_temp-vortz}).
Namely, filaments of axial vorticity are sheared in the bulk by zonal jets of alternating direction, becoming thinner, and degenerating into thermal Rossby waves towards the outer boundary.

The bottom panel demonstrates that the laterally-varying heat flux has been successfully imposed and can drastically modify the convective planform when these lateral variations are sufficiently large.
In the case with a $m=1$ and $Q^* = 3$ pattern (c), we observe that the right hemisphere is not convecting above $s \sim s_o - 1/3$ and displays only a wide  spiraling \new{arm} covering this region, a result similar to previous 3D studies \citep[see, {\it e.g.} the Fig.~4 of ][]{mound2017heat}.
For the case with a $m=2$ pattern with $Q^* = 3$ (b), we observe a similar behaviour with regions of weak convection dominated \new{exclusively by azimuthal motions} near the outer boundary, as was also seen in Fig.~\ref{fig:Comp_Hyb-3D_Raf=8e9_IHBC}(e) with $Ra_Q = 8 \times 10^9$.
The boundary perturbation does not penetrate very deep in the shell at these parameters ($s \sim s_o - 1/3$ in the $m=1$ case, and $s \sim s_o - 1/4$ in the $m=2$).
The final case (d) has been conducted at a larger forcing ($Ra_Q = 3.6 \times 10^{10}$) and displays similar, although more turbulent, features compared to the previous cases.
The region of weak convection is however smaller and limited to fluid regions above $s \sim s_o - 1/6$.
It may be that in the limit of very large $Ra_Q$ convection, the region affected by the inhomogeneous boundary conditions could shrink to a very thin layer close to the outer boundary.


Our results demonstrate that imposing a fixed heat flux at the outer boundary does not drastically change the QG-convection compared to a fixed temperature boundary condition.
However, imposing a lateral variation of heat flux at the CMB certainly can inhibit the convective motions in a region near the surface whose size depends on $Q^*$ and the supercriticality, consistent with the findings in 3D computations \citep{mound2017heat}. 

\subsection{Influence of the Prandtl number at low Rossby number}
\label{sec:res-comp_high-Ra_convection}

\new{Focusing on the Hybrid and 3D series at $Ek = 10^{-6}, \; Pr = 0.1$, we find the same limitations as in Sec.~\ref{sec:res-pizza-hyb_high-Ra}, with an apparent lack of energy in the hybrid configuration when $Ra$ is increased (see Fig.~\ref{fig:Summary_Nu-vs-Ra|Rac}).
However as $Pr$ is decreased, larger velocities are attained at smaller Nusselt numbers, and the range of agreement across the Hybrid and 3D configurations expands (up to $Ra \sim 15 \times Ra_c$ at $Pr = 0.1$). We observed similar convective patterns between the Hybrid and 3D models at all $Pr$ ($1,10^{-1},10^{-2}$) when $Ra \leq 15 \times Ra_c$.
Unfortunately, lowest Prandtl runs are extremely computationally costly to run, especially with the 3D approach, because powerful zonal flows and velocities are triggered even at low $Nu$ inducing difficulties to reach a converged power balance.
For example, for a run at $Ek = 10^{-6}, \; Pr = 0.01, \; Ra= 5 \times 10^{8} = 19.3\, Ra_c$, the convective Reynolds numbers reach $11968.4$ for a $Nu$ of $2.07$ in the 3D case; $Re_c$ values that were not even reached by $Ra=140\, Ra_c$ at $Pr=1$.
The two 3D runs at $Pr=10^{-2}$ however appear closer to the Hybrid trend of Fig.~\ref{fig:Summary_Nu-vs-Ra|Rac} with comparable $Nu$ up to $Ra/Ra_c=20$\newr{, whereas the QG model departs from the 3D trend around $Ra \sim 10\,Ra_c$ with much higher $Nu$ in the QG cases}.
Our results do thus indicate better agreement between the Hybrid and the 3D configurations at low $Pr$ \citep[as observed in][]{guervilly2019turbulent}\newr{, in contrast the QG configuration performs less well in this regime.}
We expect nonetheless the hybrid model to depart from the 3D when $Ra$ is sufficiently increased, as observed for our $Pr=0.1$ series.
The exact $Ra/Ra_c$ range of agreement when $Pr$ is decreased further below $0.1$ requires more 3D and hybrid runs in this challenging regime in order to be determined.}

\begin{figure*}
\centering{
	\includegraphics[width=\linewidth]{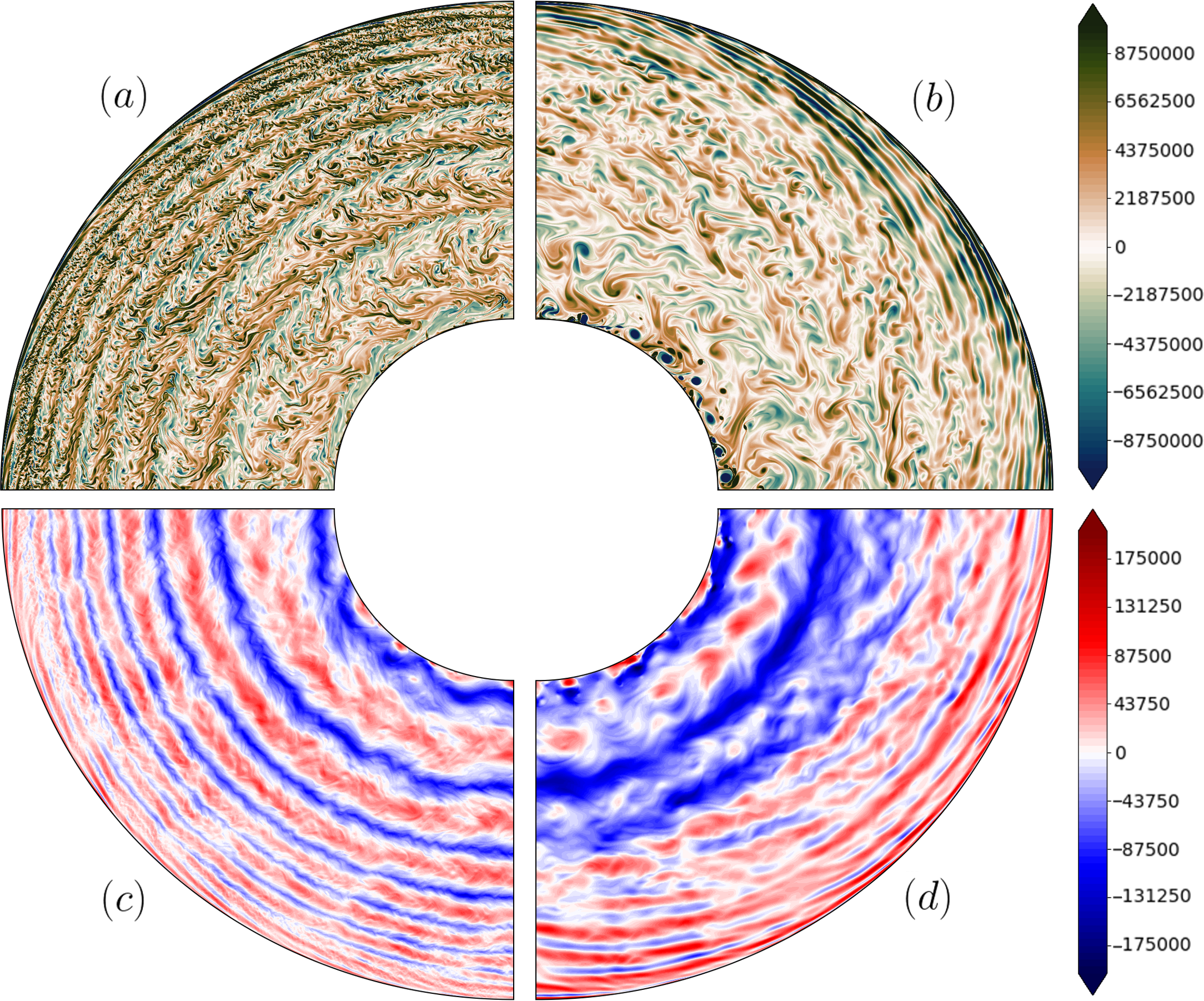}}
	\caption{
	Equatorial snapshot of the axial vorticity $\omega_z$ (a) and of the azimuthal velocity $u_\phi$ (c), for a numerical simulation with parameters $Ek = 10^{-8}$, $Pr = 1$, and $Ra = 8.99 \times 10^{12} = 142\, Ra_c$. For this case the spatial resolution is $(N_s , N_m ) = (3073, 3072)$ (left column).
	Equatorial snapshot of the axial vorticity $\omega_z$ (b) and of the azimuthal velocity $u_\phi$ (d), for a numerical simulation with parameters $Ek = 10^{-8}$, $Pr = 10^{-2}$, and $Ra = 3.37 \times 10^{10} = 4.1\, Ra_c$. The spatial resolution in that case is $(N_s , N_m ) = (3457, 3456)$ (right column).
	Respectively for the $\omega_z$ and $u_\phi$, the same colorbars are used for both cases and are saturated.
	}
	\label{fig:QG_Ek-8_Pr1-2_Ra80-200xRac}
\end{figure*}

Thus, we now take advantage of the computationally more efficient purely QG setup to study the convective flows at more extreme parameters, that is lower $Ek$ and higher $Ra$.
In particular, we investigate here the impact of the Prandtl number in this regime.
Fig.~\ref{fig:QG_Ek-8_Pr1-2_Ra80-200xRac} shows example snapshots of axial vorticity $\omega_z$ (top panel) and of azimuthal velocity $u_\phi$ (bottom panel) in the equatorial plane for two cases: $Ek = 10^{-8}, \; Pr = 1, \; Ra = 8.99 \times 10^{12} = 142\, Ra_c$ (left column) and $Ek = 10^{-8}, \; Pr = 10^{-2}, \; Ra = 3.37 \times 10^{10} = 6.1\, Ra_c$ (right column).
Despite the much higher supercriticality attained in the $Pr=1$ case, these two runs have comparable Rossby numbers: $4.10 \times 10^{-4}$ for the $Pr=1$ case and $5.97 \times 10^{-4}$ for the $Pr=10^{-2}$ case.
Both cases are purely QG calculations, and 3D temperature effects have not been included.

\begin{figure*}
\centering{
	\includegraphics[width=.97\textwidth]{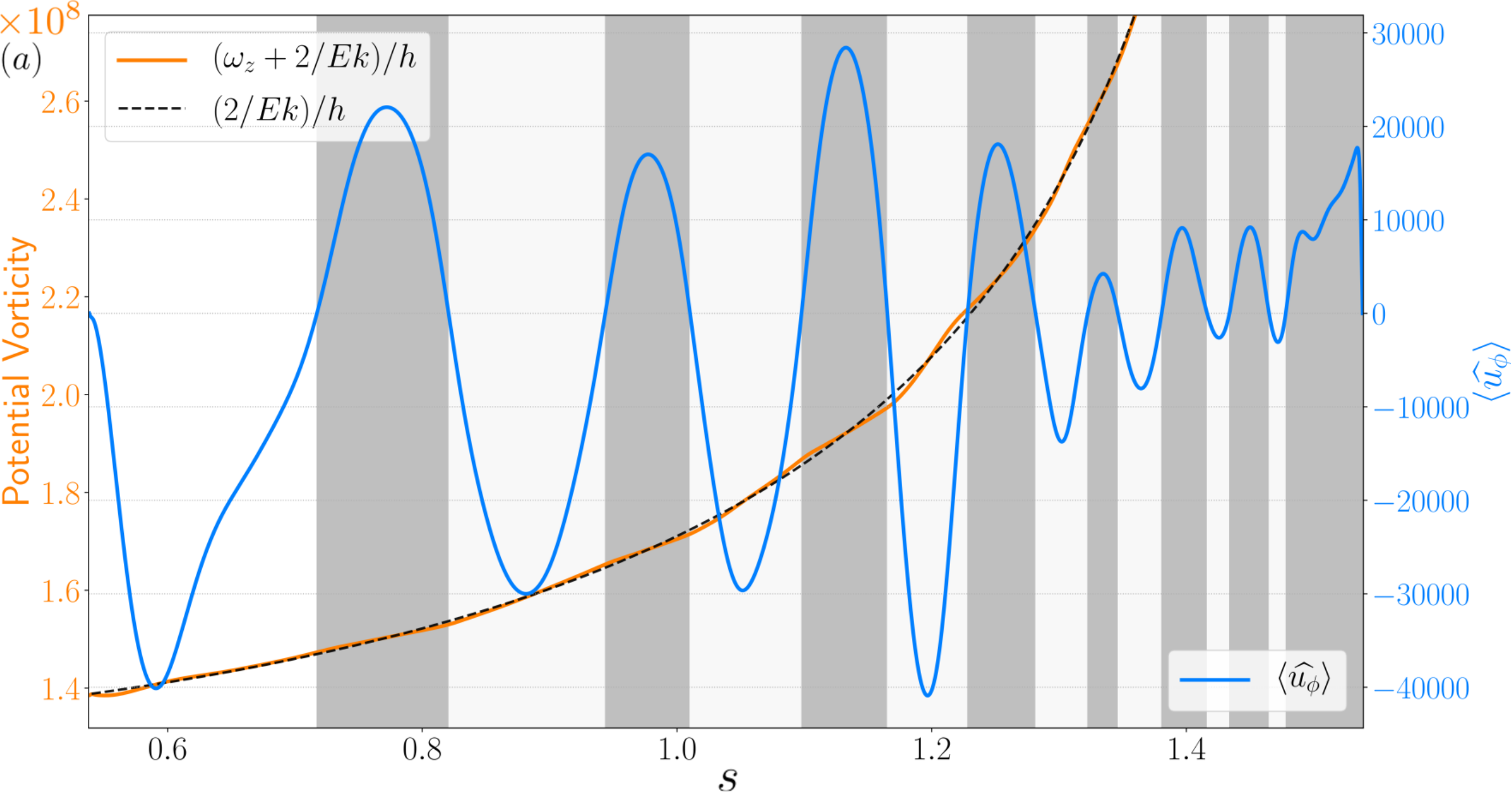} \\
	\includegraphics[width=.97\textwidth]{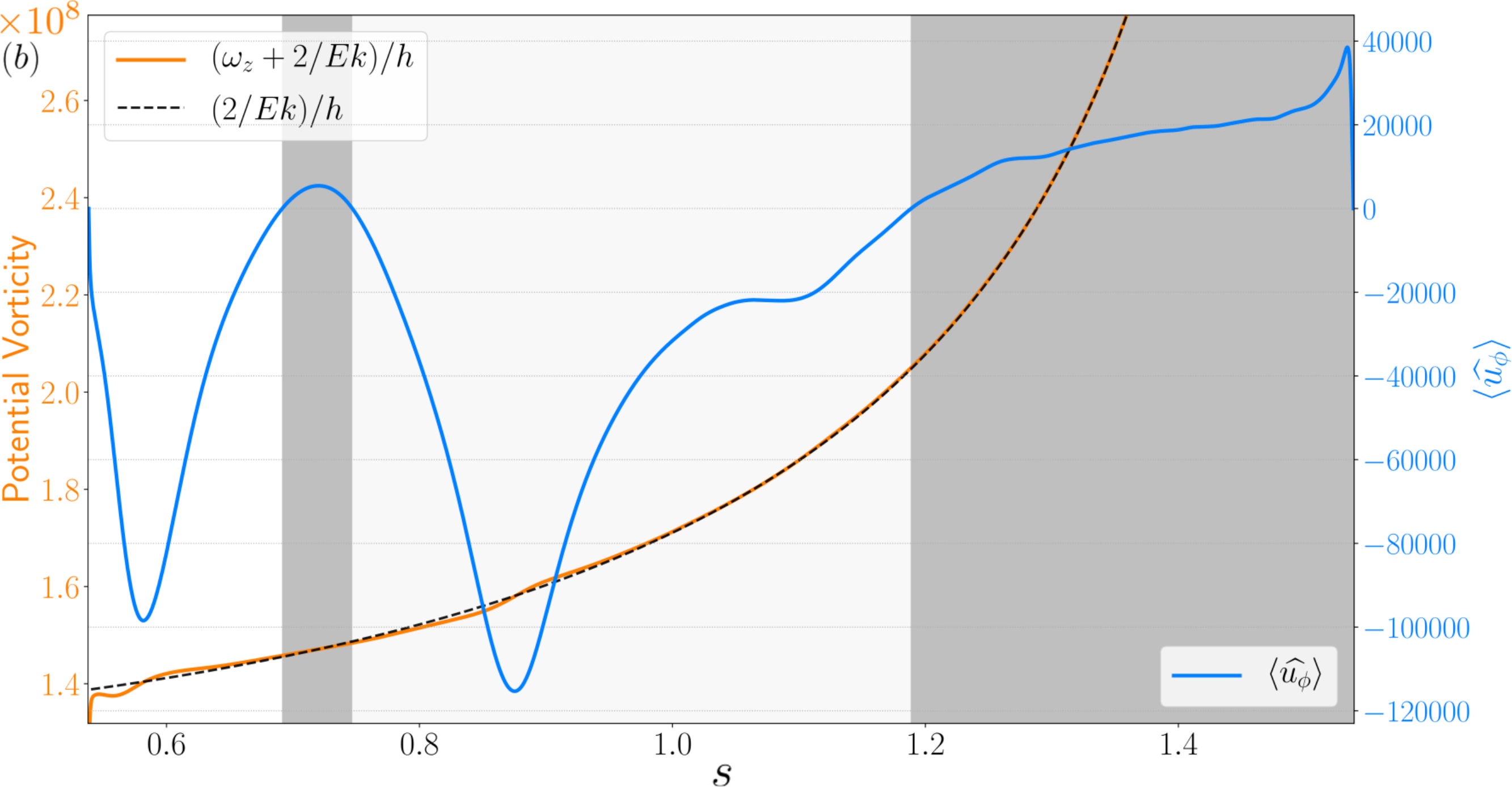}}
	\caption{
	Time averaged radial profiles of the the planetary vorticity $2 Ek^{-1}/h)$ (black-dotted curve), potential vorticity $\hat{(\omega_z + 2/Ek)/h}(s)$ (orange curve), and the azimuthal velocity $\hat{u_\phi}(s)$ (blue curve); for the numerical simulation with parameters $Ek = 10^{-8}$ $Pr = 1$, $Ra = 8.99 \times 10^{12} = 142\, Ra_c$, and a spatial resolution of $(N_s , N_m ) = (3073, 3072)$ (a), and the numerical simulation with parameters $Ek = 10^{-8}$, $Pr = 10^{-2}$, $Ra = 3.37 \times 10^{10} = 6.1\, Ra_c$ and a spatial resolution of $(N_s , N_m ) = (3457, 3456)$ (b).
	The shaded area and the white stripes correspond respectively to the fluid regions where the azimuthal velocity is positive (prograde jets) and negative (retrograde jets).
	The planetary vorticity (black-dotted curve) indicates the asymptotic behavior of the potential vorticity in a perfectly mixed shell.
	}
	\label{fig:QG-pvup_Ek-8_Pr1-2_Ra80-200xRac}
\end{figure*}

In both cases, the azimuthal velocity (c-d) displays multiple zonal jets of alternating direction (blue is retrograde and red is prograde flows), which directly translate in the axial vorticity (a-b) into alternating rings of cyclonic ($\omega_z > 0$ in red) and anticyclonic ($\omega_z < 0$ in blue) vortices.
Between two alternating jets, the vortices are streched out and sheared into azimuthally-elongated filaments, which involve a direct cascade of energy from the large to the small length-scales \citep{rhines1975waves,gastine2019pizza}.
Potential vorticity\new{, $(\omega_z + 2/Ek)/h$, is mixed} 
due to stirring by the turbulent motions, and creates these characteristic concentric jets with a typical size that is approximately predicted by the Rhines-scale $(Ro / \beta)^{1/2}$ \citep{rhines1975waves}.
Closer to the boundary, we see the \new{influence of the slope and the $\beta$-effect where the steepening of the curvature of the container impedes the radial advection of the vortices and causes the dynamics to degenerate into azimuthally elongated motions typical of} thermal Rossby waves, as already observed in \S\ref{sec:res-pizza-hyb_low-Ra}.
Figure~\ref{fig:QG-pvup_Ek-8_Pr1-2_Ra80-200xRac} additionally shows the time-averaged radial profiles of potential vorticity along with the time-averaged zonal flows.
Retrograde zonal jets where potential vorticity gradients are slightly stronger (marked by white stripes in Fig.~\ref{fig:QG-pvup_Ek-8_Pr1-2_Ra80-200xRac}), seem narrower than the regions where the gradients are weaker (corresponding to prograde jets), a result already observed by \cite{guervilly2017multiple}.
Since large supercriticalities are required to obtain well-formed potential vorticity staircases, it also appears on this figure that the case with $Pr=10^{-2}$ does not show a comparable degree of homogenisation, due to the significantly lower $Ra/Ra_c$ reached in that case, despite similar values of $Ro$.

Besides the two cases having comparable Rossby numbers and time-averaged kinetic energy spectra, the $Pr=1$ case (Fig.~\ref{fig:QG_Ek-8_Pr1-2_Ra80-200xRac}(a-c)) displays smaller eddies and thinner jets than the $Pr=10^{-2}$ case  
with a larger number of coherent jets (8 in the $Pr=1$ case compared to 3 in the $Pr=10^{-2}$ case).
Near the outer boundary the transition of the dynamics into thermal Rossby waves happens deeper in the shell in the $Pr=10^{-2}$ compared to the $Pr=1$ case.
This transition is also visible in the $\phi$-velocity where the jets lose their coherence around $s \sim s_o - 1/3$;
a direct consequence of the lower supercriticality attained in the $Pr=10^{-2}$ case ($Ra = 6.1\, Ra_c$ in that latter case, compared to $Ra = 142\, Ra_c$ for the $Pr=1$ case).

In the low Rossby regime explored here ($Ro < 6 \times 10^{-4}$), changing the Prandtl number by a factor 100 drastically modifies the form of the convective pattern.
We find that decreasing $Pr$ results in wider and fewer jets as well as larger convective structures that are maintained at a much lower supercriticality.
This effect has previously been reported by \citet{guervilly2017multiple} who suggested a weak dependence of $Ro$ on $Pr$ (see Table~\ref{tab:run_list} or the next section).

\subsection{Scaling laws for rapidly-rotating convection}
\label{sec:scaling_laws}

We now finally explore the scaling behaviour of rapidly-rotating turbulent convection in our three different model setups.
%
%
Theoretical scaling laws of rotating convection can be derived by considering the following dimensional 3D vorticity equation
\begin{align}
\label{eq:vorticity}
\dfrac{\partial {\bm \omega}}{\partial t} + \left( {\bm u} {\bm \cdot} {\bm \nabla} \right)  {\bm \omega} + 2 {\bm \Omega} {\bm \cdot} {\bm \nabla} {\bm u} = {\bm \nabla} \times \left( \alpha_T \vartheta_{3D} {\bm g} \right) + \nu \nabla^2 {\bm \omega}\,.
\end{align}
In the limit of rapid rotation, the Proudman-Taylor theorem promotes $z$ invariant flows with $l_\perp \ll l_{//}$, where $l_\perp$ and $l_{//}$ correspond to the convective flow length-scale perpendicular and parallel to the rotation axis.
Assuming that $l_{//}\sim d$, this implies that the gradients orthogonal to the axis of rotation $\nabla_\perp$ can be approximated by $1/l_\perp$, while the axial gradients $\partial/\partial z$ simply scale as $1/d$.
It also follows that $\omega \sim U_c / l_\perp$, where $U_c$ is a typical convective velocity.

In the diffusivity-free limit relevant for planetary convective cores, the dominant terms entering 
Eq.~(\ref{eq:vorticity}) involve a triple balance between  Coriolis, Inertia, and Archimedean forces 
\citep{hide1974jupiter,ingersoll1982motion,cardin1994chaotic,gillet2006quasi}
\begin{align}
\label{eq:CIA_balance_dim}
2 {\bm \Omega} {\bm \cdot} {\bm \nabla} {\bm u} \sim {\bm u} {\bm \cdot} {\bm \nabla} {\bm \omega} \sim {\bm \nabla} \times \left( \alpha_T \vartheta_{3D} {\bm g} \right)\,.
\end{align}
or in terms of scaling quantities
\begin{align}
\label{eq:CIA_balance_adim}
\Omega \dfrac{U_c}{d} \sim U_c \dfrac{\omega_c}{l_\perp} \sim \alpha_T g \dfrac{\Theta}{l_\perp}\,,
\end{align}
where $\Theta$ is a typical temperature perturbation.
The balance between Coriolis and Inertia yields
\begin{align}
\label{eq:CI_balance}
\dfrac{l_\perp}{d} \sim \left( Re_c Ek \right)^{1/2} \sim Ro^{1/2}\,.
\end{align}
This diffusion-free scaling is commonly known as the Rhines scaling \citep{rhines1975waves} and it is expected to hold in the limit of $Ek \ll 1$ when viscous effects become negligible in the bulk of the fluid \citep[{\it e.g.},][]{gastine2016scaling,guervilly2019turbulent}.
The other equality which enters Eq.~(\ref{eq:CIA_balance_adim}) coupled with the additional assumption that $\alpha_T g U_c \Theta \sim \frac{\nu^3}{d^4} Ra(Nu-1)Pr^{-2}$ \citep[see][]{jones2015TOG} yields in its dimensionless form
\begin{align}
\label{eq:CIA_balance}
Re_c \sim \left[ \dfrac{Ra}{Pr^2} (Nu -1) \right]^{2/5} Ek^{1/5}\,.
\end{align}
This equation is known as the inertial scaling of the convective velocity for rotating convection \citep[or the CIA scaling, {\it e.g} ][]{gillet2006quasi,king2013flow}.

Note that another equilibrium would hold if viscous effects would replace inertia in the vorticity balance (\ref{eq:CIA_balance_dim}).
This equilibrium is sometimes referred to as the VAC scaling, $Re_c \sim [\frac{Ra}{Pr^2}(Nu-1)] ^{1/2}\,Ek^{1/3}$, where Viscous, Archimedean and Coriolis effects are the dominant terms \citep{aubert2001systematic,king2013flow}.
We will not discuss this scaling since it does not provide a suitable interpretation of the numerical simulations in the turbulent quasi-geostrophic regime \citep{gastine2016scaling,guervilly2019turbulent,schwaiger2020relating}, as is also found with our simulations.


\begin{figure*}
\centering{
	\includegraphics[width=.99\textwidth]{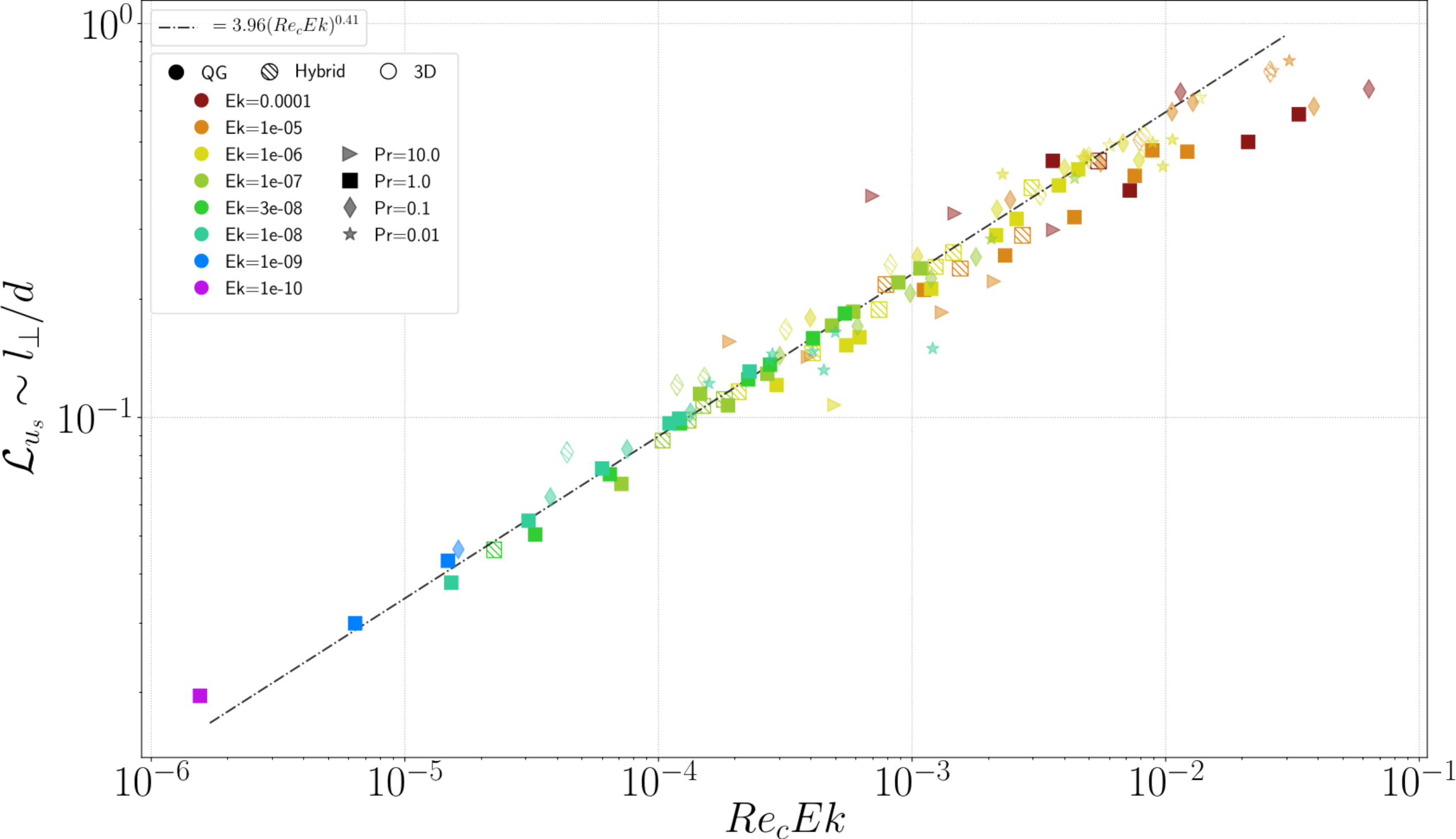}
	\vspace{0.5cm}
	\includegraphics[width=.99\textwidth]{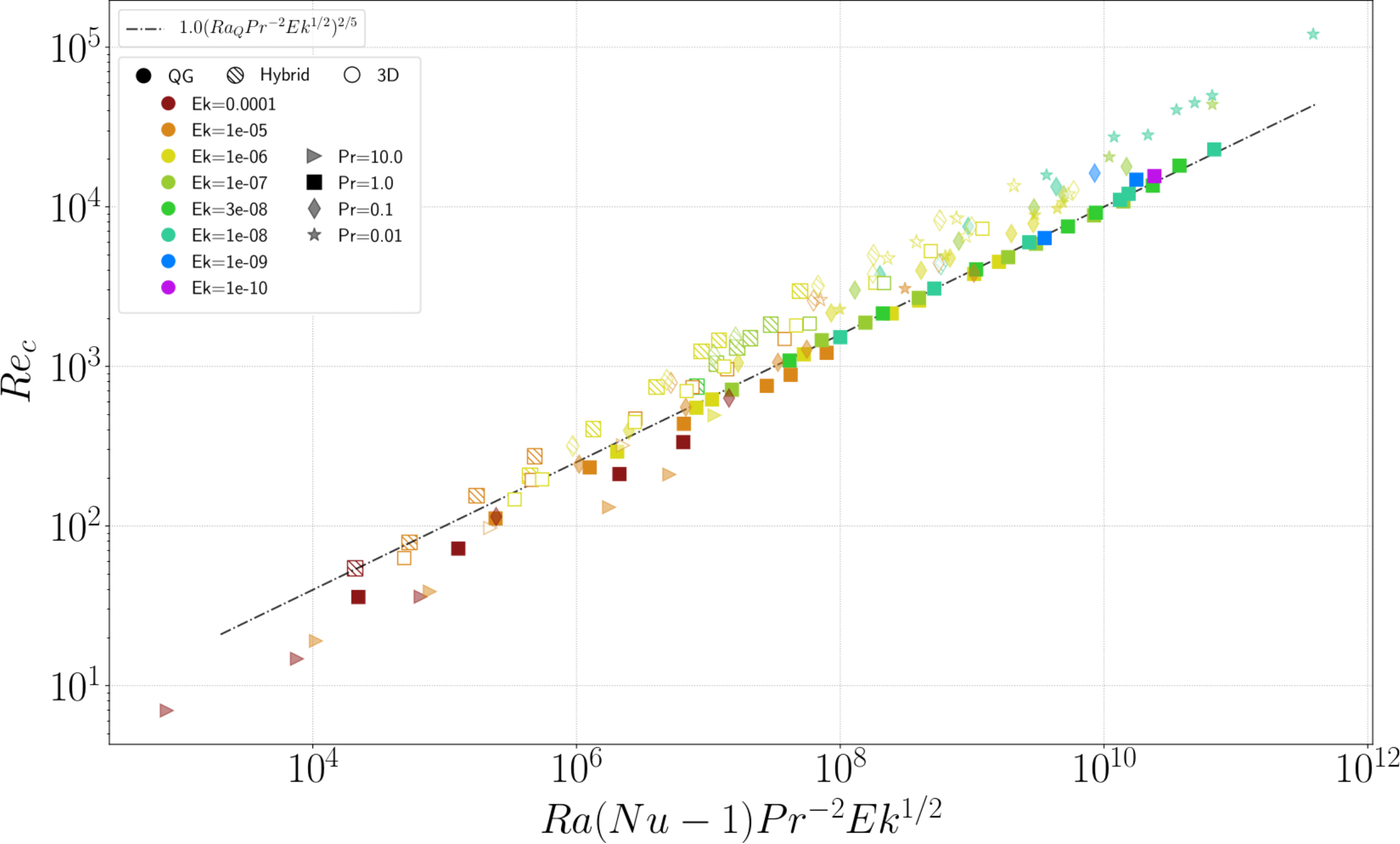}}
	\caption{
	Top: Typical length-scale of the radial velocity, $\mathcal{L}_{u_s}$, as a function of the Rossby number $Re_c\,E$.
	A best-fit power law using only the runs with $Ro \leq 10^{-3}$ is shown by the dot-dashed line.
	Bottom: Convective Reynolds number, $Re_c$, as a function of $Ra(Nu-1) Pr^{-2} Ek^{1/2}$ (Eq.~(\ref{eq:CIA_balance}). 
	Ekman and Prandtl numbers are indicated with colours and shapes respectively and the hybrid runs are indicated with the corresponding hatched symbols and the full 3D runs with empty symbols.
	The theoretically predicted scaling corresponding to the CIA balance (Eq.~(\ref{eq:CIA_balance}) is shown by the dot-dashed line.
	}
	\label{fig:Scaling-Laws_Lld-Rec_CIA}
\end{figure*}

We now analyse the relevance of the asymptotic scaling laws Eqs.~(\ref{eq:CI_balance}-\ref{eq:CIA_balance}) in the context of our ensemble of numerical simulations with fixed temperature contrast.
Figure~\ref{fig:Scaling-Laws_Lld-Rec_CIA} shows all our numerical simulations compared with the CIA scaling laws for convective velocity and length-scale.
On the top panel, the typical non-dimensional length-scale of the convection $\mathcal{L}_{u_s}$ is plotted against the Rossby number $Ro = Re_c Ek$, corresponding to (\ref{eq:CI_balance}), while the bottom figure shows $Re_c$ as a function of $Ra_Q Pr^{-2} Ek^{1/2}$, corresponding to (\ref{eq:CIA_balance}).
We observe that the Rhines scale captures well the behaviour observed in our simulations.
The majority of the points are aligned (black dot-dashed line), and departures to the theory are confined to the highest Ekman numbers ($Ek \geq 10^{-5}$)\new{. Introducing a local Rossby number $Ro_{\cal L} = Ro\,d/{\cal L}_{u_s}$ we found that all our runs have $Ro_{\cal L} < 0.1$ (not shown) which indicates that our primary assumption holds at the local level, even if the geostrophic constraint can be weaker for the highest values of $Ro_{\cal L}$ (associated with the highest $Ek$) of our dataset}.
There is no additional $Pr$ dependence since all the simulations with  $Pr \neq 1$
are close to the average scaling behaviour.
The best-fit for the whole data-set yields a power law with an exponent of $0.31$ for the $Ro$ dependency but considering only for the runs with $Ro \leq 10^{-3}$, we find a steeper slope with an exponent of $\alpha \sim 0.41$, a value comparable to that obtained in 3D parameters studies \citep{gastine2016scaling,long2020scaling} but shallower than the theoretical $1/2$ scaling.
Using QG models at Ekman numbers as low as $Ek = 10^{-11}$, \cite{guervilly2019turbulent} showed that $l_\perp /d \sim Ro^{1/2}$ is gradually approached in the low viscosity limit appropriate for planetary cores.

The bottom panel of Fig.~\ref{fig:Scaling-Laws_Lld-Rec_CIA} shows that most of our simulations agree well with the CIA theoretical scaling law -- with the power exponent $2/5$ (black dotted line) -- especially for $Ek \leq 10^{-6}$.
Cases at higher Ekman number ($Ek \geq 10^{-5}$; red and orange symbols) depart from the theory, following instead a power law with a lower exponent consistent with a deviation observed from the Rhines scaling at higher $Ek$.
The simulations conducted at different Prandtl numbers are also well aligned with the CIA power law and are parallel to the $Pr = 1$ series but shifted upward for $Pr < 1$ and downward for $Pr > 1$. 
The clear separation between the series at different Prandtl numbers suggests that there is a dependency on $Pr$ in the $Re_c$ scaling which affects the prefactor of the scaling law and is not accounted for in the CIA scaling law (\ref{eq:CIA_balance}).
The hybrid runs (hatched symbols) are offset leftwards and downwards compared to the purely QG runs, reflecting that lower Nusselt numbers and velocities are reached for the same parameters.
This shift can be understood in terms of the limitations discussed in \S~\ref{sec:res-pizza-hyb_high-Ra} with an effective lack of buoyancy power when $Ra$ is increased, yielding lower velocities -- {\it i.e.} $Re_c$ -- and lower heat transport effectiveness -- {\it i.e.} $Nu$, hence weaker $Ra(Nu-1)$ -- for the same control parameters $\left\lbrace Ek\,, \; Pr\,, \; Ra \right\rbrace$.
Note that the purely 3D runs (empty symbols) are also shifted towards the left and stand between the Hybrid and the QG cases, indicating that the QG runs are, in contrast, overpowered compared to the 3D cases.
This is likely due to the cylindrical boundary conditions in the QG case: the temperature imposed at the inner boundary is fixed for the whole column (and not only for the inner core surface) at all $Ra$, artificially supplying more thermal power to the bulk compared to the purely 3D setup (see Fig.~\ref{fig:temperature-vs-R-and-S}(b)).
Parameter studies with either QG or 3D models \citep[{\it e.g.},][]{gillet2006quasi,guervilly2010thesis,king2013flow} have reported exponents steeper than $2/5$. This has been attributed by \cite{gastine2016scaling} to the sizeable role played by viscous dissipation in the boundary layers for $Re_c < 10^4$.
Discrepancies arise at high $Ek$ and low $Ra/Ra_c$ where the VAC balance is probably more suitable and at high $Ra/Ra_c$ where the QG approximation no longer holds \citep[as has been observed in, {\it e.g.},][]{gastine2016scaling}.
The CIA scaling law hence partly captures the actual scaling behaviour of the convective velocity: fitting all the data with $Pr=1$ yields $Re_c = 0.53\, [Ra(Nu -1) Ek^{1/2}]^{0.43}$ in reasonable agreement with the theory, but the $Pr$ dependence is not well accounted for.

Despite the well-known limitations of the QG approximation, and the limitations of our hybrid method at high $Ra/Ra_c$ as documented above, we find that across our entire suite of calculations, results are broadly consistent with the Rhines and CIA scaling laws with a remaining dependence on $Pr$ for the latter.
This lends additional support to findings of previous studies that also favoured a CIA balance but focused on a weaker forcing regime in a full sphere geometry \citep{guervilly2019turbulent}, used a thinner spherical shell geometry and a different gravity profile \citep{gastine2016scaling} or used heat flux boundary conditions \citep{long2020scaling}.

\section{Summary and discussion}
\label{sec:Conclusion}

We have used QG, 3D and Hybrid models, the latter involving a QG velocity field and a 3D temperature field, to explore in a thick spherical shell the regime of strongly-driven, rapidly rotating convection, focusing on low Ekman numbers $10^{-10} \leq Ek \leq 10^{-4}$, reaching supercriticalities up to $Ra \sim 160\, Ra_c$ and considering a range of Prandtl numbers close to and below unity 
$10^{-2} \leq Pr \leq 10$, also exploring the impact of laterally-varying heat flux boundary conditions.
This work involved extending the QG convection code {\tt pizza} \citep{gastine2019pizza} to include the possibility to work with laterally-varying heat flux boundary conditions, a 3D-temperature field, and a thermal wind.

Using the hybrid QG-3D approach at parameters $Ek \leq 10^{-6}$ and $Pr \leq 1$ we are able to reproduce important aspects of convective dynamics seen in 3D models for weak to moderate supercriticalities ($Ra/Ra_c \leq 10-15$).
In that regime, the meridional temperature profile and $\phi$-averaged azimuthal velocity are well retrieved although, as for purely QG models, the dynamical behaviour in the hybrid model deviates from the 3D models close to the outer boundary.
When $Ra$ is further increased, we find our hybrid model develops much weaker convective flows compared with the 3D configuration in all cases.
\new{\newr{In contrast, t}he range of agreement between the hybrid QG-3D and the 3D configurations increases when \newr{$Pr = 10^{-2}$} \citep[as suggested by][]{guervilly2019turbulent} \newr{while the QG model departs from the 3D around $Ra \sim 10\,Ra_c$. We expect that the hybrid model will eventually diverge from the full 3D model when $Ra/Ra_c$ is sufficiently high, but}
the exact value of the diverging point remains to be determined at $Pr \leq 10^{-2}$ \newr{and has not been yet numerically reached despite reaching large $Re$ values}.}
This is to some extent expected since 3D effects become important when the thermal forcing increases \citep{calkins2013three}: the appearance of non-QG, equatorially anti-symmetric, axial flows that do not vary linearly along the rotation axis -- breaking the underlying classical QG assumption -- and the associated missing correlations between $u_z$ and $T$, are part of this discrepancy.\new{  We found these effects account for up to $46\%$ of the missing convective power for a case at $Ek=10^{-6}$, $Pr=1$, $Ra = 53\, Ra_c$}.
The thick shell geometry studied and the \new{omission of the} dynamics inside the tangent cylinder, \new{as well as the enforced linearity of $u_z$,} may also play a role in the less vigorous convection found in the hybrid model.  By construction in our hybrid setup the vertical motions remain weak compared to the horizontal motions.
Alternative formulations of QG-type models have recently been proposed that aim to better represent all flow components \citep{gerick2020pressure,jackson2020plesio}; these could perhaps provide a means to improve on the results presented here.

\new{In theory the Hybrid method can be $3-5$ times faster than a 3D model when using the same resolution -- because compute time for Legendre transforms associated with the velocity field are saved. However, we did not find major computational advantages in using the hybrid method at $Pr=1$ mainly because the $z$-interpolation scheme requires the resolution to be high. At $Pr<1$, the hybrid model becomes more advantageous; since the 3D grid involves only the temperature field it can be much coarser than the 2D grid associated with the velocity field.}
The hybrid QG-3D approach studied here is nevertheless suitable for studying the rapidly-rotating regime of convection at moderate forcing ({\it i.e.} $Ra/Ra_c \leq 15$) \new{at all $Pr$} and 
\new{we envision that the hybrid method could become even more attractive in terms of computational resources, even at $Pr=1$, if the accuracy of the interpolation methods can be improved without sacrificing too much speed.}

At the relatively low Ekman numbers considered here imposing a fixed heat flux condition at the outer boundary had little impact on the convective dynamics, compared to using a fixed temperature condition.
However imposing lateral variations in the heat flux at the outer boundary can result in regions close to the boundary where the convection is inhibited \new{or suppressed, and characterised by spiral arms where only azimuthal motions are possible}, even at high supercriticalities, provided $Q^*$ -- the peak-to-peak relative amplitude of the flux perturbation -- is sufficiently large.
We also observe enhanced convective patterns in our QG models attached to the fluid regions with higher heat-flux.
Such alternating regions of inhibited and enhanced convection  are well known from previous studies in 3D \citep[{\it e.g.},][]{mound2017heat} and are evident in our 3D comparison calculations carried out in a similar regime.
When comparing the 3D, QG and hybrid models using inhomogneous thermal boundary conditions, we find that the QG and 3D configurations are qualitatively similar whereas the hybrid model again seems to lack buoyancy power.
Despite the absence of small-scale convection found in the hybrid setup, the basic heat anomaly pattern at the outer boundary and the upwelling/downwelling system under the enhanced flux regions are similar to those found in the 3D cases, suggesting that the dynamics is relatively well captured at mid-to-low latitudes.
For $Q^*$ in the range from $2-5$, and for the relatively high supercriticalities explored here the underlying convection deep in the shell is not greatly affected.

In general, we find that 
azimuthal shearing of axial vorticity dominates the convective dynamics in the bulk and leads to the formation of multiple zonal jets of alternating sign when $Ek \leq 10^{-7}$, $Re \gg 1$ and $Ro \ll 1$, as reported in previous QG studies \citep[{\it e.g.},][]{guervilly2017multiple}.
When decreasing the ratio of diffusivities such that $Pr \leq 10^{-1}$, we find the QG-dynamics is not fundamentally modified: multiple zonal jets still dominate in the bulk but a lower supercriticality is \new{required} for the same $Ro$ number, leading to the formation of fewer and wider zonal jets at $Pr<1$.

Regarding scaling laws, our data set follows reasonably well the inertial scaling of rotating convection, which relies on a triple force balance between buoyancy, Coriolis force and inertia \citep[{\it e.g}, ][]{cardin1994chaotic,gillet2006quasi}. 
The convective flow length-scale $l_\perp / d$ gradually approaches the asymptotic Rhines scaling $l_\perp \sim Ro^{1/2}$ at low $Ek$ \citep{rhines1975waves}, albeit with an exponent lower than $1/2$ for the $Ek$ considered here.
For the velocity scaling, we find that the simulations with $Pr=1$ follow reasonably well the theoretical CIA scaling $Re_c \sim [Ra(Nu -1) Pr^{-2} Ek^{1/2}]^{2/5}$ with a retrieved exponent equal to $0.43$.
We find a clear dependence of the velocity scaling behaviour on the Prandtl number, which is not well described by the classical inertial scaling of rotating convection.
The 3D runs stand between the results of the QG and the hybrid setups, suggesting that the QG configuration produces too much convective power.
This is most likely a consequence of applying temperature boundary conditions in cylindrical geometry which involves the crude approximation of a fixed temperature on the whole tangent cylinder at $s_i$, $T_{2D}(s_i) \sim 0.445$ for all $Ra$ \new{and may partly compensate for the lack of ageostrophic components in the QG models}.
On the other hand, the hybrid configuration clear\new{ly} lacks convective power at $Ra/Ra_c \ge 10-15$.
Our results are overall consistent with other recent parameter studies in different geometries and studying different ranges of the control parameters \citep[{\it e.g.},][]{gastine2016scaling,guervilly2019turbulent}.
The Prandtl number dependence seems to mainly affect the prefactors of the scaling laws, suggesting there is no fundamental change in the dynamics, at least for the parameter range explored here.

This study is a first step towards a more general hybrid QG-3D approach to Earth's core dynamics that will include the crucial effects of a 3D magnetic field on the QG-convection. 

\section*{Acknowledgements}

\new{We thank two anonymous reviewers for constructive comments that helped improve the manuscript.}  Nathana\"{e}l Schaeffer and C\'{e}line Guervilly are thanked for helpful discussions and for sharing their code for the $z$-averaging functions and for the $z$-integration of the thermal wind.
We also thank Tobias Schwaiger for providing us with his non-magnetic data base of simulations.
We additionally thank J\'{e}r\'{e}mie Vidal for helping with the {\tt SINGE} code and the convergence of the onset of convection for 3D runs at $Ek \leq 10^{-6}$.
3D global numerical computations using the \texttt{MagIC} code were performed at S-CAPAD, IPGP and using HPC resources from the IDRIS Jean Zay CPU cluster (Grant 2021-A0070410095).
QG and hybrid QG-3D numerical computations using the \texttt{pizza} code were performed at DTU Space, using the Humboldt and Larmor CPU clusters. 
OB and CCF were supported by the European Research Council (ERC) under the European
Union’s Horizon 2020 research and innovation program (grant agreement No. 772561).

\section*{Data availability}

All our data are available upon reasonable request to the corresponding author and some key parameters from our whole data-set are already included in the present article.
The codes used (pizza and MagIC) are both freely available (at \url{http://www.github.com/magic-sph/pizza} and \url{http://www.github.com/magic-sph/magic}, respectively) under the GNU GPL v3 license. 

\bibliography{artbib}

\begin{thebibliography}{57}
\expandafter\ifx\csname natexlab\endcsname\relax\def\natexlab#1{#1}\fi

\bibitem[Ascher et~al.(1995)Ascher, Ruuth, \& Wetton]{ascher1995implicit}
Ascher, U.~M., Ruuth, S.~J., \& Wetton, B.~T., 1995.
\newblock Implicit-explicit methods for time-dependent partial differential
  equations, {\it SIAM Journal on Numerical Analysis\/}, {\bf 32}(3), 797--823.

\bibitem[Ascher et~al.(1997)Ascher, Ruuth, \& Spiteri]{ascher1997implicit}
Ascher, U.~M., Ruuth, S.~J., \& Spiteri, R.~J., 1997.
\newblock Implicit-explicit runge-kutta methods for time-dependent partial
  differential equations, {\it Applied Numerical Mathematics\/}, {\bf 25}(2-3),
  151--167.

\bibitem[Aubert(2015)]{aubert2015geomagnetic}
Aubert, J., 2015.
\newblock Geomagnetic forecasts driven by thermal wind dynamics in the
  {E}arth's core, {\it Geophys. J. Int.\/}, {\bf 203}(3), 1738--1751.

\bibitem[Aubert et~al.(2001)Aubert, Brito, Nataf, Cardin, \&
  Masson]{aubert2001systematic}
Aubert, J., Brito, D., Nataf, H.-C., Cardin, P., \& Masson, J.-P., 2001.
\newblock A systematic experimental study of rapidly rotating spherical
  convection in water and liquid gallium, {\it Physics of the Earth and
  Planetary Interiors\/}, {\bf 128}(1-4), 51--74.

\bibitem[Aubert et~al.(2003)Aubert, Gillet, \&
  Cardin]{aubert2003quasigeostrophic}
Aubert, J., Gillet, N., \& Cardin, P., 2003.
\newblock Quasigeostrophic models of convection in rotating spherical shells,
  {\it Geochemistry, Geophysics, Geosystems\/}, {\bf 4}(7), 1--19.

\bibitem[Aurnou et~al.(2015)Aurnou, Calkins, Cheng, Julien, King, Nieves,
  Soderlund, \& Stellmach]{aurnou2015rotating}
Aurnou, J., Calkins, M., Cheng, J., Julien, K., King, E., Nieves, D.,
  Soderlund, K.~M., \& Stellmach, S., 2015.
\newblock Rotating convective turbulence in earth and planetary cores, {\it
  Physics of the Earth and Planetary Interiors\/}, {\bf 246}, 52--71.

\bibitem[Busse(1970)]{busse1970thermal}
Busse, F.~H., 1970.
\newblock Thermal instabilities in rapidly rotating systems, {\it Journal of
  Fluid Mechanics\/}, {\bf 44}(3), 441--460.

\bibitem[Calkins et~al.(2012)Calkins, Aurnou, Eldredge, \&
  Julien]{calkins2012influence}
Calkins, M.~A., Aurnou, J.~M., Eldredge, J.~D., \& Julien, K., 2012.
\newblock The influence of fluid properties on the morphology of core
  turbulence and the geomagnetic field, {\it Earth and Planetary Science
  Letters\/}, {\bf 359}, 55--60.

\bibitem[Calkins et~al.(2013)Calkins, Julien, \& Marti]{calkins2013three}
Calkins, M.~A., Julien, K., \& Marti, P., 2013.
\newblock Three-dimensional quasi-geostrophic convection in the rotating
  cylindrical annulus with steeply sloping endwalls, {\it Journal of Fluid
  Mechanics\/}, {\bf 732}, 214--244.

\bibitem[Calkins et~al.(2015)Calkins, Hale, Julien, Nieves, Driggs, \&
  Marti]{calkins2015equivalence}
Calkins, M.~A., Hale, K., Julien, K., Nieves, D., Driggs, D., \& Marti, P.,
  2015.
\newblock The asymptotic equivalence of fixed heat flux and fixed temperature
  thermal boundary conditions for rapidly rotating convection, {\it Journal of
  Fluid Mechanics\/}, {\bf 784}, R2.

\bibitem[Cardin \& Olson(1994)]{cardin1994chaotic}
Cardin, P. \& Olson, P., 1994.
\newblock Chaotic thermal convection in a rapidly rotating spherical shell:
  consequences for flow in the outer core, {\it Physics of the earth and
  planetary interiors\/}, {\bf 82}(3-4), 235--259.

\bibitem[Chandrasekhar(1961)]{chandrasekhar1961hydrodynamic}
Chandrasekhar, S., 1961.
\newblock {\it Hydrodynamic and Hydrodynamic stability\/}, OUP.

\bibitem[Clart{\'e} et~al.(2021)Clart{\'e}, Schaeffer, Labrosse, \&
  Vidal]{clarte2021effects}
Clart{\'e}, T.~T., Schaeffer, N., Labrosse, S., \& Vidal, J., 2021.
\newblock The effects of a robin boundary condition on thermal convection in a
  rotating spherical shell, {\it Journal of Fluid Mechanics\/}, {\bf 918}.

\bibitem[Dormy et~al.(2004)Dormy, Soward, Jones, Jault, \&
  Cardin]{dormy2004onset}
Dormy, E., Soward, A., Jones, C., Jault, D., \& Cardin, P., 2004.
\newblock The onset of thermal convection in rotating spherical shells, {\it
  Journal of Fluid Mechanics\/}, {\bf 501}, 43--70.

\bibitem[Gastine(2019)]{gastine2019pizza}
Gastine, T., 2019.
\newblock pizza: an open-source pseudo-spectral code for spherical
  quasi-geostrophic convection, {\it Geophysical Journal International\/}, {\bf
  217}(3), 1558--1576.

\bibitem[Gastine et~al.(2016)Gastine, Wicht, \& Aubert]{gastine2016scaling}
Gastine, T., Wicht, J., \& Aubert, J., 2016.
\newblock Scaling regimes in spherical shell rotating convection, {\it Journal
  of Fluid Mechanics\/}, {\bf 808}, 690--732.

\bibitem[Gerick et~al.(2020)Gerick, Jault, Noir, \& Vidal]{gerick2020pressure}
Gerick, F., Jault, D., Noir, J., \& Vidal, J., 2020.
\newblock Pressure torque of torsional alfv{\'e}n modes acting on an
  ellipsoidal mantle, {\it Geophysical Journal International\/}, {\bf 222}(1),
  338--351.

\bibitem[Gibbons et~al.(2007)Gibbons, Gubbins, \& Zhang]{gibbons2007convection}
Gibbons, S., Gubbins, D., \& Zhang, K., 2007.
\newblock Convection in rotating spherical fluid shells with inhomogeneous heat
  flux at the outer boundary, {\it Geophysical and Astrophysical Fluid
  Dynamics\/}, {\bf 101}(5-6), 347--370.

\bibitem[Gillet \& Jones(2006)]{gillet2006quasi}
Gillet, N. \& Jones, C., 2006.
\newblock The quasi-geostrophic model for rapidly rotating spherical convection
  outside the tangent cylinder, {\it Journal of Fluid Mechanics\/}, {\bf 554},
  343--369.

\bibitem[Glatzmaier(1984)]{glatzmaier1984numerical}
Glatzmaier, G.~A., 1984.
\newblock Numerical simulations of stellar convective dynamos. i. the model and
  method, {\it Journal of Computational Physics\/}, {\bf 55}(3), 461--484.

\bibitem[Glatzmaier \& Roberts(1996)]{glatzmaier1996evolutionary}
Glatzmaier, G.~A. \& Roberts, P.~H., 1996.
\newblock An anelastic evolutionary geodynamo simulation driven by
  compositional and thermal convection, {\it Physica D: Nonlinear Phenomena\/},
  {\bf 97}(1), 81--94.

\bibitem[Goluskin(2016)]{goluskin2016internally}
Goluskin, D., 2016.
\newblock {\it Internally heated convection and Rayleigh-B{\'e}nard
  convection\/}, Springer.

\bibitem[Gubbins et~al.(2003)Gubbins, Dehant, Creager, Karato, \&
  Zatman]{gubbins2003thermal}
Gubbins, D., Dehant, V., Creager, K., Karato, S., \& Zatman, S., 2003.
\newblock Thermal core-mantle interactions: theory and observations, {\it
  Earth’s core: dynamics, structure, rotation\/}, pp. 163--179.

\bibitem[Guervilly(2010)]{guervilly2010thesis}
Guervilly, C., 2010.
\newblock Dynamos numériques planétaires générées par cisaillement en
  surface ou chauffage interne, {\it Universit{\'e} de Grenoble\/}, {\bf
  France}.

\bibitem[Guervilly \& Cardin(2016)]{guervilly2016subcritical}
Guervilly, C. \& Cardin, P., 2016.
\newblock Subcritical convection of liquid metals in a rotating sphere using a
  quasi-geostrophic model, {\it Journal of Fluid Mechanics\/}, {\bf 808},
  61--89.

\bibitem[Guervilly \& Cardin(2017)]{guervilly2017multiple}
Guervilly, C. \& Cardin, P., 2017.
\newblock Multiple zonal jets and convective heat transport barriers in a
  quasi-geostrophic model of planetary cores, {\it Geophysical Journal
  International\/}, {\bf 211}(1), 455--471.

\bibitem[Guervilly et~al.(2019)Guervilly, Cardin, \&
  Schaeffer]{guervilly2019turbulent}
Guervilly, C., Cardin, P., \& Schaeffer, N., 2019.
\newblock Turbulent convective length scale in planetary cores, {\it Nature\/},
  {\bf 570}(7761), 368--371.

\bibitem[Hide(1974)]{hide1974jupiter}
Hide, R., 1974.
\newblock Jupiter and saturn, {\it Proceedings of the Royal Society of London.
  A. Mathematical and Physical Sciences\/}, {\bf 336}(1604), 63--84.

\bibitem[Ingersoll \& Pollard(1982)]{ingersoll1982motion}
Ingersoll, A.~P. \& Pollard, D., 1982.
\newblock Motion in the interiors and atmospheres of jupiter and saturn: Scale
  analysis, anelastic equations, barotropic stability criterion, {\it
  Icarus\/}, {\bf 52}(1), 62--80.

\bibitem[Jackson \& Maffei(2020)]{jackson2020plesio}
Jackson, A. \& Maffei, S., 2020.
\newblock Plesio-geostrophy for earth’s core: I. basic equations, inertial
  modes and induction, {\it Proceedings of the Royal Society A: Mathematical,
  Physical and Engineering Sciences\/}.

\bibitem[{Jones}(2015)]{jones2015TOG}
{Jones}, C.~A., 2015.
\newblock {8.05 Thermal and Compositional Convection in the Outer Core}, in
  {\em Treatise on Geophysics (Second Edition)\/}, pp. 115 --159, ed. Schubert,
  G., Elsevier, Oxford, second edition edn.

\bibitem[Julien et~al.(2012)Julien, Knobloch, Rubio, \& Vasil]{julien2012heat}
Julien, K., Knobloch, E., Rubio, A.~M., \& Vasil, G.~M., 2012.
\newblock Heat transport in low-rossby-number rayleigh-b{\'e}nard convection,
  {\it Physical review letters\/}, {\bf 109}(25), 254503.

\bibitem[King \& Aurnou(2013)]{king2013turbulent}
King, E.~M. \& Aurnou, J.~M., 2013.
\newblock Turbulent convection in liquid metal with and without rotation, {\it
  Proceedings of the National Academy of Sciences\/}, {\bf 110}(17),
  6688--6693.

\bibitem[King \& Buffett(2013)]{king2013flow}
King, E.~M. \& Buffett, B.~A., 2013.
\newblock Flow speeds and length scales in geodynamo models: The role of
  viscosity, {\it Earth Planet. Sc. Lett.\/}, {\bf 371}, 156--162.

\bibitem[Labb{\'e} et~al.(2015)Labb{\'e}, Jault, \&
  Gillet]{labbe2015magnetostrophic}
Labb{\'e}, F., Jault, D., \& Gillet, N., 2015.
\newblock On magnetostrophic inertia-less waves in quasi-geostrophic models of
  planetary cores, {\it Geophysical \& Astrophysical Fluid Dynamics\/}, {\bf
  109}(6), 587--610.

\bibitem[Long et~al.(2020)Long, Mound, Davies, \& Tobias]{long2020scaling}
Long, R., Mound, J., Davies, C., \& Tobias, S., 2020.
\newblock Scaling behaviour in spherical shell rotating convection with
  fixed-flux thermal boundary conditions, {\it Journal of Fluid Mechanics\/},
  {\bf 889}.

\bibitem[Maffei et~al.(2017)Maffei, Jackson, \&
  Livermore]{maffei2017characterization}
Maffei, S., Jackson, A., \& Livermore, P.~W., 2017.
\newblock Characterization of columnar inertial modes in rapidly rotating
  spheres and spheroids, {\it Proceedings of the Royal Society A: Mathematical,
  Physical and Engineering Sciences\/}, {\bf 473}(2204), 20170181.

\bibitem[Marti et~al.(2016)Marti, Calkins, \& Julien]{marti2016computationally}
Marti, P., Calkins, M., \& Julien, K., 2016.
\newblock A computationally efficient spectral method for modeling core
  dynamics, {\it Geochemistry, Geophysics, Geosystems\/}, {\bf 17}(8),
  3031--3053.

\bibitem[Mound \& Davies(2017)]{mound2017heat}
Mound, J.~E. \& Davies, C.~J., 2017.
\newblock Heat transfer in rapidly rotating convection with heterogeneous
  thermal boundary conditions, {\it Journal of Fluid Mechanics\/}, {\bf 828},
  601--629.

\bibitem[Muite(2010)]{muite2010numerical}
Muite, B., 2010.
\newblock A numerical comparison of chebyshev methods for solving fourth order
  semilinear initial boundary value problems, {\it Journal of computational and
  applied mathematics\/}, {\bf 234}(2), 317--342.

\bibitem[Rhines(1975)]{rhines1975waves}
Rhines, P.~B., 1975.
\newblock Waves and turbulence on a beta-plane, {\it Journal of Fluid
  Mechanics\/}, {\bf 69}(3), 417--443.

\bibitem[{Roberts} \& {King}(2013)]{roberts2013genesis}
{Roberts}, P.~H. \& {King}, E.~M., 2013.
\newblock {On the genesis of the Earth's magnetism}, {\it Reports on Progress
  in Physics\/}, {\bf 76}(9), 096801.

\bibitem[Rossby(1969)]{rossby1969study}
Rossby, H., 1969.
\newblock A study of b{\'e}nard convection with and without rotation, {\it
  Journal of Fluid Mechanics\/}, {\bf 36}(2), 309--335.

\bibitem[Sahoo \& Sreenivasan(2020)]{sahoo2020convection}
Sahoo, S. \& Sreenivasan, B., 2020.
\newblock Convection in a rapidly rotating cylindrical annulus with laterally
  varying boundary heat flux, {\it Journal of Fluid Mechanics\/}, {\bf 883}.

\bibitem[Sakuraba \& Roberts(2009)]{sakuraba2009generation}
Sakuraba, A. \& Roberts, P.~H., 2009.
\newblock Generation of a strong magnetic field using uniform heat flux at the
  surface of the core, {\it Nature Geoscience\/}, {\bf 2}(11), 802--805.

\bibitem[Schaeffer(2013)]{schaeffer2013efficient}
Schaeffer, N., 2013.
\newblock Efficient spherical harmonic transforms aimed at pseudospectral
  numerical simulations, {\it Geochemistry, Geophysics, Geosystems\/}, {\bf
  14}(3), 751--758.

\bibitem[Schaeffer \& Cardin(2005)]{schaeffer2005quasigeostrophic}
Schaeffer, N. \& Cardin, P., 2005.
\newblock Quasigeostrophic model of the instabilities of the stewartson layer
  in flat and depth-varying containers, {\it Physics of Fluids\/}, {\bf
  17}(10), 104111.

\bibitem[Schaeffer \& Cardin(2006)]{schaeffer2006quasi}
Schaeffer, N. \& Cardin, P., 2006.
\newblock Quasi-geostrophic kinematic dynamos at low magnetic prandtl number,
  {\it Earth and Planetary Science Letters\/}, {\bf 245}(3-4), 595--604.

\bibitem[Schaeffer et~al.(2017)Schaeffer, Jault, Nataf, \&
  Fournier]{schaeffer2017turbulent}
Schaeffer, N., Jault, D., Nataf, H.-C., \& Fournier, A., 2017.
\newblock Turbulent geodynamo simulations: a leap towards {E}arth's core, {\it
  Geophys. J. Int.\/}, {\bf 211}(1), 1--29.

\bibitem[Schwaiger et~al.(2021)Schwaiger, Gastine, \&
  Aubert]{schwaiger2020relating}
Schwaiger, T., Gastine, T., \& Aubert, J., 2021.
\newblock Relating force balances and flow length scales in geodynamo
  simulations, {\it Geophysical Journal International\/}, {\bf 224}(3),
  1890--1904.

\bibitem[Sheyko et~al.(2018)Sheyko, Finlay, Favre, \& Jackson]{sheyko2018scale}
Sheyko, A., Finlay, C., Favre, J., \& Jackson, A., 2018.
\newblock Scale separated low viscosity dynamos and dissipation within the
  earth’s core, {\it Nature Scientific reports\/}, {\bf 8}(1), 1--7.

\bibitem[Stellmach \& Hansen(2008)]{stellmach2008efficient}
Stellmach, S. \& Hansen, U., 2008.
\newblock An efficient spectral method for the simulation of dynamos in
  cartesian geometry and its implementation on massively parallel computers,
  {\it Geochemistry, Geophysics, Geosystems\/}, {\bf 9}(5).

\bibitem[{Stellmach} et~al.(2014){Stellmach}, {Lischper}, {Julien}, {Vasil},
  {Cheng}, {Ribeiro}, {King}, \& {Aurnou}]{stellmach2014approaching}
{Stellmach}, S., {Lischper}, M., {Julien}, K., {Vasil}, G., {Cheng}, J.~S.,
  {Ribeiro}, A., {King}, E.~M., \& {Aurnou}, J.~M., 2014.
\newblock {Approaching the Asymptotic Regime of Rapidly Rotating Convection:
  Boundary Layers versus Interior Dynamics}, {\it Phys.~Rev.~Lett.\/}, {\bf
  113}(25), 254501.

\bibitem[Valdettaro et~al.(2007)Valdettaro, Rieutord, Braconnier, \&
  Frayss{\'e}]{valdettaro2007convergence}
Valdettaro, L., Rieutord, M., Braconnier, T., \& Frayss{\'e}, V., 2007.
\newblock Convergence and round-off errors in a two-dimensional eigenvalue
  problem using spectral methods and arnoldi--chebyshev algorithm, {\it Journal
  of computational and applied mathematics\/}, {\bf 205}(1), 382--393.

\bibitem[Vidal \& Schaeffer(2015)]{vidal2015singe}
Vidal, J. \& Schaeffer, N., 2015.
\newblock {Quasi-geostrophic modes in the Earth's fluid core with an outer
  stably stratified layer}, {\it Geophysical Journal International\/}, {\bf
  202}(3), 2182--2193.

\bibitem[Wicht(2002)]{wicht2002inner}
Wicht, J., 2002.
\newblock Inner-core conductivity in numerical dynamo simulations, {\it Physics
  of the Earth and Planetary Interiors\/}, {\bf 132}(4), 281--302.

\bibitem[{Yadav} et~al.(2016){Yadav}, {Gastine}, {Christensen}, {Duarte}, \&
  {Reiners}]{yadav2016effect}
{Yadav}, R.~K., {Gastine}, T., {Christensen}, U.~R., {Duarte}, L.~D.~V., \&
  {Reiners}, A., 2016.
\newblock {Effect of shear and magnetic field on the heat-transfer efficiency
  of convection in rotating spherical shells}, {\it Geophysical Journal
  International\/}, {\bf 204}(2), 1120--1133.

\end{thebibliography}

\bibliographystyle{gji}


\onecolumn

\appendix
\section{Numerical simulations with fixed temperature contrast}
\label{sec:Append-A-Results}

\begin{center}
\begin{longtable}{rrlrrrrrc}
\caption{Summary of the numerical simulations with fixed temperature contrast computed in this study. All models have been computed with $\eta = r_i/r_o = 0.35$. $Ra$ is the Rayleigh number (the supercriticality $Ra = \cdot \times Ra_c$ is also provided), $Pr$ is the Prandtl number, $Nu$ is the Nusselt number, $Re_c$ is the convective Reynolds number, $Re_\text{zon}$ is the zonal Reynolds number, $\mathcal{L}_{u_s}$ is the typical length-scale for the cylindrical radial velocity field and $(N_s, N_m)/(N_r, \ell_{\text{max}})$ are the grid-size for the run.}
\label{tab:run_list} \\
\hline
Method  & $Ra$	& $= \cdot \times Ra_c$	& $Pr$	& $Nu$	& $Re_c$    & $Re_\text{zon}$	& $\mathcal{L}_{u_s}$	& $(N_s, N_m)/(N_r, \ell_{\text{max}})$ \\
\hline
\endfirsthead

\hline 
Method  & $Ra$	& $= \cdot \times Ra_c$	& $Pr$	& $Nu$	& $Re_c$    & $Re_\text{zon}$	& $\mathcal{L}_{u_s}$ & $(N_s, N_m)/(N_r, \ell_{\text{max}})$ \\
\hline
\endhead

\hline
\multicolumn{9}{c}{Continued on next page $\ldots$} \\
\hline
\endfoot

\hline
\hline
\endlastfoot

	& 	& 	&   & 	& $Ek = 1 \times 10^{-4}$	& 	&   &   \\
QG  & $4.83 \times 10^{6}$	& $= 4.8 \times$	& $10$	& $2.62$	& $7.0$    & $0.5$	& $3.651 \times 10^{-1}$	& $(193,192)/(-,-)$ \\
QG  & $1.12 \times 10^{7}$	& $= 10.2 \times$	& $10$	& $7.76$	& $14.7$   & $1.3$	& $3.296 \times 10^{-1}$	& $(193,192)/(-,-)$ \\
QG  & $3.87 \times 10^{7}$	& $= 35.1 \times$	& $10$	& $18.02$	& $36.0$   & $6.6$	& $2.993 \times 10^{-1}$	& $(193,192)/(-,-)$ \\
QG  & $2.43 \times 10^{6}$	& $= 3.4 \times$	& $1$	& $1.91$	& $35.9$   & $14.7$	& $4.481 \times 10^{-1}$	& $(97,96)/(-,-)$ \\
Hybrid  & $2.00 \times 10^{6}$	& $= 6.6 \times$	& $1$	& $1.46$	& $54.4$   & $27.2$	& $4.479 \times 10^{-1}$	& $(97,96)/(97,96)$ \\
QG  & $4.83 \times 10^{6}$	& $= 6.8 \times$	& $1$	& $3.63$	& $72.2$   & $40.0$	& $3.769 \times 10^{-1}$	& $(97,96)/(-,-)$ \\
QG  & $1.93 \times 10^{7}$	& $= 27.1 \times$	& $1$	& $11.95$	& $211.5$   & $191.7$	& $5.006 \times 10^{-1}$	& $(97,96)/(-,-)$ \\
QG  & $3.87 \times 10^{7}$	& $= 54.3 \times$	& $1$	& $17.69$	& $335.2$   & $339.4$	& $5.884 \times 10^{-1}$	& $(97,96)/(-,-)$ \\
QG  & $4.30 \times 10^{5}$	& $= 1.7 \times$	& $0.1$	& $1.05$	& $44.6$    & $28.3$	& $7.412 \times 10^{-1}$	& $(193,192)/(-,-)$ \\
QG  & $9.66 \times 10^{5}$	& $= 3.3 \times$	& $0.1$	& $1.25$	& $114.4$   & $120.5$	& $6.699 \times 10^{-1}$	& $(193,192)/(-,-)$ \\
QG  & $4.83 \times 10^{6}$	& $= 16.6 \times$	& $0.1$	& $3.97$	& $633.0$   & $912.2$	& $6.823 \times 10^{-1}$	& $(193,192)/(-,-)$ \\
	& 	& 	&   & 	& $Ek = 1 \times 10^{-5}$	& 	&   &   \\
QG  & $1.12 \times 10^{8}$	& $= 6.2 \times$	& $10$	& $3.97$	& $19.0$    & $0.9$ & $1.559 \times 10^{-1}$	& $(1537,1536)/(-,-)$ \\
QG  & $2.25 \times 10^{8}$	& $= 12.4 \times$	& $10$	& $11.86$	& $38.8$    & $2.7$	& $1.424 \times 10^{-1}$	& $(1537,1536)/(-,-)$ \\
3D  & $5.00 \times 10^{8}$	& $= 30.8 \times$	& $10$	& $15.03$	& $97.1$   & $13.1$	& $-$	& $(-,-)/(193,170)$ \\
QG  & $1.12 \times 10^{9}$	& $= 61.8 \times$	& $10$	& $50.66$	& $131.0$   & $16.1$	& $1.849 \times 10^{-1}$	& $(1537,1536)/(-,-)$ \\
3D  & $2.00 \times 10^{9}$	& $= 123.4 \times$	& $10$	& $36.63$	& $319.8$   & $53.0$	& $-$	& $(-,-)/(256,213)$ \\
QG  & $2.25 \times 10^{9}$	& $= 123.5 \times$	& $10$	& $72.15$	& $210.2$   & $38.3$	& $2.215 \times 10^{-1}$	& $(1537,1536)/(-,-)$ \\
3D  & $4.00 \times 10^{7}$	& $= 3.8 \times$	& $1$	& $1.39$	& $62.9$   & $21.9$	& $-$	& $(-,-)/(97,170)$ \\
Hybrid  & $4.64 \times 10^{7}$	& $= 4.7 \times$	& $1$	& $1.37$	& $78.8$    & $27.5$	& $2.172 \times 10^{-1}$	& $(193,128)/(193,128)$ \\
QG  & $5.21 \times 10^{7}$	& $= 5.0 \times$	& $1$	& $2.48$	& $111.5$   & $31.2$	& $2.107 \times 10^{-1}$	& $(193,192)/(-,-)$ \\
Hybrid  & $9.28 \times 10^{7}$	& $= 9.4 \times$	& $1$	& $1.60$	& $154.9$   & $57.4$	& $2.385 \times 10^{-1}$	& $(193,128)/(193,128)$ \\
3D  & $1.00 \times 10^{8}$	& $= 9.5 \times$	& $1$	& $2.45$	& $194.5$   & $53.8$	& $-$	& $(-,-)/(97,213)$ \\
QG  & $1.04 \times 10^{8}$	& $= 9.9 \times$	& $1$	& $4.82$	& $232.8$   & $81.5$	& $2.578 \times 10^{-1}$	& $(193,192)/(-,-)$ \\
Hybrid  & $1.86 \times 10^{8}$	& $= 18.9 \times$	& $1$	& $1.82$	& $272.9$   & $131.2$	& $2.899 \times 10^{-1}$	& $(193,128)/(193,128)$ \\
3D  & $2.00 \times 10^{8}$	& $= 19.0 \times$	& $1$	& $5.42$	& $471.3$   & $161.0$	& $-$	& $(-,-)/(97,256)$ \\
QG  & $2.09 \times 10^{8}$	& $= 19.9 \times$	& $1$	& $10.88$	& $436.8$   & $214.1$	& $3.225 \times 10^{-1}$	& $(193,192)/(-,-)$ \\
3D  & $3.00 \times 10^{8}$	& $= 28.6 \times$	& $1$	& $8.99$	& $741.1$   & $338.3$	& $-$	& $(-,-)/(121,288)$ \\
3D  & $4.00 \times 10^{8}$	& $= 38.1 \times$	& $1$	& $11.99$	& $959.3$   & $499.5$	& $-$	& $(-,-)/(121,288)$ \\
QG  & $4.18 \times 10^{8}$	& $= 39.8 \times$	& $1$	& $21.99$	& $775.1$   & $528.3$	& $4.107 \times 10^{-1}$	& $(193,192)/(-,-)$ \\
QG  & $5.21 \times 10^{8}$	& $= 49.7 \times$	& $1$	& $26.52$	& $885.7$   & $687.5$	& $4.765 \times 10^{-1}$	& $(193,192)/(-,-)$ \\
3D  & $7.00 \times 10^{8}$	& $= 66.7 \times$	& $1$	& $18.09$	& $1485.4$   & $1031.8$	& $-$	& $(-,-)/(161,426)$ \\
QG  & $8.36 \times 10^{8}$	& $= 79.6 \times$	& $1$	& $30.82$	& $1216.9$  & $1206.1$	& $4.729 \times 10^{-1}$	& $(385,384)/(-,-)$ \\
QG  & $1.08 \times 10^{7}$	& $= 3.0 \times$	& $0.1$	& $1.31$	& $243.8$   & $230.9$	& $3.568 \times 10^{-1}$	& $(193,192)/(-,-)$ \\
\new{Hybrid}  & $4.00 \times 10^{7}$	& $= 11.3 \times$	& $0.1$	& $1.41$	& $789.1$   & $1067.7$	& $5.062 \times 10^{-1}$	& $(193,240)/(129,144)$ \\
QG  & $2.16 \times 10^{7}$	& $= 6.0 \times$	& $0.1$	& $1.99$	& $554.4$   & $665.8$	& $4.432 \times 10^{-1}$	& $(193,192)/(-,-)$ \\
QG  & $4.31 \times 10^{7}$	& $= 11.9 \times$	& $0.1$	& $3.47$	& $1059.5$  & $1493.5$	& $5.969 \times 10^{-1}$	& $(385,384)/(-,-)$ \\
QG  & $5.39 \times 10^{7}$	& $= 14.9 \times$	& $0.1$	& $4.27$	& $1280.3$  & $1876.1$	& $6.310 \times 10^{-1}$	& $(385,384)/(-,-)$ \\
QG  & $2.25 \times 10^{8}$	& $= 62.0 \times$	& $0.1$	& $15.52$	& $3841.3$  & $6703.9$	& $6.159 \times 10^{-1}$	& $(769,768)/(-,-)$ \\
3D  & $2.50 \times 10^{8}$	& $= 70.8 \times$	& $0.1$	& $8.05$	& $4426.5$   & $5711.5$	& $-$	& $(-,-)/(145,426)$ \\
Hybrid  & $2.50 \times 10^{8}$	& $= 70.8 \times$	& $0.1$	& $1.82$	& $2584.0$   & $4611.2$	& $7.549 \times 10^{-1}$	& $(449,480)/(321,352)$ \\
QG  & $1.12 \times 10^{7}$	& $= 5.7 \times$	& $0.01$	& $1.87$	& $3074.0$  & $5093.7$	& $8.048 \times 10^{-1}$	& $(769,768)/(-,-)$ \\
\new{Hybrid}  & $1.00 \times 10^{7}$	& $= 5.1 \times$	& $0.01$	& $1.22$	& $2614.7$   & $3995.6$	& $7.580 \times 10^{-1}$	& $(513,512)/(321,352)$ \\
	& 	& 	&   &   & $Ek = 1 \times 10^{-6}$	& 	&   &  \\
QG  & $2.25 \times 10^{10}$	& $= 67.6 \times$	& $10$	& $50.47$	& $493.7$   & $89.0$	& $1.075 \times 10^{-1}$	& $(4097,4096)/(-,-)$ \\
3D  & $8.00 \times 10^{8}$	& $= 4.3 \times$	& $1$	& $1.42$	& $146.7$   & $33.1$	& $-$	& $(-,-)/(129,256)$ \\
3D  & $1.00 \times 10^{9}$	& $= 5.3 \times$	& $1$	& $1.55$	& $195.9$   & $41.7$	& $-$	& $(-,-)/(129,256)$ \\
Hybrid  & $1.00 \times 10^{9}$	& $= 6.1 \times$	& $1$	& $1.44$	& $206.8$   & $35.4$	& $1.164 \times 10^{-1}$	& $(257,256)/(257,256)$ \\
QG  & $1.12 \times 10^{9}$	& $= 6.3 \times$	& $1$	& $2.81$	& $292.2$   & $48.5$	& $1.208 \times 10^{-1}$	& $(385,384)/(-,-)$ \\
3D  & $2.00 \times 10^{9}$	& $= 10.6 \times$	& $1$	& $2.38$	& $448.1$   & $97.1$	& $-$	& $(-,-)/(129,341)$ \\
Hybrid  & $ 2.00 \times 10^{9}$	& $= 12.3 \times$	& $1$	& $1.67$	& $404.4$   & $86.9$	& $1.456 \times 10^{-1}$	& $(257,256)/(257,256)$ \\
QG  & $2.00 \times 10^{9}$	& $= 11.3 \times$	& $1$	& $5.05$	& $550.4$   & $126.6$	& $1.547 \times 10^{-1}$	& $(385,384)/(-,-)$ \\
QG  & $2.25 \times 10^{9}$	& $= 12.7 \times$	& $1$	& $5.75$	& $620.7$   & $153.6$	& $1.598 \times 10^{-1}$	& $(385,384)/(-,-)$ \\
3D  & $2.80 \times 10^{9}$	& $= 14.9 \times$	& $1$	& $3.44$	& $700.6$   & $176.9$	& $-$	& $(-,-)/(161,426)$ \\
3D  & $3.50 \times 10^{9}$	& $= 18.6 \times$	& $1$	& $4.77$	& $997.7$   & $266.6$	& $-$	& $(-,-)/(181,426)$ \\
Hybrid  & $ 4.00 \times 10^{9}$	& $= 24.5 \times$	& $1$	& $2.01$	& $740.7$   & $256.0$	& $1.879 \times 10^{-1}$	& $(257,256)/(257,256)$ \\
QG  & $4.49 \times 10^{9}$	& $= 25.3 \times$	& $1$	& $12.79$	& $1190.0$  & $539.7$	& $2.121 \times 10^{-1}$	& $(385,384)/(-,-)$ \\
3D  & $5.50 \times 10^{9}$	& $= 29.3 \times$	& $1$	& $9.43$	& $1803.9$   & $1014.4$	& $-$	& $(-,-)/(201,426)$ \\
Hybrid  & $ 7.80 \times 10^{9}$	& $= 47.9 \times$	& $1$	& $2.14$	& $1239.3$  & $571.8$	& $2.401 \times 10^{-1}$	& $(385,384)/(385,384)$ \\
QG  & $8.99 \times 10^{9}$	& $= 50.7 \times$	& $1$	& $28.31$	& $2145.5$  & $1419.2$	& $2.902 \times 10^{-1}$	& $(577,576)/(-,-)$ \\
3D  & $1.00 \times 10^{10}$	& $= 53.2 \times$	& $1$	& $19.46$	& $3339.9$   & $2405.2$	& $-$	& $(-,-)/(321,682)$ \\
Hybrid  & $ 1.00 \times 10^{10}$	& $= 61.3 \times$	& $1$	& $2.20$	& $1455.9$  & $829.5$	& $2.622 \times 10^{-1}$	& $(513,512)/(513,341)$ \\
QG  & $1.12 \times 10^{10}$	& $= 63.4 \times$	& $1$	& $36.27$	& $2584.4$  & $1845.5$	& $3.192 \times 10^{-1}$	& $(577,576)/(-,-)$ \\
3D  & $1.60 \times 10^{10}$	& $= 85.1 \times$	& $1$	& $31.36$	& $5272.2$   & $4102.7$	& $-$	& $(-,-)/(433,682)$ \\
QG  & $1.80 \times 10^{10}$	& $= 101.4 \times$	& $1$	& $58.65$	& $3793.5$  & $3109.7$	& $3.882 \times 10^{-1}$	& $(577,576)/(-,-)$ \\
QG  & $2.25 \times 10^{10}$	& $= 126.7 \times$	& $1$	& $72.09$	& $4520.3$  & $3957.2$	& $4.265 \times 10^{-1}$	& $(769,768)/(-,-)$ \\
Hybrid  & $ 2.60 \times 10^{10}$	& $= 159.5 \times$	& $1$	& $2.90$	& $2963.4$  & $2090.2$	& $3.829 \times 10^{-1}$	& $(513,512)/(513,341)$ \\
3D  &  $2.66 \times 10^{10}$	& $= 141.5 \times$	& $1$	& $45.94$	& $7285.0$   & $7945.6$	& $-$	& $(-,-)/(577,853)$ \\
Hybrid  & $ 1.03 \times 10^{8}$	& $= 2.0 \times$	& $0.1$	& $1.09$	& $317.5$   & $227.5$	& $1.671 \times 10^{-1}$	& $(385,384)/(257,171)$ \\
QG  & $1.16 \times 10^{8}$	& $= 2.2 \times$	& $0.1$	& $1.22$	& $396.8$   & $285.3$	& $1.790 \times 10^{-1}$	& $(385,384)/(-,-)$ \\
Hybrid  & $ 2.06 \times 10^{8}$	& $= 4.1 \times$	& $0.1$	& $1.23$	& $821.6$   & $684.4$	& $2.438 \times 10^{-1}$	& $(385,384)/(257,171)$ \\
QG  & $2.31 \times 10^{8}$	& $= 4.4 \times$	& $0.1$	& $1.73$	& $1049.2$  & $855.3$	& $2.564 \times 10^{-1}$	& $(385,384)/(-,-)$ \\
QG  & $4.63 \times 10^{8}$	& $= 8.8 \times$	& $0.1$	& $2.84$	& $2161.3$  & $2044.0$	& $3.378 \times 10^{-1}$	& $(385,384)/(-,-)$ \\
QG  & $9.26 \times 10^{8}$	& $= 17.7 \times$	& $0.1$	& $5.43$	& $3990.9$  & $4472.9$	& $4.296 \times 10^{-1}$	& $(577,576)/(-,-)$ \\
3D  & $1.00 \times 10^{9}$	& $= 19.7 \times$	& $0.1$	& $2.77$	& $3790.7$   & $3458.6$	& $-$	& $(-,-)/(257,309)$ \\
Hybrid  & $ 1.00 \times 10^{9}$	& $= 19.7 \times$	& $0.1$	& $1.68$	& $3196.6$  & $3663.4$	& $3.678 \times 10^{-1}$	& $(385,384)/(257,171)$ \\
QG  & $1.15 \times 10^{9}$	& $= 22.1 \times$	& $0.1$	& $6.87$	& $4768.8$  & $5706.2$	& $4.569 \times 10^{-1}$	& $(577,576)/(-,-)$ \\
QG  & $1.85 \times 10^{9}$	& $= 35.3 \times$	& $0.1$	& $11.72$	& $6791.4$  & $9383.2$	& $4.970 \times 10^{-1}$	& $(769,768)/(-,-)$ \\
3D  & $2.00 \times 10^{9}$	& $= 39.4 \times$	& $0.1$	& $5.95$	& $7575.3$   & $8782.8$	& $-$	& $(-,-)/(321,426)$ \\
Hybrid  & $ 2.00 \times 10^{9}$	& $= 39.4 \times$	& $0.1$	& $1.89$	& $4975.7$  & $6699.7$	& $4.566 \times 10^{-1}$	& $(385,384)/(257,256)$ \\
QG  & $2.25 \times 10^{9}$	& $= 42.9 \times$	& $0.1$	& $13.95$	& $7845.2$  & $11884.2$	& $4.495 \times 10^{-1}$	& $(769,768)/(-,-)$ \\
3D  & $5.00 \times 10^{9}$	& $= 98.5 \times$	& $0.1$	& $12.77$	& $12743.8$   & $22752.8$	& $-$	& $(-,-)/(385,1024)$ \\
Hybrid  & $5.00 \times 10^{9}$	& $= 98.5 \times$	& $0.1$	& $2.14$	& $8225.4$  & $13536.3$	& $5.201 \times 10^{-1}$	& $(769,768)/(385,384)$ \\
QG  & $4.85 \times 10^{7}$	& $= 1.8 \times$	& $0.01$	& $1.20$	& $2274.7$  & $2943.8$	& $4.144 \times 10^{-1}$	& $(577,576)/(-,-)$ \\
QG  & $9.71 \times 10^{7}$	& $= 3.6 \times$	& $0.01$	& $1.65$	& $4930.7$  & $7871.4$	& $4.566 \times 10^{-1}$	& $(577,576)/(-,-)$ \\
\new{Hybrid}  & $1.00 \times 10^{8}$	& $= 3.9 \times$	& $0.01$	& $1.23$	& $4749.7$  & $7399.9$	& $4.566 \times 10^{-1}$	& $(353,384)/(225,256)$ \\
\new{Hybrid}  & $1.30 \times 10^{8}$	& $= 5.0 \times$	& $0.01$	& $1.29$	& $6020.8$  & $9797.3$	& $4.908 \times 10^{-1}$	& $(513,512)/(225,256)$ \\
QG  & $1.94 \times 10^{8}$	& $= 7.1 \times$	& $0.01$	& $2.54$	& $8888.8$  & $17077.8$	& $4.993 \times 10^{-1}$	& $(1025,1024)/(-,-)$ \\
\new{3D}  & $2.00 \times 10^{8}$	& $= 7.7 \times$	& $0.01$	& $1.45$	& $6538.2$  & $11456.1$	& $-$	& $(-,-)/(321,682)$ \\
\new{Hybrid}  & $2.00 \times 10^{8}$	& $= 7.7 \times$	& $0.01$	& $1.38$	& $8233.1$  & $14835.2$	& $4.583 \times 10^{-1}$	& $(961,1008)/(241,288)$ \\
QG  & $2.25 \times 10^{8}$	& $= 8.9 \times$	& $0.01$	& $2.97$	& $9761.0$  & $19614.7$	& $4.339 \times 10^{-1}$	& $(1025,1024)/(-,-)$ \\
QG  & $2.43 \times 10^{8}$	& $= 8.2 \times$	& $0.01$	& $3.00$	& $10636.1$ & $21312.7$	& $5.072 \times 10^{-1}$	& $(1537,1536)/(-,-)$ \\
\new{Hybrid}  & $4.00 \times 10^{8}$	& $= 15.5 \times$	& $0.01$	& $1.52$	& $13556.1$  & $24142.9$	& $6.494 \times 10^{-1}$	& $(1249,1344)/(289,384)$ \\
\new{3D}  & $5.00 \times 10^{8}$	& $= 19.3 \times$	& $0.01$	& $2.07$	& $11968.4$  & $28074.3$	& $-$	& $(-,-)/(321,1024)$ \\
	& 	& 	&   & 	& $Ek = 1 \times 10^{-7}$	& 	&   &  \\	
QG  & $2.43 \times 10^{10}$	& $= 7.4 \times$	& $1$	& $2.97$	& $714.3$   & $115.3$	& $6.781 \times 10^{-2}$	& $(577,576)/(-,-)$ \\
Hybrid  & $5.00 \times 10^{10}$	& $= 15.1 \times$	& $1$	& $1.73$	& $1038.9$  & $580.6$	& $8.735 \times 10^{-2}$	& $(577,672)/(577,384)$ \\
QG  & $4.83 \times 10^{10}$	& $= 14.8 \times$	& $1$	& $5.74$	& $1459.0$  & $848.3$	& $1.148 \times 10^{-1}$	& $(769,768)/(-,-)$ \\
\new{Hybrid}  & $6.45 \times 10^{10}$	& $= 19.4 \times$	& $1$	& $1.81$	& $1315.6$  & $887.7$	& $9.837 \times 10^{-2}$	& $(769,768)/(769,384)$ \\
3D  & $6.50 \times 10^{10}$	& $= 19.6 \times$	& $1$	& $3.85$	& $1853.8$   & $910.8$	& $-$	& $(-,-)/(433,682)$ \\
QG  & $6.50 \times 10^{10}$	& $= 19.9 \times$	& $1$	& $8.55$	& $1879.9$  & $1362.0$	& $1.073 \times 10^{-1}$	& $(769,768)/(-,-)$ \\
Hybrid  & $ 8.00 \times 10^{10}$	& $= 24.1 \times$	& $1$	& $1.82$	& $1501.6$  & $1048.2$	& $1.1068 \times 10^{-1}$	& $(769,768)/(769,512)$ \\
QG  & $9.66 \times 10^{10}$	& $= 29.6 \times$	& $1$	& $13.91$	& $2685.4$  & $2343.9$	& $1.290 \times 10^{-1}$	& $(769,768)/(-,-)$ \\
3D  & $1.00 \times 10^{11}$	& $= 30.1 \times$	& $1$	& $7.79$	& $3325.4$   & $2798.3$	& $-$	& $(-,-)/(513,896)$ \\
Hybrid  & $1.00 \times 10^{11}$	& $= 30.1 \times$	& $1$	& $1.94$	& $1821.0$  & $1382.9$	& $1.112 \times 10^{-1}$	& $(769,512)/(769,512)$ \\
QG  & $1.93 \times 10^{11}$	& $= 59.3 \times$	& $1$	& $31.71$	& $4833.0$  & $5217.6$	& $1.711 \times 10^{-1}$	& $(769,768)/(-,-)$ \\
QG  & $2.43 \times 10^{11}$	& $= 74.5 \times$	& $1$	& $40.75$	& $5865.3$  & $6659.6$	& $1.857 \times 10^{-1}$	& $(1153,1152)/(-,-)$ \\
QG  & $3.87 \times 10^{11}$	& $= 118.6 \times$	& $1$	& $69.78$	& $8834.0$  & $10483.3$	& $2.202 \times 10^{-1}$	& $(1537,1536)/(-,-)$ \\
QG  & $4.83 \times 10^{11}$	& $= 148.2 \times$	& $1$	& $92.81$	& $10796.3$ & $12937.1$	& $2.390 \times 10^{-1}$	& $(2049,2048)/(-,-)$ \\
Hybrid  & $2.20 \times 10^{9}$	& $= 2.0 \times$	& $0.1$	& $1.16$	& $1182.5$  & $703.5$	& $1.208 \times 10^{-1}$	& $(769,768)/(513,256)$ \\
Hybrid  & $2.70 \times 10^{9}$	& $= 2.5 \times$	& $0.1$	& $1.19$	& $1517.0$  & $925.8$	& $1.259 \times 10^{-1}$	& $(769,768)/(513,256)$ \\
QG  & $4.99 \times 10^{9}$	& $= 6.0 \times$	& $0.1$	& $1.82$	& $3005.5$  & $1994.4$	& $1.434 \times 10^{-1}$	& $(769,768)/(-,-)$ \\
QG  & $1.12 \times 10^{10}$	& $= 13.6 \times$	& $0.1$	& $3.23$	& $6091.6$  & $5945.9$	& $1.704 \times 10^{-1}$	& $(769,768)/(-,-)$ \\
QG  & $2.00 \times 10^{10}$	& $= 24.1 \times$	& $0.1$	& $5.67$	& $9833.7$ & $11422.3$	& $2.066 \times 10^{-1}$	& $(769,768)/(-,-)$ \\
QG  & $2.49 \times 10^{10}$	& $= 30.1 \times$	& $0.1$	& $7.28$	& $11883.6$ & $14389.5$	& $2.254 \times 10^{-1}$	& $(1025,1024)/(-,-)$ \\
QG  & $4.00 \times 10^{10}$	& $= 48.3 \times$	& $0.1$	& $12.71$	& $17857.6$ & $23170.1$	& $2.555 \times 10^{-1}$	& $(1537,1536)/(-,-)$ \\
QG  & $2.25 \times 10^{9}$	& $= 5.8 \times$	& $0.01$	& $2.54$	& $20527.5$ & $30576.4$	& $2.840 \times 10^{-1}$	& $(1537,1536)/(-,-)$ \\
QG  & $5.62 \times 10^{9}$	& $= 14.5 \times$	& $0.01$	& $4.73$	& $43765.4$ & $65138.1$	& $4.050 \times 10^{-1}$	& $(4609,4608)/(-,-)$ \\
    & 	& 	& 	& 	& $Ek = 3 \times 10^{-8}$	& 	&   &   \\
Hybrid  & $1.07 \times 10^{11}$	& $= 6.8 \times$	& $1$	& $1.44$	& $748.5$   & $259.5$	& $4.613 \times 10^{-2}$	& $(769,768)/(769,256)$ \\
QG  & $1.20 \times 10^{11}$	& $= 7.9 \times$	& $1$	& $2.98$	& $1088.2$  & $364.3$	& $5.043 \times 10^{-2}$	& $(769,768)/(-,-)$ \\
QG  & $2.40 \times 10^{11}$	& $= 15.7 \times$	& $1$	& $6.05$	& $2144.8$  & $1667.4$	& $7.173 \times 10^{-2}$	& $(1025,1024)/(-,-)$ \\
QG  & $4.81 \times 10^{11}$	& $= 31.5 \times$	& $1$	& $13.88$	& $4052.7$  & $4268.4$	& $9.658 \times 10^{-2}$	& $(1025,1024)/(-,-)$ \\
QG  & $9.62 \times 10^{11}$	& $= 62.9 \times$	& $1$	& $33.00$	& $7528.6$  & $9158.5$	& $1.249 \times 10^{-1}$	& $(1025,1024)/(-,-)$ \\
QG  & $1.20 \times 10^{12}$	& $= 78.6 \times$	& $1$	& $42.72$	& $9179.8$  & $11560.0$	& $1.361 \times 10^{-1}$	& $(1025,1024)/(-,-)$ \\
QG  & $1.92 \times 10^{12}$	& $= 125.8 \times$	& $1$	& $71.35$	& $13560.0$ & $18168.4$	& $1.589 \times 10^{-1}$	& $(1537,1536)/(-,-)$ \\
QG  & $2.40 \times 10^{12}$	& $= 157.3 \times$	& $1$	& $90.78$	& $18102.3$ & $24309.7$	& $1.836 \times 10^{-1}$	& $(2305,2304)/(-,-)$ \\
	& 	& 	&   & 	& $Ek = 1 \times 10^{-8}$	& 	&   &  \\
QG  & $5.21 \times 10^{11}$	& $= 8.2 \times$	& $1$	& $2.92$	& $1523.8$  & $889.8$	& $3.807 \times 10^{-2}$	& $(1025,1024)/(-,-)$ \\
QG  & $1.04 \times 10^{12}$	& $= 16.5 \times$	& $1$	& $5.95$	& $3073.2$  & $2900.9$	& $5.469 \times 10^{-2}$	& $(1537,1536)/(-,-)$ \\
QG  & $2.09 \times 10^{12}$	& $= 32.9 \times$	& $1$	& $13.99$	& $5994.2$  & $7201.6$	& $7.423 \times 10^{-2}$	& $(1537,1536)/(-,-)$ \\
QG  & $4.18 \times 10^{12}$	& $= 66.0 \times$	& $1$	& $32.65$	& $11052.2$ & $15298.2$	& $9.662 \times 10^{-2}$	& $(2049,2048)/(-,-)$ \\
QG  & $4.49 \times 10^{12}$	& $= 71.0 \times$	& $1$	& $35.20$	& $12059.8$ & $16988.9$	& $9.931 \times 10^{-2}$	& $(3073,3072)/(-,-)$ \\
QG  & $8.99 \times 10^{12}$	& $= 142.0 \times$	& $1$	& $77.22$	& $22869.3$ & $34026.5$	& $1.308 \times 10^{-1}$	& $(3073,3072)/(-,-)$ \\
QG  & $5.62 \times 10^{10}$	& $= 4.1 \times$	& $0.1$	& $1.36$	& $3747.8$  & $1887.6$	& $6.280 \times 10^{-2}$	& $(1537,1536)/(-,-)$ \\
Hybrid  & $1.00 \times 10^{11}$	& $= 4.2 \times$	& $0.1$	& $1.58$	& $4358.6$  & $2942.0$	& $8.157 \times 10^{-2}$	& $(1537,1536)/(513,171)$ \\
QG  & $1.12 \times 10^{11}$	& $= 8.1 \times$	& $0.1$	& $1.83$	& $7519.1$  & $5751.9$	& $8.310 \times 10^{-2}$	& $(1537,1536)/(-,-)$ \\
QG  & $2.25 \times 10^{11}$	& $= 16.2 \times$	& $0.1$	& $2.93$	& $13375.8$ & $14941.6$	& $1.030 \times 10^{-1}$	& $(2049,2048)/(-,-)$ \\
QG  & $1.12 \times 10^{10}$	& $= 2.0 \times$	& $0.01$	& $1.33$	& $15817.8$ & $12622.1$	& $1.220 \times 10^{-1}$	& $(2049,1536)/(-,-)$ \\
QG  & $2.00 \times 10^{10}$	& $= 3.6 \times$	& $0.01$	& $1.60$	& $27367.1$ & $24838.9$	& $1.399 \times 10^{-1}$	& $(2049,2048)/(-,-)$ \\
QG  & $2.25 \times 10^{10}$	& $= 4.1 \times$	& $0.01$	& $1.96$	& $28161.1$ & $24492.0$	& $1.448 \times 10^{-1}$	& $(3073,2048)/(-,-)$ \\
QG  & $3.37 \times 10^{10}$	& $= 6.1 \times$	& $0.01$	& $2.05$	& $40509.8$ & $43890.5$	& $1.470 \times 10^{-1}$	& $(3457,3456)/(-,-)$ \\
QG  & $3.93 \times 10^{10}$	& $= 7.1 \times$	& $0.01$	& $2.24$	& $44887.0$ & $54466.6$	& $1.319 \times 10^{-1}$	& $(3457,3456)/(-,-)$ \\
QG  & $4.49 \times 10^{10}$	& $= 8.1 \times$	& $0.01$	& $2.47$	& $49801.2$ & $61299.8$	& $1.647 \times 10^{-1}$	& $(4097,4096)/(-,-)$ \\
QG  & $1.12 \times 10^{11}$	& $= 20.3 \times$	& $0.01$	& $4.43$	& $120663.5$    & $206311.7$	& $1.496 \times 10^{-1}$	& $(9217,9216)/(-,-)$ \\
	& 	& 	&   & 	& $Ek = 1 \times 10^{-9}$	& 	&   &  \\
QG  & $2.25 \times 10^{13}$	& $= 17.6 \times$	& $1$	& $5.99$	& $6366.9$  & $7671.7$	& $3.002 \times 10^{-2}$	& $(3073,2048)/(-,-)$ \\
QG  & $4.49 \times 10^{13}$	& $= 35.2 \times$	& $1$	& $11.75$	& $16231.2$ & $24728.6$	& $5.514 \times 10^{-2}$	& $(3073,3072)/(-,-)$ \\
QG  & $2.25 \times 10^{12}$	& $= 9.4 \times$	& $0.1$	& $2.20$	& $16253.8$ & $7816.3$	& $4.624 \times 10^{-2}$	& $(4609,4096)/(-,-)$ \\
	& 	& 	&   & 	& $Ek = 1 \times 10^{-10}$	& 	&   &  \\
QG  & $4.83 \times 10^{14}$	& $= 17.6 \times$	& $1$	& $5.99$	& $15563.3$ & $17876.4$	& $1.964 \times 10^{-2}$	& $(6145,6144)/(-,-)$ \\
\end{longtable}
\end{center}

\section{Numerical simulations with inhomogeneous heat flux boundary conditions}
\label{sec:Append-B-res-HFB}

\begin{center}
\begin{longtable}{rlccrrrrrc}
\caption{Summary of the numerical simulations with inhomogeneous heat flux boundary conditions computed in this study. All models have been computed with $\eta = r_i/r_o = 0.35$. $Ra_Q$ is the flux-based Rayleigh number, $Y_{(m)/(\ell,m)}$ is the mode $m$ (QG) or $(\ell,m)$ (Hybrid or 3D) of the imposed lateral flux variations ($Y_0$ or $Y_{0,0}$ indicates no lateral variations), $Q^*$ is the relative amplitude of the lateral flux variations, $Pr$ is the Prandtl number, $Nu_\Delta$ is the Nusselt number based on the temperature contrast in the shell, $Re_c$ is the convective Reynolds number, $Re_\text{zon}$ is the zonal Reynolds number, $\mathcal{L}_{u_s}$ is the typical length-scale for the cylindrical radial velocity field and $(N_s, N_m)/(N_r, \ell_{\text{max}})$ are the grid-size for the run.}
\label{tab:run_ihbc} \\
\hline
Method  & $Ra_Q  $	& $Y_{(m)/(\ell,m)}$	& $Q^*$	& $Pr$	& $Nu_\Delta$	& $Re_c$    & $Re_\text{zon}$	& $\mathcal{L}_{u_s}$	& $(N_s, N_m)/(N_r, \ell_{\text{max}})$ \\
\hline
\endfirsthead

\hline 
Method  & $Ra_Q  $	& $Y_{(m)/(l,m)}$	& $Q^*$	& $Pr$	& $Nu_\Delta$	& $Re_c$  & $Re_\text{zon}$	& $\mathcal{L}_{u_s}$	& $(N_s, N_m)/(N_r, \ell_{\text{max}})$ \\
\hline
\endhead

\hline
\multicolumn{10}{c}{Continued on next page $\ldots$} \\
\hline
\endfoot

\hline
\hline
\endlastfoot

	& 	& 	&   & 	& $Ek = 1 \times 10^{-4}$	& 	&   &   &   \\
Hybrid  & $1.00 \times 10^{7}$	& $Y_{2,2}$	& $3$	& $1$	& $1.59$	& $135.1$   & $68.3$	& $3.142 \times 10^{0}$	& $(65,96)/(65,96)$ \\
	& 	& 	&   & 	& $Ek = 1 \times 10^{-5}$	& 	&   &   &   \\
Hybrid  & $2.00 \times 10^{8}$	& $Y_{2,2}$	& $3$	& $1$	& $1.67$	& $425.4$   & $87.7$	& $4.488 \times 10^{-1}$	& $(129,128)/(129,128)$ \\
Hybrid  & $4.00 \times 10^{8}$	& $Y_{2,2}$	& $3$	& $1$	& $1.71$	& $543.1$   & $110.0$	& $6.283 \times 10^{-1}$	& $(129,128)/(129,128)$ \\
	& 	& 	&   & 	& $Ek = 1 \times 10^{-6}$	& 	&   &   &   \\
QG  & $2.00 \times 10^{9}$	& $Y_{0}$	& $-$	& $1$	& $4.20$	& $460.2$   & $119.5$	& $2.417 \times 10^{-1}$	& $(385,384)/(-,-)$ \\
Hybrid  & $2.00 \times 10^{9}$	& $Y_{2,2}$	& $3$	& $1$	& $2.11$	& $484.6$   & $195.7$	& $1.963 \times 10^{-1}$	& $(289,320)/(289,320)$ \\
QG  & $2.00 \times 10^{9}$	& $Y_{2}$	& $3$	& $1$	& $30.32$	& $144.4$   & $192.5$	& $8.491 \times 10^{-2}$	& $(385,384)/(-,-)$ \\
QG  & $4.00 \times 10^{9}$	& $Y_{0}$	& $-$	& $1$	& $4.25$	& $927.8$   & $300.8$	& $1.745 \times 10^{-1}$	& $(385,384)/(-,-)$ \\
QG  & $4.00 \times 10^{9}$	& $Y_{1}$	& $3$	& $1$	& $13.16$	& $611.9$   & $135.4$	& $1.366 \times 10^{-1}$	& $(513,512)/(-,-)$ \\
3D  & $4.00 \times 10^{9}$	& $Y_{2,2}$	& $3$	& $1$	& $4.33$	& $735.6$   & $256.8$	& $-$	& $(-,-)/(257,341)$ \\
Hybrid  & $4.00 \times 10^{9}$	& $Y_{2,2}$	& $3$	& $1$	& $1.59$	& $773.5$   & $429.4$	& $1.496 \times 10^{-1}$	& $(337,384)/(337,384)$ \\
QG  & $4.00 \times 10^{9}$	& $Y_{2}$	& $3$	& $1$	& $8.86$	& $640.3$   & $120.2$	& $1.745 \times 10^{-1}$	& $(513,512)/(-,-)$ \\
QG  & $8.00 \times 10^{9}$	& $Y_{0}$	& $-$	& $1$	& $5.88$	& $1257.9$  & $656.8$	& $2.417 \times 10^{-1}$	& $(513,512)/(-,-)$ \\
3D  & $8.00 \times 10^{9}$	& $Y_{2,2}$	& $3$	& $1$	& $7.15$	& $974.7$   & $352.4$	& $-$	& $(-,-)/(321,341)$ \\
Hybrid  & $8.00 \times 10^{9}$	& $Y_{2,2}$	& $3$	& $1$	& $2.03$	& $1414.7$  & $489.6$	& $2.618 \times 10^{-1}$	& $(385,416)/(385,416)$ \\
QG  & $8.00 \times 10^{9}$	& $Y_{2}$	& $3$	& $1$	& $13.92$	& $893.8$   & $238.4$	& $1.496 \times 10^{-1}$	& $(513,512)/(-,-)$ \\
\new{Hybrid} & $1.26 \times 10^{10}$	& $Y_{2,2}$	& $3$	& $1$	& $1.90$	& $1306.3$  & $1162.2$	& $2.417 \times 10^{-1}$	& $(417,448)/(417,448)$ \\
\new{QG}  & $2.60 \times 10^{10}$	& $Y_{2}$	& $3$	& $1$	& $21.59$	& $1456.0$  & $835.3$	& $1.745 \times 10^{-1}$	& $(513,512)/(-,-)$ \\
\new{QG}  & $3.60 \times 10^{10}$	& $Y_{0}$	& $-$	& $1$	& $10.26$	& $231.0$  & $186.2$	& $1.653 \times 10^{-1}$	& $(513,512)/(-,-)$ \\
QG  & $3.60 \times 10^{10}$	& $Y_{2}$	& $3$	& $1$	& $22.61$	& $1914.1$  & $957.0$	& $3.142 \times 10^{-1}$	& $(513,512)/(-,-)$ \\
QG  & $3.60 \times 10^{10}$	& $Y_{2}$	& $5$	& $1$	& $59.32$	& $720.5$  & $185.5$	& $1.963 \times 10^{-1}$	& $(1537,1536)/(-,-)$ \\
	& 	& 	&   & 	& $Ek = 1 \times 10^{-7}$	& 	&   &   &   \\
QG  & $1.62 \times 10^{11}$	& $Y_{0}$	& $-$	& $1$	& $5.88$	& $2736.8$  & $2336.3$	& $1.745 \times 10^{-1}$	& $(1537,1536)/(-,-)$ \\
QG  & $1.62 \times 10^{11}$	& $Y_{2}$	& $2$	& $1$	& $7.18$	& $2392.5$    & $2067.1$	& $1.257 \times 10^{-1}$	& $(1537,1536)/(-,-)$ \\
\new{QG}  & $1.62 \times 10^{11}$	& $Y_{2}$	& $5$	& $1$	& $17.85$	& $1538.2$    & $822.0$	& $8.055 \times 10^{-2}$	& $(1537,1536)/(-,-)$ \\
\new{QG}  & $3.20 \times 10^{11}$	& $Y_{0}$	& $-$	& $1$	& $10.86$	& $2699.1$  & $2280.1$	& $1.653 \times 10^{-1}$	& $(1537,1536)/(-,-)$ \\
\new{QG}  & $3.20 \times 10^{11}$	& $Y_{1}$	& $3$	& $1$	& $119.83$	& $1150.9$  & $385.8$	& $9.240 \times 10^{-2}$	& $(1537,1536)/(-,-)$ \\
\new{QG}  & $3.20 \times 10^{11}$	& $Y_{2}$	& $3$	& $1$	& $24.42$	& $1752.4$  & $1117.0$	& $1.527 \times 10^{-1}$	& $(1537,1536)/(-,-)$ \\
\end{longtable}
\end{center}

\section{Onset of convection}
\label{sec:Append-C-rac}

\begin{figure}
\centerline{
    \includegraphics[width=0.87\linewidth]{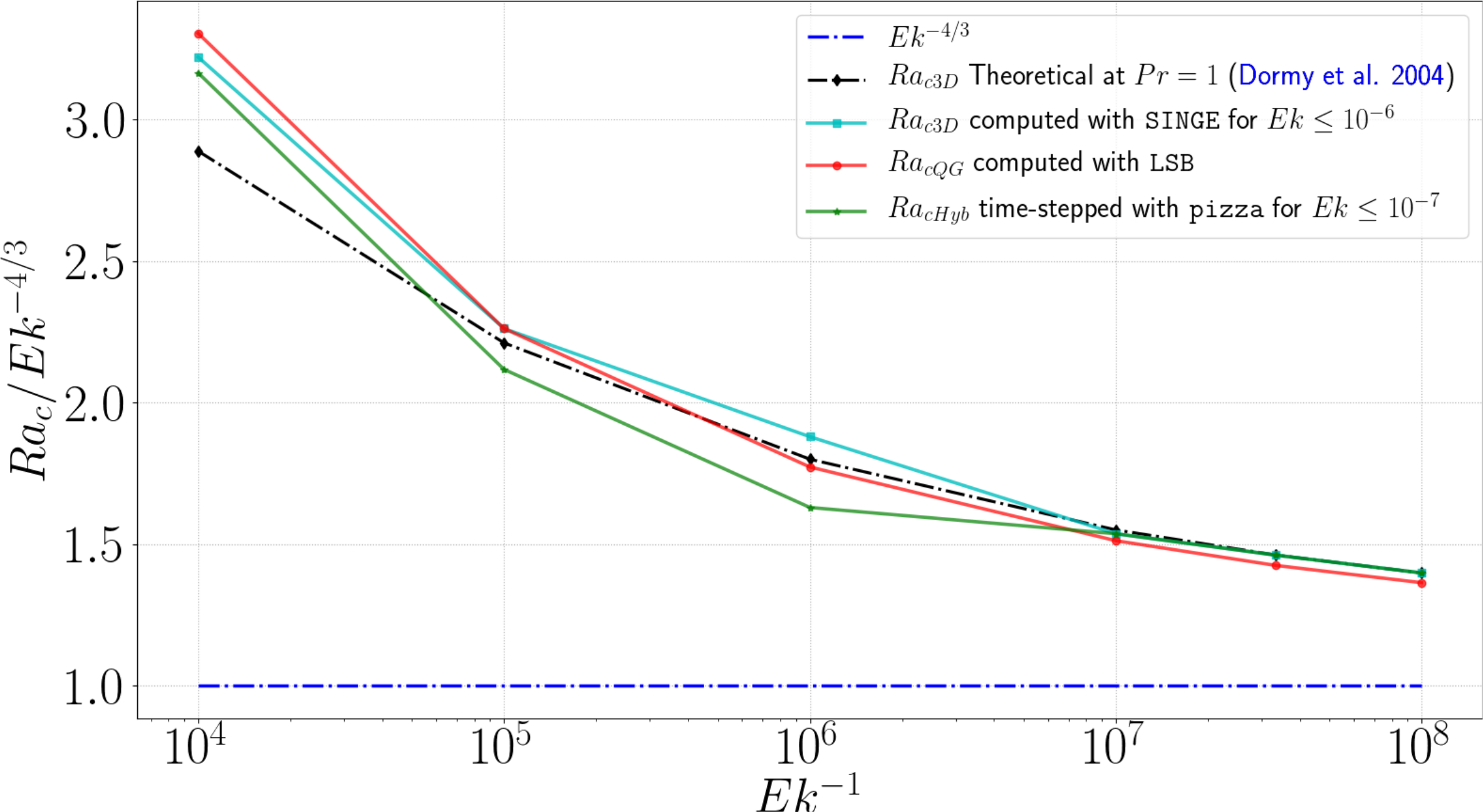}}
    \caption[Behaviour of $Ra_c$ {\it versus} $Ek^{-1}$ at $Pr=1$ for our $3$ models.]{
    \new{Evolution of the critical Rayleigh number $Ra_c$ as a function of the inverse Ekman number $Ek^{-1}$ at $Pr=1$ with fixed temperature difference across the shell, for our QG (using {\tt LSB}, in red), Hybrid (time-stepped with {\tt pizza} for $Ek \leq 10^{-6}$, in green) and 3D (using {\tt SINGE} for $Ek \leq 10^{-7}$, in cyan) models, and compared with an analytical solution in the $Ek \rightarrow 0$ limit \citep[][, in black]{dormy2004onset}.}
    }
    \label{fig:scaling_Rac-vs-Ek}
\end{figure}

We compute the onset of rotating convection using the open source software \texttt{SINGE}\footnote{\url{https://bitbucket.org/vidalje/singe}} \citep{vidal2015singe} for the 3D configuration and
the {\tt Linear Solver Builder package} \citep[{\tt LSB}, ][]{valdettaro2007convergence} for the QG setup.
In absence of a dedicated linear solver for the hybrid QG-3D configuration, we make the assumption that the critical Rayleigh number for this setup is the same as in the 3D configuration\new{, except for $3$ cases at $Ek=\lbrace 10^{-4}\,, 10^{-5}\,, 10^{-6} \rbrace$ and $Pr=1$, for which, we have determined the onset by time-integrating the nonlinear equations (\ref{eq:zonal_QG}-\ref{eq:heat_3D}-\ref{eq:momentum_QG-hyb}) using {\tt pizza} with an initial sectorial temperature perturbation and by bracketing the Rayleigh number until the critical value is attained.
Figure~\ref{fig:scaling_Rac-vs-Ek} displays the $Ra_c$ values obtained for the $3$ methods at $Pr=1$ and $10^{-4} \leq Ek \leq 10^{-8}$.}

Both {\tt SINGE} and {\tt LSB} codes solve for the generalized eigenvalue problems formed by the linearized Navier-Stokes and temperature equations. They seek normal modes of the form
\[
f(r,\theta,\phi)=\mathcal{F}(r,\theta) \exp(\lambda t + i m \phi)\,,
\]
in the 3D configuration and of the form
\[
g(s,\phi)=\mathcal{G}(s) \exp(\lambda t + i m \phi)\,,
\]
in the QG setup.
Starting at a given $Ra$, the critical Rayleigh number $Ra_c$ for a given azimuthal wavenumber $m_c$ is attained when $\Re(\lambda)=0$. Note that for the 3D configuration, it becomes numerically demanding to determine the onset of convection using a linear solver for $Ek < 10^{-7}$ for $Pr \geq 1$ and for $Ek < 10^{-6}\,; \; Pr < 1$.
For the cases with $Pr=1$, we then resort to using the asymptotic expansion derived by \cite{dormy2004onset} for spherical shells with differential heating (see their Eq.~3.25a).
For the remaining configurations the leading-order asymptotic scaling for the onset of rotating convection $Ra_c \sim Ek^{-4/3}$ is employed.
Table~\ref{tab:run_rac} summarises the critical Rayleigh numbers $Ra_c$ and azimuthal wavenumbers $m_c$ for the different setups.

\begin{center}
\begin{longtable}{crcrrc}
\caption{Summary of the critical Rayleigh numbers $Ra_c$ and critical azimuthal wavenumbers $m_c$ computed for our different setups.
We have \new{computed the onset of the hybrid method for $3$ configurations} but have otherwise assumed that hybrid QG-3D and purely 3D runs have the same $Ra_c$.}
\label{tab:run_rac} \\
\hline
$Ek$	& $Pr$	& Setup	& $Ra_c$    & $m_c$	& Computation Method \\
\hline
\endfirsthead

\hline 
$Ek$	& $Pr$	& Setup	& $Ra_c$    & $m_c$	& Computation Method \\
\hline
\endhead

\hline
\multicolumn{6}{c}{Continued on next page $\ldots$} \\
\hline
\endfoot

\hline
\hline
\endlastfoot
$1 \times 10^{-4}$	& $10$   & QG	& $1.101 \times 10^6$   & $8$	& {\tt LSB} \\
"	& $1$   & QG	& $7.121 \times 10^5$   & $8$	& {\tt LSB} \\
"	& $1$   & \new{Hybrid}	& $6.82 \times 10^5$    &$8$	& Time-integrated with {\tt pizza} \\
"	& $1$   & 3D	& $6.94 \times 10^5$    &$7$	& {\tt SINGE} \\
"	& $0.1$   & QG	& $2.911 \times 10^5$   & $6$	& {\tt LSB} \\
"	& $0.01$   & QG	& $1.663 \times 10^5$   & $4$	& {\tt LSB} \\
$1 \times 10^{-5}$	& $10$   & QG	& $1.819 \times 10^7$   & $18$	& {\tt LSB} \\
"	& $10$   & 3D $\&$ Hybrid	& $1.621 \times 10^7$   & $17$	& {\tt SINGE} \\
"	& $1$   & QG	& $1.050 \times 10^7$   & $16$	& {\tt LSB} \\
"	& $1$   & \new{Hybrid}	& $9.83 \times 10^6$    &$16$	& Time-integrated with {\tt pizza} \\
"	& $1$   & 3D	& $1.05 \times 10^7$    & $15$	& {\tt SINGE} \\
"	& $0.1$   & QG	& $3.623 \times 10^6$   & $11$	& {\tt LSB} \\
"	& $0.1$   & 3D $\&$ Hybrid	& $3.53 \times 10^6$    & $11$	& {\tt SINGE} \\
"	& $0.01$   & QG	& $1.967 \times 10^6$   & $6$	& {\tt LSB} \\
"	& $0.01$   & 3D $\&$ Hybrid	& $1.952 \times 10^6$   & $6$	& {\tt SINGE} \\
$1 \times 10^{-6}$	& $10$   & QG	& $3.322 \times 10^8$   & $38$	&{\tt LSB} \\
"	& $1$   & QG	& $1.773 \times 10^8$   & $33$	& {\tt LSB} \\
"	& $1$   & \new{Hybrid}	& $1.63 \times 10^6$    &$33$	& Time-integrated with {\tt pizza} \\
"	& $1$   & 3D	& $1.88 \times 10^8$    & $31$	& {\tt SINGE} \\
"	& $0.1$   & QG	& $5.24 \times 10^7$    & $23$	& {\tt LSB} \\
"	& $0.1$   & 3D $\&$ Hybrid	& $5.076 \times 10^7$   & $23$	& {\tt SINGE} \\
"	& $0.01$   & QG	& $2.728 \times 10^7$   & $12$	& {\tt LSB} \\
"	& $0.01$   & 3D $\&$ Hybrid	& $2.587 \times 10^7$   & $12$	& {\tt SINGE} \\
$1 \times 10^{-7}$	& $10$   & QG	& $6.446 \times 10^9$   & $80$	& {\tt LSB} \\
"	& $1$   & QG	& $3.259 \times 10^9$   & $69$	& {\tt LSB} \\
"	& $1$   & 3D $\&$ Hybrid	& $3.321 \times 10^9$   & $67$	& {\tt SINGE} \\
"	& $0.1$   & QG	& $8.287 \times 10^8$   & $47$	& {\tt LSB} \\
"	& $0.1$   & 3D $\&$ Hybrid	& $2.219 \times 10^8$   & $-$	& $Ek^{4/3}$ extrapolation from $Ek = 1 \times 10^{-6}\,; \; Pr=0.1$ \\
"	& $0.01$   & QG	& $3.868 \times 10^8$   & $26$	& {\tt LSB} \\
$3 \times 10^{-8}$	& $1$   & QG	& $1.529 \times 10^{10}$    & $102$	& {\tt LSB} \\
"	& $1$   & 3D $\&$ Hybrid	& $1.568 \times 10^{10}$    & $-$	& \cite{dormy2004onset} \\
$1 \times 10^{-8}$	& $1$   & QG	& $6.332 \times 10^{10}$    & $145$	& {\tt LSB} \\
"	& $1$   & 3D $\&$ Hybrid	& $6.492 \times 10^{10}$    & $-$	& \cite{dormy2004onset} \\
"	& $0.1$   & QG	& $1.383 \times 10^{10}$    & $98$	& {\tt LSB} \\
"	& $0.01$   & QG	& $5.548 \times 10^9$   & $56$	& {\tt LSB} \\
"	& $0.1$   & 3D $\&$ Hybrid	& $4.781 \times 10^9$   & $-$	& $Ek^{4/3}$ extrapolation from $Ek = 1 \times 10^{-6}\,; \; Pr=0.1$ \\
$1 \times 10^{-9}$	& $1$   & QG	& $1.276 \times 10^{12}$    & $310$	& {\tt LSB} \\
"	& $0.1$   & QG	& $2.396 \times 10^{11}$    & $205$	& {\tt LSB} \\
$1 \times 10^{-10}$	& $1$   & QG	& $2.749 \times 10^{13}$    & $-$	& $Ek^{4/3}$ extrapolation from $Ek = 1 \times 10^{-9}\,; \; Pr=1$ \\
\end{longtable}
\twocolumn
\end{center}

\section{Code Validation, benchmarks and scaling}
\label{sec:validation}

\subsection{Benchmark of the $z$-integral functions}
\label{sec:3D-z-func-bench}

In this section we describe numerical tests we have carried out in order to validate our hybrid QG-3D extension of {\tt pizza}.
In linking the QG and 3D parts of the code there are two key aspects that require validation: (i) the $z$-averaging (\ref{eq:z_average}) of the 3D-buoyancy to obtain its contribution on the QG-grid; (ii) the $z$-integration of the 3D-temperature field to compute the thermal wind contribution (\ref{eq:thw-u_phi3D}) on the 3D-grid.

In order to discuss the validation and the accuracy of our numerical schemes, we define a relative error estimate, $e_{\text{rel}}$, as
\begin{align}
\label{eq:3D-analytic_error-rel}
e_{\text{rel}}(f) = \displaystyle\left[ \dfrac{ \left\lbrace (f_{\text{ref}} - f)^2 \right\rbrace_s }{ \left\lbrace f_{\text{ref}}^2 \right\rbrace_s } \right]^{1/2}\,,
\end{align}
where the brackets in the above equation correspond to an average over the annulus
\begin{align}
\label{eq:annulus_average}
\left\lbrace f \right\rbrace_s \equiv \dfrac{1}{\cal V} \displaystyle\int_{0}^{2\pi}\int_{s_i}^{s_o} f h(s)\, s \,\mathrm{d}s\,\mathrm{d}\phi\,.
\end{align}

\subsubsection{Analytical benchmark of $z$-averaging}

Our computation of the $z$-averaged buoyancy term in equation (\ref{eq:momentum_QG-hyb}) relies on a simple bilinear interpolation and a basic averaging: $N_z = 2 N_s$ points are regularly distributed between $h$ and $-h$ for each cylindrical radius $s$ on the QG-grid and the $4$ nearest points of the 3D-grid are used to interpolate the value of the field at the corresponding $z$ points; then a simple summation divided by the number of points is performed along the $z$-direction at each location in the equatorial plane to obtain the average.
The bilinear interpolation is expected to have an accuracy of order $2$ and the basic summing an accuracy of order $1$.
We choose a simple and fast order $1-2$ scheme because retaining the speed and efficiency of the QG approach is our priority and it is readily parallelized.
It has also been chosen in previous studies \citep{guervilly2010thesis,guervilly2016subcritical}.

We tested our parallel implementation of this $z$-averaging scheme by comparison with an analytical solution. We considered the following 3D field
\begin{align}
\label{eq:3D-analytic_zavgfunc}
f'(r, \theta, \phi) = z \,\sin(\pi z) \,, \; \text{ with } \; z = r \, \cos \theta\,,
\end{align}
whose $z$-integral is
\begin{align}
\label{eq:3D-analytic_zavgfunc_int}
f(s, \phi) = \int_{-h}^{h} f'(r, \theta, \phi) \mathrm{d}z = \left[ \dfrac{\sin(\pi z)}{\pi^2} -\dfrac{z\cos(\pi z)}{\pi} \right]_{-h}^{h}\,.
\end{align}
Then we compute the relative error value, $e_{\text {rel}}$, obtained between our scheme and this analytical solution while testing different grid-sizes, to test the accuracy and the convergence of our scheme.
We set $N_r = N_s = N_z /2 $, $N_m = N_{\phi 3D} = 2 \times N_\theta$, and we vary the grid-size from $(N_s, N_m) = (16, 16)$ to $(N_s, N_m) = (512, 512)$.
The results are displayed as a function of the grid resolution in Figure~\ref{fig:comp-zavg-zthw_Analytic-zsinpiz_conv} (blue curve). The maximum value of the averaged function $f$ is $0.338396$.
As expected, we find that the accuracy varies approximately as $1.2$-$1.4$ times the grid resolution.

The accuracy ($\sim \left\lbrace 1 \times 10^{-1}\,, \; 4 \times 10^{-3} \right\rbrace$) and the convergence of our scheme (order $1.2 - 1.4$) validates our implementation of Eq.~(\ref{eq:momentum_QG-hyb}).

\subsubsection{Analytical benchmark of $z$-integration for thermal wind}

Turning to the implementation of the thermal wind (\ref{eq:thw-u_phi3D}), our implementation relies on a 2-neighbours interpolation scheme and performs a $z$-integration between any point, $(r, \theta)$, on the 3D grid and the half-height of the column of fluid above this point, $h$, by interpolating and summing the $N_z$ points across the $\theta$ lines directly above the position of interest.
Because only 2 points are involved in the interpolation and the integration is again a basic sum, we expect the scheme to have an accuracy of order $1$.
In order to test the accuracy of this procedure we compare against the integral of the same analytical function considered in the previous test (Eq.~\ref{eq:3D-analytic_zavgfunc_int}) but here evaluate the analytic integral between $z$ and $h$.
We again compute $e_{rel}$ compared to the analytical solution (\ref{eq:3D-analytic_zavgfunc_int}), fix $N_r = N_s = N_z /2 $, $N_m = N_{\phi 3D} = 2 \times N_\theta$ and vary the grid-size from $(N_s, N_m) = (16, 16)$ to $(N_s, N_m) = (1024, 1024)$.

\begin{figure}
\centerline{
    \includegraphics[width=0.97\linewidth]{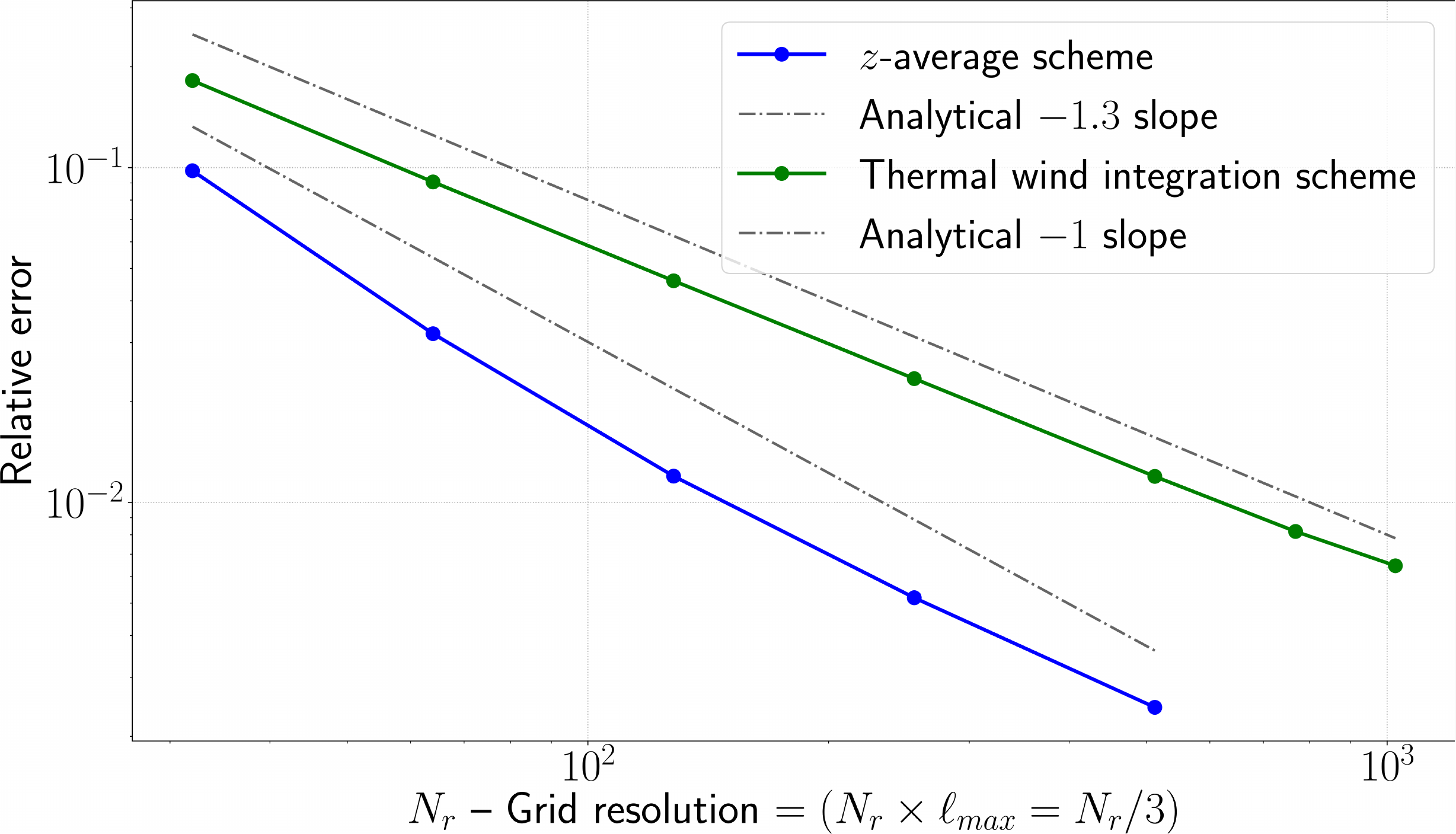}}
    \caption[Convergence of the error of a $z$-integration schemes applied to an analytical solution.]{
    Convergence of the relative error for the two $z$-integration schemes applied to an analytical field as a function of the resolution: the scheme used for the $z$-average of the buoyancy (blue curve) and the $z$-integration scheme used for the thermal wind (green curve).
    The dotted grey-curve displays the order of convergence of each scheme.
    The analytical function used for this test is $z\, \sin(\pi z)$ (Eq.~\ref{eq:3D-analytic_zavgfunc_int}).
    }
    \label{fig:comp-zavg-zthw_Analytic-zsinpiz_conv}
\end{figure}

The results are displayed in Fig.~\ref{fig:comp-zavg-zthw_Analytic-zsinpiz_conv} (green curve) and we can observe that the accuracy at the lowest resolution is $\sim 2 \times 10^{-1}$ and $7 \times 10^{-3}$ at the highest resolution. The maximum value of the integrated function $f$ is $0.462040$.
We find that the scheme for the $z$-integration of the thermal-wind contribution has the expected accuracy of order $1$ (see the slope), validating our implementation of the scheme.

\label{lastpage}

\end{document}